\documentclass[]{jfm} 
\usepackage{graphicx}
\usepackage{newtxtext}
\usepackage{newtxmath}
\usepackage{natbib}
\usepackage{hyperref}
\hypersetup{
    colorlinks = true,
    urlcolor   = blue,
    citecolor  = blue,
}

\newcommand{\RomanNumeralCaps}[1]
\usepackage{floatrow}
\usepackage[dvipsnames]{xcolor}
\usepackage{todonotes}
\usepackage{amssymb}
\usepackage{amsmath}
\usepackage{bm}
\usepackage{afterpage}
\usepackage[utf8]{inputenc}
\usepackage[position=top]{subfig}
\usepackage{caption}
\usepackage{booktabs}
\usepackage{float}
\usepackage{multirow}
\usepackage{mathtools,tikz}
\usepackage[math]{cellspace} %  increase the spacing between rows of the matrix
\DeclareRobustCommand\sampleline[1]{%
  \tikz\draw[#1] (0,0) (0,\the\dimexpr\fontdimen22\textfont2\relax)
  -- (2em,\the\dimexpr\fontdimen22\textfont2\relax);%
}
\usepackage{url}
\usepackage{soul}
\usepackage{cleveref}
\usepackage{booktabs}
\allowdisplaybreaks
\DeclareSymbolFont{yhlargesymbols}{OMX}{yhex}{m}{n} 
\DeclareMathAccent{\yhwidehat}{\mathord}{yhlargesymbols}{"62}

\title{Scale interactions and energy transfer in the turbulent  wake of a bluff body}   
\author{Jinyuan Liu\aff{1} \and Sutanu Sarkar\aff{1,}\corresp{\email{ssarkar@ucsd.edu}} }
\affiliation{\aff{1}Mechanical and Aerospace Engineering, University of California San Diego, La Jolla 92093, CA} 
\begin{document}
\floatsetup[figure]{style=plain,subcapbesideposition=top}
\maketitle

\begin{abstract}

Turbulent bluff-body wakes exemplify the coexistence of large-scale coherent structures and fine-scale turbulence -- two ends of a wide dynamical range of scales connected through the turbulent cascade. In this work, we study the multiscale dynamics in the high-Reynolds-number  wake behind a circular disk extending  to 140 disk diameters downstream. One-point and two-point statistics are first examined, including the budget and spectra of the turbulent kinetic energy (TKE). Streamwise advection is found to contribute the most to the TKE balance, while the dissipation rate does not follow the classical equilibrium scaling, $\varepsilon \nsim \mathcal{U}^3/\mathcal{L}$, where $\mathcal{U}$ and $\mathcal{L}$ are streamwise-local characteristic velocity and length scales. The largest scales are represented by the three-dimensional coherent modes extracted using spectral proper orthogonal decomposition, whereas the TKE and Reynolds shear stress spectra exhibit inertial-range scalings. A filtering-based triple decomposition further separates the fluctuations into large- and small-scale components and partitions the kinetic energy, with respective spatial transports at each scale and an inter-scale transfer in between. The inter-scale fluxes indicate a statistical forward cascade and  follow the classical $\mathcal{U}^3/\mathcal{L}$ scaling, while their radial profiles become self-similar. 
The disequilibrium between inter-scale flux and dissipation is shown to arise from  non-negligible streamwise advection at the sub-filter scale. 
Finally, the observed anti-correlation between the dissipation coefficient and the local Taylor Reynolds number, approximately $C_\varepsilon  = \varepsilon \mathcal{L}/\mathcal{U}^3 \sim Re_\lambda^{-1}$, is shown to originate from a similar correlation in the coarse-grained, locally averaged statistics. The results suggest that the instantaneous cascade-dissipation disequilibrium is intrinsic to turbulence and becomes apparent when 
 large-scale unsteadiness and  length-scale growth prevent statistical equilibrium.
% from being established. 
%\jy{(word limit 250 OK.)} 

\end{abstract}

%%%%%%%%%%%%%%%%%%%%%%%%%%%%%%%%%%%%%%%%%%%%%%%%%%%%%%
% \tableofcontents
% \input{./sections_v3/sec1.tex}
% \input{./sections_v3/sec2.tex}
% \input{./sections_v3/sec3.tex}
% \input{./sections_v3/sec4.tex}
% \input{./sections_v3/sec5.tex}
% \input{./sections_v3/sec6.tex}
% \input{./sections_v3/sec7.tex}

% \tableofcontents

% \clearpage

\section{Introduction} 

% [Paragraph 1: Wakes in nature.]

% \jy{[Intro completely re-written.]}

In the classical Kolmogorov picture of  
%homogeneous 
turbulence, at a sufficiently high $Re$, the memory of the large scales, which depends on the initial or boundary conditions, gets gradually lost during a down-scale energy cascade. As a consequence, the small scales become universal and independent of the large scales, despite feeding on their energy. In statistically homogeneous turbulence, there is an equilibrium between the averaged dissipation rate $\varepsilon$ and the inertial-range inter-scale energy transfer rate ${\it \Pi}$, both scaling as large-eddy energy divided by the large-eddy turnover time, $\mathcal{U}^2/(\mathcal{L}/\mathcal{U})=\mathcal{U}^3/\mathcal{L}$, where $\mathcal{U}$ and $\mathcal{L}$ are characteristic velocity and length scales. Such an equilibrium leads to a constancy of the dissipation coefficient $C_{\varepsilon} = \varepsilon  \mathcal{L}/ \mathcal{U}^3$ when the  Taylor Reynolds number ($Re_{\lambda}$) is sufficiently high \citep{taylor1935statistical,tennekes1972first,sreenivasan1984scaling,frisch1996turbulence,vassilicos2015dissipation}, despite the precise value of $C_{\varepsilon}$ being  flow-dependent \citep{sreenivasan1998update}.

The equilibrium dissipation scaling
 %underpins a great amount of 
is a foundation of turbulence theories and models. In turbulent axisymmetric wakes,  it leads to the classical, single power-law, self-similar evolution of the far wake \citep{tennekes1972first,george1989self}. Most notably, the wake centreline velocity deficit decays as $x^{-2/3}$ and the wake width grows as $x^{1/3}$, where $x$ is the streamwise distance, regardless of the wake generator. In the more recent theory of \cite{george1989self}, the self-similarity of dominant terms in the transport equation of the turbulent kinetic energy (TKE) was included, requiring an explicit scaling of the dissipation rate $\varepsilon$ to close an algebraic system that determines the power laws of different quantities. When the equilibrium scaling $C_{\varepsilon} ={\rm const.}$ is applied, the classical power laws are recovered. Recently, accumulated experimental and numerical evidence has identified power laws other than the classical prediction, in wakes past objects of various shapes, such as fractal-shaped two-dimensional plates \citep{nedic2013axisymmetric,dairay2015non}, a circular disk \citep{chongsiripinyo2020decay}, a sphere \citep{de2014large,saunders2022decay}, and a slender body of revolution \citep{ortiz2021high}.  It was also found that the dissipation follows non-equilibrium scalings in the wakes \citep{nedic2013axisymmetric,dairay2015non,ortiz2021high}, where instead of being a constant, $C_{\varepsilon}$ depends on the large-scale Reynolds number or the Taylor Reynolds number \citep{vassilicos2015dissipation}.  {However, there has been a scatter in the reported non-equilibrium scalings and the reason for $\varepsilon$ to be non-equilibrium has yet to be fully understood. Whether non-equilibrium originates from a non-classical cascade scaling (${\it \Pi} \nsim \mathcal{U}^3/\mathcal{L}$) or a small-scale imbalance ($\varepsilon \ne {\it \Pi}$), remains to be answered. }

The non-equilibrium behaviour of dissipation was first identified in the near field of grid turbulence by \cite{seoud2007dissipation} in the range $x_{\rm peak}<x<3x_{\rm peak}$, where $x_{\rm peak}$ is the streamwise location of the maximum TKE. They found that $C_{\varepsilon} \propto Re_{\lambda}^{-1}$ for $100<Re_{\lambda}<1000$, instead of being independent of $Re_\lambda$ as expected at high Reynolds numbers. Further experiments by \cite{valente2012universal,hearst2014decay} measured into the far field  and showed that grid turbulence returns to the classical $C_{\varepsilon}={\rm const.}$ scaling by $x/L_0=25$, where $L_0$ is the characteristic grid scale. Later, \cite{goto2015energy} found in forced isotropic turbulence  that there is a coexistence of   $C_{\varepsilon}(t) \propto Re_{\lambda}^{-1}(t)$ in the transient sense and   $C_{\varepsilon} \approx Re_{\lambda}^0$ (as $Re_{\lambda}$ becomes asymptotically large) in the time-averaged sense, uncovering that deviations from the equilibrium were concealed by the global -- in this case temporal -- average. In systems with external spatial or temporal variations, such as decaying isotropic turbulence or forced isotropic turbulence with unsteady forcing \citep{valente2014origin,goto2015energy,ghira2026non}, or the wakes of \cite{nedic2013axisymmetric,dairay2015non,portela2018turbulence,ortiz2021high}, non-equilibrium dissipation remains visible in global statistics too. 

In the present work, the relation between local non-equilibrium and  global equilibrium/non-equilibrium will be explored  in the turbulent wake past a circular disk, which possesses lateral/streamwise inhomogeneity and temporal non-stationarity at specific large-scale frequencies and length scales due to the coherent structures (CS).  In the disk wake, the most dominant CS are (i) the helical (azimuthal wavenumber $m=1$), vortex-shedding structure which oscillates at a temporal frequency of $f=0.135$ and (ii) the double-helical ($m=2$), low-frequency ($f\approx 0$) structure \citep{fuchs1979large,berger1990coherent,johansson2002proper,nidhan2020spectral}. They are referred to as the VS and the DH structures. The CS represent the largest eddies, contain a significant fraction of the TKE and the Reynolds shear stress, and are hence able to influence the evolution of the mean wake. Meanwhile, they also interact with the coexisting broadband turbulence that has its own range of space/time scales. The influence of the CS on the energy cascade, and thus the not-so-large scales, remains unclear. A central motivating question of the present work is to characterise the reach of the  CS  in  scale space -- its extent and its consequences -- in a canonical high-$Re$ flow.

In efforts to quantify the inter-scale energy transfer and explore its balance with dissipation, the generalised Kolmogorov  equation \citep{hill2002exact}, also known as the structure function approach, has been used extensively in turbulent free-shear flows, such as  jets \citep{burattini2005scale,cimarelli2021spatially}, wakes \citep{thiesset2013scale,thiesset2014dynamical,portela2017turbulence,noriega2026turbulent}, and grid turbulence \citep{hearst2014scale,valente2015energy}. Derived from the Navier--Stokes equations, the two-point generalised Kolmogorov equation has physical-space terms analogous to those of the TKE equation, while also  providing the scale-by-scale energy balance at each scale $l$. Equally popular is the filtering approach that quantifies the energy exchange between the filtered and the residual scales, which is ideal for homogeneous geometries and essential to large-eddy simulation (LES) applications, but has received less attention in inhomogeneous free-shear flows. Apart from the above two commonly used methods, there are also other approaches such as modal based inter-scale transfer, such as the ones formulated in \cite{baj2017interscale} and \cite{kinjangi2023characterization}, who applied  orthogonal modal decompositions to define the coherent and the random components of the flow, and focused on the energy transfer among specific coherent modes and between the coherent and the random motions. 

Previous work on inter-scale transfer in turbulent wakes mostly concentrated on the very-near or relatively near wake \citep{thiesset2014dynamical,portela2017turbulence,baj2017interscale,kinjangi2023characterization,noriega2026turbulent}, where the flow is highly anisotropic due to the large-scale CS and the shear production is still strong, and less on the far wake except \cite{thiesset2013scale}. They used either the structure function approach \citep{thiesset2013scale,thiesset2014dynamical,portela2017turbulence,noriega2026turbulent} or the modal decomposition approach \citep{baj2017interscale,kinjangi2023characterization}. In this work, we will focus on the fully self-similar, turbulent far wake region {\em up to 140 diameters } downstream of the wake generator and take the filtering approach, leveraging the flow's  statistical azimuthal homogeneity and temporal stationarity. The long downstream distance allows the anisotropic CS to decay and for the equilibrium, if any, to develop. The filtering, which is conducted in  Fourier space for the homogeneous directions,  constitutes a triple decomposition (TD) of the flow  into its mean and the large- and small-scale parts of the fluctuations. The energy exchange among the mean, so-defined large scale, and small scale, as well as the spatial transports in each part will be evident from the decomposed kinetic energy equations. We note that the filtering (TD) approach provides balances at the filtered and the sub-filter scales, with inter-scale fluxes transferring energy in between,  {which differs from  the scale-by-scale balance explored in the cited literature}. The scale dependence of the  transfer between the large and small parts will also be analysed by changing the filter size.

The rest of this paper is organised as follows. Section \ref{nume} introduces the numerical procedures and the database; sections \ref{onepoint} and \ref{twopoint} present the one-point and two-point statistics where  all scale ranges are characterised; section \ref{scale_decomp} further decomposes the fluctuations into large- and small-scale components and investigates the inter-scale energy transfer; section \ref{cg} identifies spatially local scales of instantaneous TKE and dissipation; section \ref{conclusion} summarises the work.

\section{Numerical simulations} \label{nume} 

The flow is governed by the incompressible Navier-Stokes equations:
% in a non-dimensional form as: 
\begin{align}
  \frac{\partial u_i}{\partial x_i} & = 0, \\ 
  \frac{\partial u_i}{\partial t} + u_j \frac{\partial u_i }{\partial x_j} & = 
  -  \frac{\partial p}{\partial x_i} 
  + \frac{1}{Re}  \frac{\partial^2 u_i}{\partial x^2_j}, \label{eqn:momemtum}     
\end{align} where $u_i$ are velocity components in Cartesian coordinates  $x_i$. The normalising units for $u_i, x_i, t, p$ are the free-stream velocity $U_{\infty}$, the wake-generator diameter normal to the stream $D$, the convective time unit $D/U_{\infty}$, and the dynamic pressure unit $\rho U^2_{\infty}$, respectively. The molecular kinematic viscosity is $\nu$, whose non-dimensionalisation is the inverse of the Reynolds number, $Re=U_{\infty}D/\nu$. Unless otherwise specified, all variables are non-dimensionalised as outlined above.

In the present axisymmetric wakes, it is advantageous to solve the governing equations in  cylindrical coordinates 
%since turbulent fluctuations confined in the wake, allowing a cluster of radial grids near the centerline and a stretching outwards to efficiently distribute the resolution. Moreover, a cylindrical representation repects the statistical axis-symmetry of the flow.
 %The corresponding coordinates and velocity components are 
 corresponding to  $(x,r,\theta)$ and $(u_x,u_r,u_{\theta})$. The governing equations are solved on a staggered grid. The spatial discretisation is second-order, finite difference and the time-stepper is a third-order, explicit, low-storage Runge--Kutta (RK3) scheme. Each RK3 sub-time-step contains a prediction and a correction step, as typical in a projection formulation. The resulting pressure Poisson equation is solved using an efficient direct solver invoking Fourier transform in the periodic azimuthal direction and a block-tridiagonal matrix inversion in the two inhomogeneous directions. The wake generator is treated with a sharp-interface, immersed boundary method (IBM) \citep{balaras2004modeling,yang2006embedded}. The capability of this numerical formulation in solving high-Reynolds number, turbulent wakes was tested and validated in \cite{chongsiripinyo2020decay,ortiz2021high}, where more details about the solver can be found.

A two-stage simulation strategy is utilised to simulate the flow into the far wake. There is a crucial numerical stiffness associated with solving the entire flow in a single domain: the boundary-layer and near-wake grid size is much smaller than the Kolmogorov length scale of the far wake.
%as the Kolmogorov scale increases downstream, there is an increasingly wider spread of scale resolution requirements in the domain. Although the spatial resolution can be stretched to maintain a roughly constant resolution with respect to the Kolmogorov scale, the time step in an explicit time-stepping scheme is constrained by the smallest grid, both for numerical stability and accuracy considerations. 
An adequate split  of the single domain in $x$-direction into two enables a much larger time step in the second (far wake) domain relative to  the near wake without sacrificing accuracy. This so-called hybrid simulation method has been validated in turbulent, bluff- and slender-body wakes  \citep{vandine2018hybrid,ortiz2021high}. 

A hybrid simulation consists of a body-inclusive (BI) portion, where the wake generator is resolved 
%with IBM, 
and a body-exclusive (BE) portion, which draws on the BI stage for  time-accurate inflow data. The present work focuses on the evolution of the far wake, hence only the BE simulation is conducted. The  inflow for the BE stage  is obtained from a large-eddy, BI simulation of the wake of a circular disk with thickness $0.01 D$ and at $Re=50\, 000$ \citep{chongsiripinyo2020decay}. The sampling location is $x/D=10$, which is sufficiently far from the wake generator to allow intense near-wake turbulence to decay and for spatial and temporal turbulent scales to grow.

In the present BE simulation, higher resolution than \cite{chongsiripinyo2020decay} is used to enable direct numerical simulation at $(\Delta x, \Delta r) \le 4 \eta$, where $\eta(x)$ is the 
%radial maximum of 
local Kolmogorov scale. The numerical parameters are listed in table \ref{table:num_paras} and {the spectra to be shown later will demonstrate the adequacy of the chosen grid}. The Taylor Reynolds number $Re_{\lambda}$ exceeds $O(100)$ throughout the entire domain until $x/D=140$, guaranteeing inertial turbulence. The outflow is an Orlanski-type convective boundary that allows free propagation of vortices out of the domain without distorting features inside. The radial boundary is sufficiently far (10 times wider than the largest wake width at $x=140$) to minimise blockage effects. The radial boundary conditions are axial ($\partial_r=0$) at the centreline ($r=0$) and Neumann in the far field ($r=L_r$). In the azimuthal direction, periodicity is enforced.  After the initial transient of two flow-throughs ($2(L_{x,o}-L_{x,i})/U_{\infty}$), statistics are sampled over a timespan of $504 D/U_{\infty}$ (approximately 68 vortex-shedding cycles).

\begin{table}[t!]
  \caption{Simulation parameters. Here $L_{x,i}$ and $L_{x,o}$ are the locations of the inlet and the outlet, while $L_r$ and $L_{\theta}$ are the radial and azimuthal domain size, respectively. There are $N_x,N_r,N_{\theta}$ grid points in the $(x,r,\theta)$ directions, respectively, and $N_t$ snapshots are uniformly sampled over a time interval $T$ to  compute the statistics. } 
  \begin{tabular}{c c c c c c c c c c}
    % \hline 
    \toprule
     $L_{x,i}$ & $L_{x,o}$ & $L_r$ & $L_{\theta}$ & $N_x$ & $N_r$ & $N_{\theta}$  & $Re$ & $N_t$ & $TU_{\infty}/D$  \\
    % \hline 
    \midrule 
      10 & 140 & 45 & $2\pi$ & 5120 & 454 & 512  & 50\,000 & 7000 & 504  \\
    % spheroid  & 10 & $2\pi$ & 9  & 120 & 532 & 256 & 8192 & 100\,000 & 172\,000  & \\
    \bottomrule
  \end{tabular} 
  \label{table:num_paras}
\end{table}

\section{One-point statistics}  \label{onepoint}

One-point statistics, which include mean flow, TKE, Reynolds stress, dissipation, and other TKE budget terms,   are reported here to provide context for the scale decompositions discussed in the later sections. These statistics demonstrate that the wake   exhibits self-similar behaviour with anomalous power-law exponents. 

Fluctuations, e.g., $ u'_i = u_i - \overline{u}_i$, 
%\begin{equation}
%  u'_i = u_i - \overline{u}_i, 
%\end{equation}
denote the departure from the ensemble average $\overline{u}_i$, where $i=x,r,\theta$ in the cylindrical coordinates. All statistics are functions of $x$ and $r$ in this flow. 
 %Pressure fluctuations are defined similarly. 
 The wake defect velocity at each $x$ is defined as $(U_{\infty} - \bar{u}(x))/U_{\infty}$ and its scale (radial maximum) is $U_d$. The momentum half width $L_d$ is defined such that $(U_{\infty} - \bar{u}(r=L_d))/U_{\infty} = U_d/2$. Similarly, the TKE scale $K_s(x)$ is defined as the radial maximum of the TKE,
  \begin{equation}
 k= \frac{1}{2}\left(\overline{u_x'^2}+\overline{u_r'^2}+\overline{u_{\theta}'^2}\right), 
\end{equation} and the TKE length scale $L_k$ is such that $k(r=L_k)=K_s/2$. The associated velocity scale of TKE (the root-mean-square velocity) is $U_k=\sqrt{2K_s/3}$.

\begin{figure}[t!]
  % \centering 
  {\includegraphics[width=0.85\linewidth]{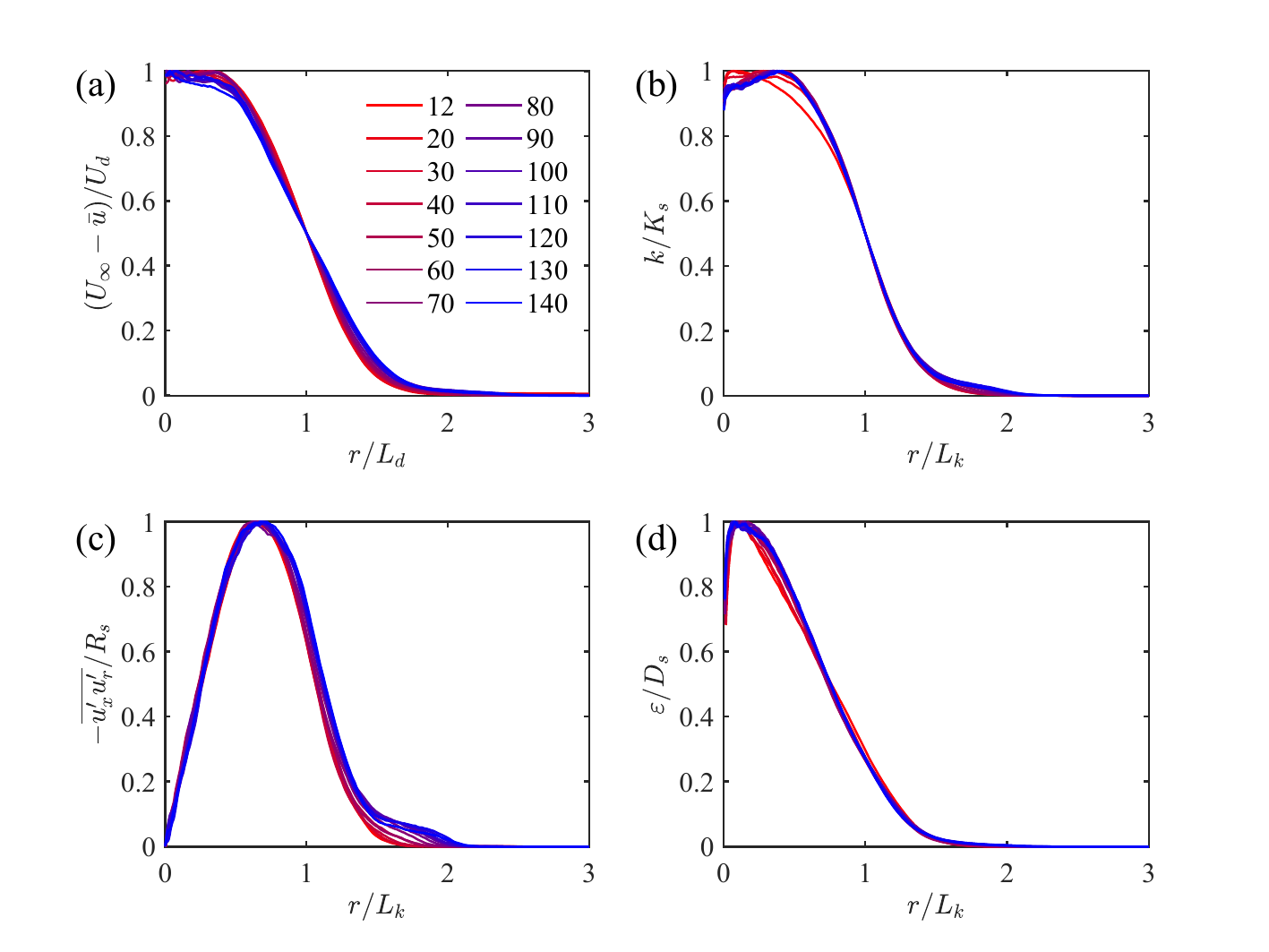}} \\
  \vspace{-0.25cm}
  {\includegraphics[width=0.85\linewidth]{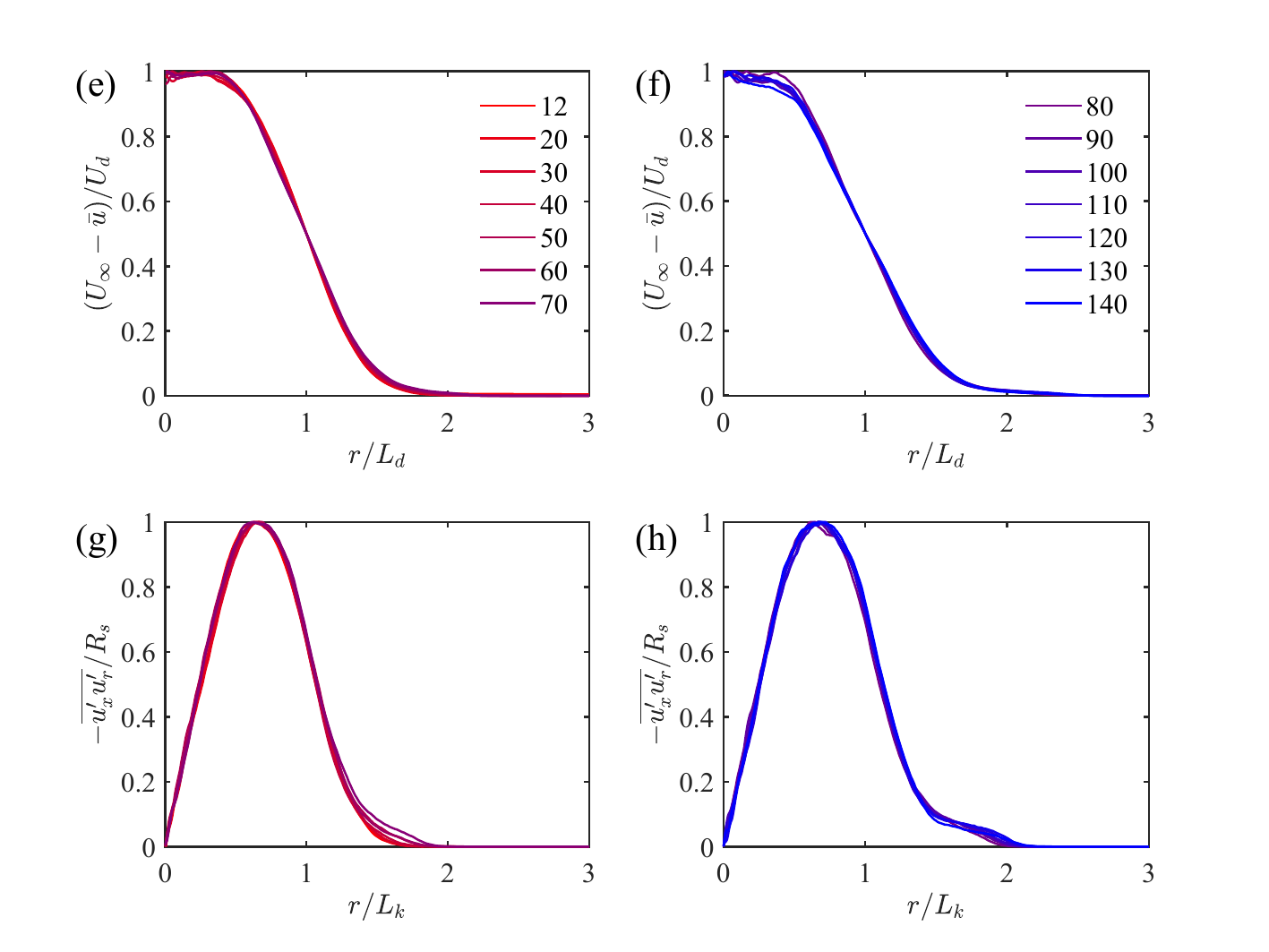}} 
  \caption{Self-similar profiles: (a-d) for the entire domain, (e,g) $12\le x \le 70$, and (f,h) $80\le x \le 140$. From red to blue, $x$ increases from 12 to 140. The length scales $L_d$ and $L_k$ denote the half-widths based on the profiles of $U_d$ and $k$, and are defined in the text. The characteristic scales $U_d$, $K_s$, $R_s$, and $D_s$ are respective radial maxima at each $x$ location.   }   
  \label{fig:selfsim_profs} 
\end{figure} 

\begin{figure}[thb]
  \centering
  {\includegraphics[width=0.9\linewidth]{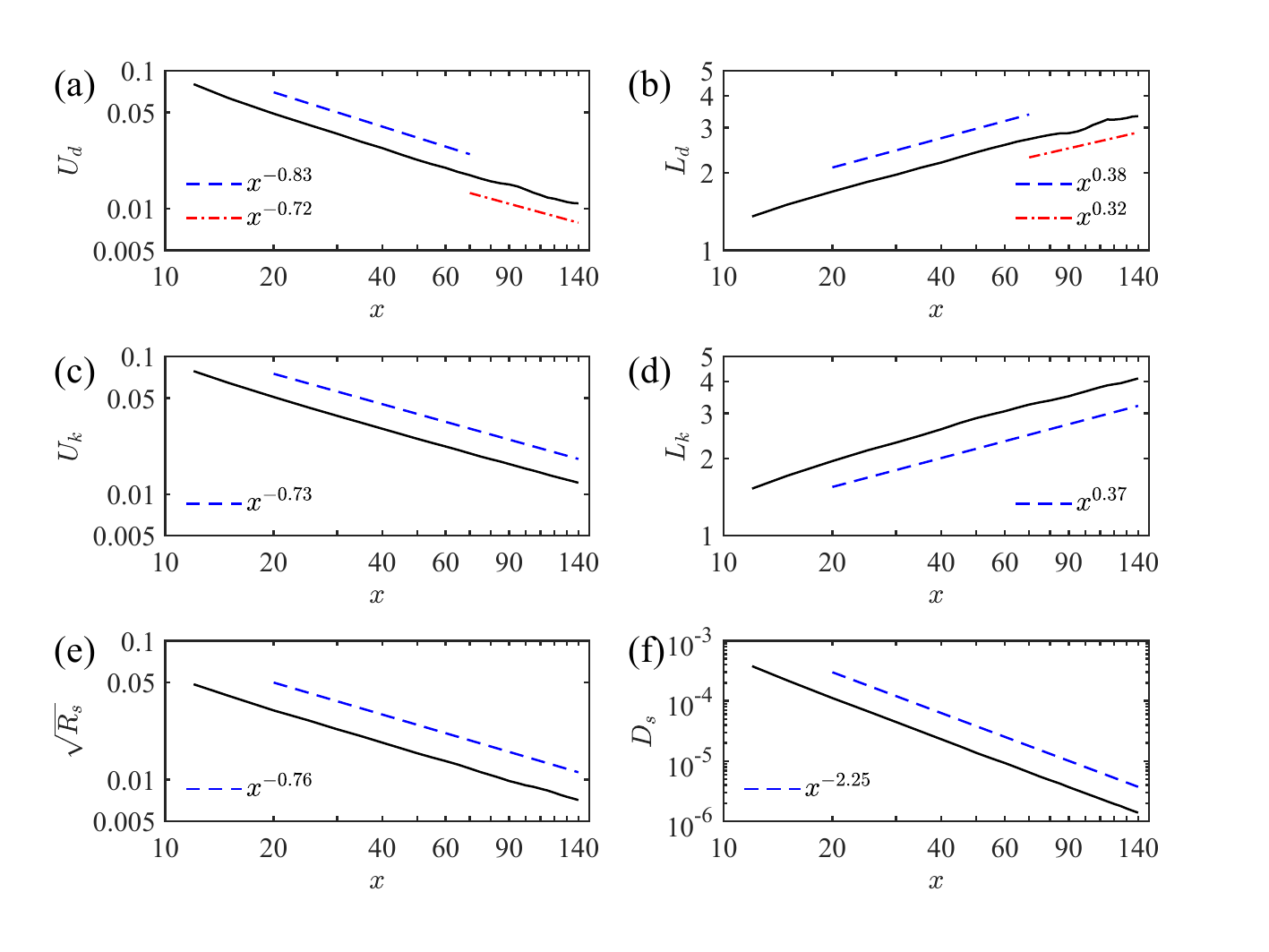}} 
  \caption{Characteristic scales: (a,b) velocity and length scales of momentum deficit, $U_d$ and $L_d$; (c,d) velocity and length scales of the TKE,  $U_k=\sqrt{2K_s/3}$ and $L_k$; (e) velocity scale of the Reynolds shear stress, $\sqrt{R_s}$, and (f) dissipation scale,  $D_s$. The dashed and dashed-dotted lines in each plot denote the best power-law fits as listed in table \ref{table:fitting} with the ends of each line denoting the fitting range. }
  \label{char_scales}
\end{figure}

The flow has reached complete self-similarity beyond $x=30$ where, as shown in figure \ref{fig:selfsim_profs}(a-d), the defect velocity, TKE, Reynolds shear stress, and TKE dissipation become self-similar. The self-similar evolution continues for another $100D$ until the  domain outlet. This complete one-point  self-similarity has also been observed in the turbulent far wakes of bluff \citep{obligado2016nonequilibrium}, slender \citep{ortiz2021high}, and fractal-shaped \citep{dairay2015non} wake generators. {In the following sections \ref{twopoint}-\ref{scale_decomp}, we will show that two-point statistics, such as spectra, correlation function, coherent structures, and inter-scale fluxes also become self-similar in the far wake.} 

Self-similarity is typically accompanied by power laws. As shown in figure \ref{char_scales}, the streamwise evolutions of the characteristic scales exhibit power laws no later than the onset of self-similarity. The power-law slopes in figure \ref{char_scales} result from the best linear fit in the range denoted by the extent of each dashed line and are summarised in table \ref{table:fitting}. The fitting quality, $R^2$, is also given and is above 0.99 for all the dashed lines shown in figure \ref{char_scales}. Actually, in terms of self-similar profile collapse and power law fits, turbulent quantities have a lower variability compared to mean-flow quantities, since the latter have the imprint of very low-frequency modes which will be shown in section \ref{spod}.

Interestingly,  $U_d$ appears to follow a two-stage evolution, with $U_d\propto x^{-0.83}$ in $20<x<70$ and $U_d \propto x^{-0.72}$ in $70<x<140$. Such a  shift  of the power-law exponent is accompanied by the slight variation in the radial profiles of the defect velocity and the Reynolds stress, as shown in figure \ref{fig:selfsim_profs}(e-h). When split, the collapse of profiles at different $x$ is better compared to (a-d). Similarly, \cite{dairay2015non} found in their simulation of a fractal plate wake in a domain up to $x=110$ that the one-point statistics are better described in two stages, $U_d \propto x^{-0.94}$ in $10<x<50$ and $U_d \propto x^{-0.86}$ in $55<x<100$,  qualitatively agreeing with  the slowdown of $U_d$ decay in the far wake observed in the present wake. They attribute the decay rate change of $U_d$ to the change of the non-equilibrium scaling of the dissipation rate. However, we observe no significant change of the power laws of $U_k,L_k$ or $D_s$ and they all show better fitting robustness than $U_d$ and $L_d$. Hence, we believe the power law change of the mean wake is due to the dynamics of the largest scales, instead of the smallest ones. As will be shown in section \ref{spod}, there is an exchange of dominance between the coherent modes, from the $m=1$ VS structure to the $m=2$ DH structure, at approximately $x=60$. The second stage $U_d \propto x^{-0.72}$ is somewhat close to the classical scaling $U_d \sim x^{-2/3}$ \citep{tennekes1972first} but it will be shown in section \ref{cg} that the dissipation does not follow an equilibrium scaling {anywhere in the present wake}. Whether the power-law exponent at larger $x$  will asymptote to -2/3  or a new exponent will emerge is left  to future work.

\begin{table}[t!]
  \caption{Power law fits. Greek letters $\alpha,\beta,\gamma$ denote the exponents of the best linear fits (in  $\log$-$\log$ space) of characteristic velocity, length, and dissipation scales. The corresponding 
  % coefficients of determination 
  $R^2$ values are also provided. } 
  \small 
  \begin{tabular}{c c c c c c c}
    % \hline 
    \toprule
    Fitting range & $U_d \propto x^{\alpha}$ & $L_d \propto x^{\beta}$ & $U_k \propto x^{\alpha}$ & $L_k \propto x^{\beta}$ & $R_s^{1/2} \propto x^{\alpha}$ & $D_s \propto x^{\gamma}$     \\
    % \hline 
    \midrule 
    \multirow{2}{*}{$20<x<140$} & $\alpha=-0.7736$ & $\beta=0.3474$ & $\alpha=-0.7318$ & $\beta=0.3744$ & $\alpha=-0.7648$ & $\gamma=-2.2526$ \\
                                & $R^2=0.9977$    & $R^2=0.9950$   & $R^2=0.9997$     & $R^2=0.9973$   & $R^2=0.9997$ & $R^2=0.9999$ \\ 
                               \midrule 
    \multirow{2}{*}{$20<x<70$}  & $\alpha=-0.8262$ & $\beta=0.3806$ & $\alpha=-0.7487$ & $\beta=0.4073$ & $\alpha=-0.7691$ & $\gamma=-2.2719$ \\
                                & $R^2=0.9997$    & $R^2=0.9996$   & $R^2=0.9998$     & $R^2=0.9991$   & $R^2=0.9999$ & $R^2=0.9999$ \\ 
                               \midrule 
    \multirow{2}{*}{$70<x<140$} & $\alpha=-0.7160$ & $\beta=0.3235$ & $\alpha=-0.7089$ & $\beta=0.3459$ & $\alpha=-0.7368$ & $\gamma=-2.2213$ \\
                                & $R^2=0.9916$    & $R^2=0.9705$   & $R^2=0.9999$     & $R^2=0.9973$   & $R^2=0.9982$ & $R^2=0.9999$ \\ 
    \bottomrule
  \end{tabular} 
  \label{table:fitting}
\end{table} 

The TKE transport equation is
 \begin{equation}
  0  = \underbrace{ - \left( \frac{\partial k}{\partial t}+ \bar{u}_i \frac{\partial k}{\partial x_i} \right)}_\text{$A$} 
   \underbrace{- \frac{\partial}{\partial x_i} \left(  \overline{p'u'_i}  + \overline{k' u'_i}  - \frac{1}{Re} \frac{\partial k}{\partial x_i} \right) }_\text{$T=T_p+T_t+T_v$} 
  \underbrace{-\overline{u'_i u'_j} \frac{\partial \bar{u}_i}{\partial x_j}}_\text{$P$} 
   \underbrace{-\frac{1}{Re} \overline{\frac{\partial u'_i}{\partial x_j} \frac{\partial u'_i}{\partial x_j}}}_\text{-$\varepsilon$} \label{tke}
\end{equation} 
where $A,T,P,\varepsilon$ are advection, transport, production, and dissipation, respectively, while the transport is split into pressure (p), turbulent (t), and viscous (v) components. Here $k'= u'_i u'_i/2$ denotes  the instantaneous perturbation energy.  
%T. In a spatially evolving wake (present), TKE is independent of time and $A=\bar{u}_j \partial_j k$, whereas in a temporally evolving wake (streamwise periodic), $A = \partial_t k$. 
Full expressions used to compute  each term in the cylindrical coordinates  are provided in Appendix \ref{tke_full}.

\begin{figure}[t!]
  \centering 
  \captionsetup[subfloat]{farskip=-2pt,captionskip=0pt}
  \subfloat{{\includegraphics[width=0.5\linewidth]{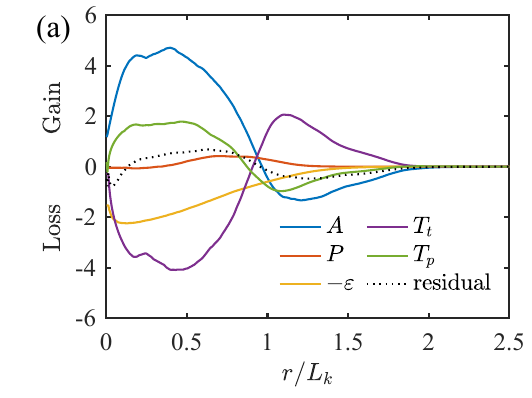}}}% 
  \subfloat{{\includegraphics[width=0.5\linewidth]{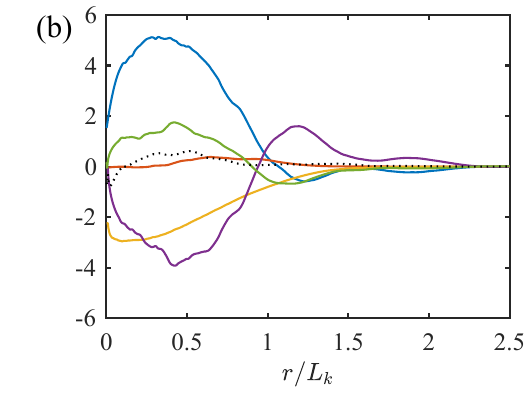}}}% 
  \caption{TKE balance in the wake:  (a) $x=30$, (b) $x=90$. The vertical axis is scaled with the local $U_k^3/L_k$ and the horizontal axis with the local $L_k$. } 
  \label{fig:tke_bud} 
\end{figure} 

TKE budget terms are shown in figure \ref{fig:tke_bud} at $x=30$ and 90. All terms agree qualitatively well with the experimental measurement in an axisymmetric wake at $x/D=100$ \citep{uberoi1970turbulent}. Advection is an energy source in the core of the wake, while it is negative at the edge of the wake. Production is small in the far wake compared to other terms and is almost negligible. 
%This agrees the fact the shear parameter indicates weak shear effects (figure \ref{fig:scale-sepa}c-d).
 Turbulent transport $T_t$ is negative in the core of the wake ($r<L_k$) and positive outside, indicating an outward transport, and has  larger magnitude and opposite sign relative to the pressure transport $T_p$. 
 %Meanwhile, the magnitude of $T_t$ is larger than $T_p$ despite the opposite sign.
  The location $r=L_k$ roughly divides the wake into a fully turbulent wake core and the intermittently turbulent wake edge, beyond which $\varepsilon$ is small. 

The dominant balance throughout the far wake is among $A,T_t, T_p$, and $\varepsilon$.
%, and there is no change of dominant balance in the wake or further simplification possible. 
The diminishing magnitude of production also means it need not be included in a theory of self-similar far wakes. In terms of area-integrated contributions, spatial transports almost vanish, leaving  $A$ balancing $\varepsilon$. The turbulent transport $T_t$ has vanishing spatial integral but, importantly, it participates in inter-space and inter-scale transfer, as per  section \ref{scale_decomp}. One observes that in figure \ref{fig:tke_bud}(a-b), the other budget terms roughly scale with $U_k^3/L_k$, but $\varepsilon$ clearly does not. {The scaling for $\varepsilon$ is left for sections \ref{scale_decomp}-\ref{cg}.} A self-similar profile collapse of terms other than $\varepsilon$ has also been attempted, but there is no clear evidence that they have reached self-similarity in the wake (not shown), unlike what was assumed in \cite{george1989self}.

\section{The scale space: spectra, correlation functions, and coherent modes}   \label{twopoint}

We proceed with the scale-space analysis as follows.
First, the energy spectra across  Fourier modes ($m$ and $f$) and the two-point, single-time  correlation functions are analysed at various locations to investigate the range of scales and  energy density distribution.  CS are defined as dominant flow modes that are more energetic than other modes and  contain a significant portion of the TKE.
%and have a strong influence on the flow evolution. 
Following the Fourier decomposition, a spectral proper orthogonal decomposition (SPOD) expands the fluctuations in all spatial ($x,r,\theta$) and temporal ($t$) directions with orthonormal bases {that are ranked according to the energy content.} The most dominant structures are extracted in terms of  3-D global SPOD  modes. In the next section (section \ref{scale_decomp}), a triple decomposition based on spectral filtering is used to isolate the {ensemble mean, the coherent, and the residual components of the fluctuations.} This approach enables the analysis of individual  energetics {as well as the energy exchange between scale ranges.}

\subsection{Spectra and two-point correlation functions}  

This section's analysis of spectra and two-point correlations will show that,  for the entire streamwise domain up to $x = 140$, the wake remains at reasonably high local $Re$, exhibiting spectra with inertial-range Kolmogorov power-law scaling, scale-dependent anisotropy that decreases with wavenumber, and a reasonable scale separation between integral  and dissipation length scales. 
  
The fluctuations are transformed to Fourier space in the statistically homogeneous directions $\theta$ and $t$. Due to strict periodicity of $2\pi$ in the  azimuthal direction, the Fourier transform yields integer  wavenumbers $m$. 
% Assume the azimuthal Fourier expansion to be \begin{equation}
  % u'_i(x,r,\theta) = \sum \hat{u}_i(x,r,m) e^{{\rm i} m \theta}.  
% \end{equation}
In the temporal direction,
% although the flow is statistically stationary, the signal in any arbitrary interval is not periodic. In order to perform a temporal Fourier transform, 
the time series is truncated into overlapped blocks, and a Hamming window is applied on each block to enforce periodicity, following the standard Welch procedure \citep{welch1967use}. An overlap ratio of $50\%$ is used, resulting in $N_{blk}=27$ independent Fourier realisations (blocks) with $N_{FFT}=512$ snapshots in each block and an overlap of $N_{ovlp}=256$ between two sequential blocks. Each block is counted as an independent Fourier realisation and an ensemble average is taken over all blocks. The thus-obtained temporal frequency ($f$) is  numerically equal to the Strouhal number, $St=fD/U_{\infty}$  since $D$ and $U_{\infty}$ were  used for non-dimensionalisation of the examined data. 

The complex Fourier modes for each velocity component  are \begin{equation}
  \hat{u}_i(x,r,m,f) = \mathcal{F}(u'_i(x,r,\theta,t)), 
\end{equation} where $\mathcal{F}(\cdot)$ denotes the Fourier transform, and the spectral density of TKE is \begin{equation}
  \Phi_k(x,r,m,f) = \frac{\frac{1}{2}(\hat{u}^{}_x \hat{u}^*_x + \hat{u}^{}_r \hat{u}^*_r + \hat{u}^{}_\theta \hat{u}^*_\theta)}{\Delta m \Delta f} , 
\end{equation} where the asterisk denotes complex conjugation. The  cospectral density of the Reynolds shear stress is defined similarly as \begin{equation}
  \Phi_{x r}(x,r,m,f) = \frac{ \Re\{-\hat{u}^{}_x \hat{u}^*_r\}}{\Delta m \Delta f} . 
\end{equation} 

The one-dimensional spectra $E_k(m; x, r)$ and $E_k(f; x, r)$ result from integrating $\Phi_k$ in the other homogeneous direction. The integration of $\Phi_k$ over the entire $(m,f)$ space recovers the one-point TKE ($k(x,r)$) after correcting for the windowing effect by multiplying the energy correction factor specific to the Hamming window. This agreement has been verified numerically.

\begin{figure}[thb]
  \centering 
  \captionsetup[subfloat]{farskip=-2pt,captionskip=0pt}
  \subfloat{{\includegraphics[width=0.5\linewidth]{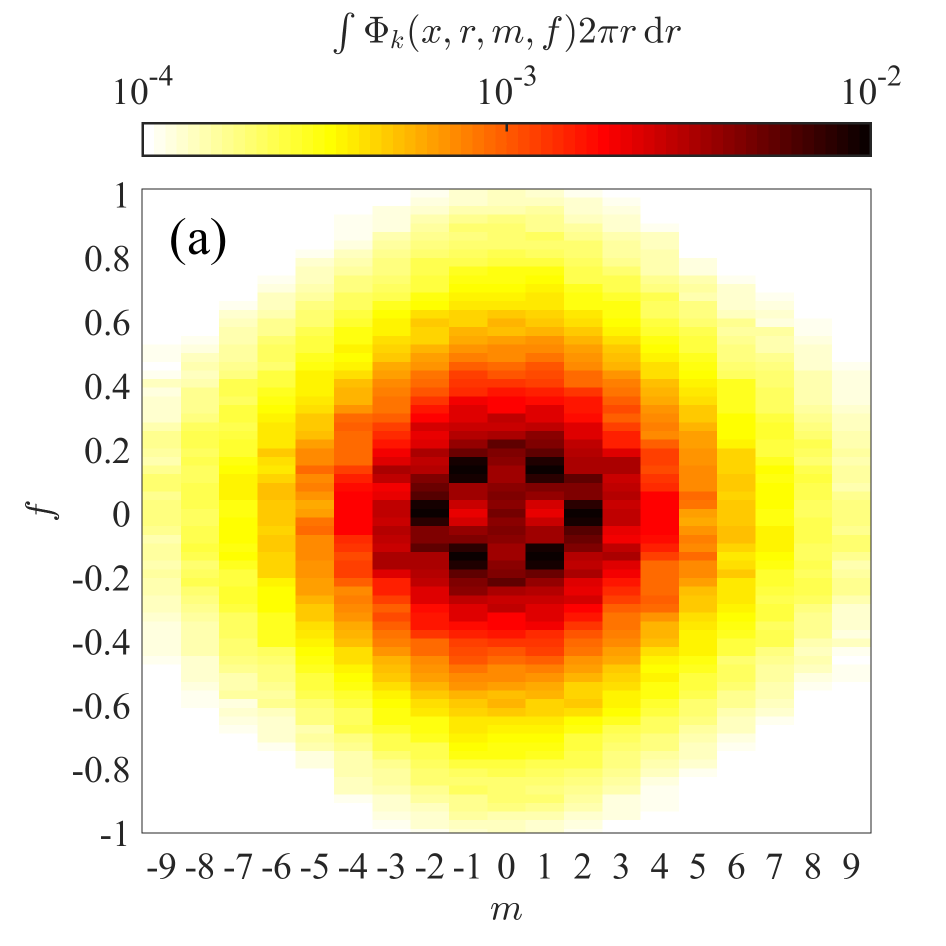}}}% 
  \subfloat{{\includegraphics[width=0.5\linewidth]{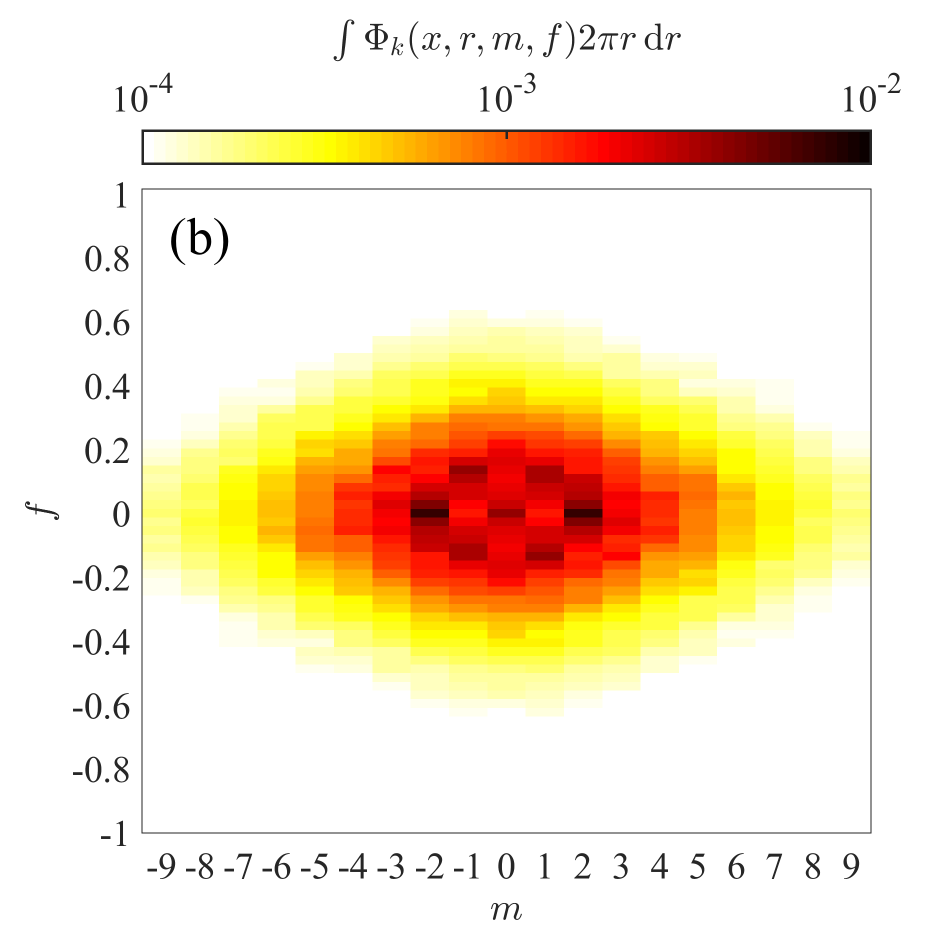}}}% 
  \caption{Radially integrated TKE spectra: (a) $x=30$ and (b) $x=90$. } 
  \label{fig:spec_2d} 
\end{figure}

The radial integral of $\Phi_k(m,f; x, r)$ at a given $x$ is a function of $(m,f)$. Figure \ref{fig:spec_2d}(a,b) shows these 2-D TKE spectra at $x=30$ and 90, respectively. It can be seen that the spectra possess Hermitian symmetry. The most dominant modes at $x=30$ are $(|m|,|f|)=(1,0.135)$ and $(|m|,|f|)=(2,0)$. The latter results from nonlinear self-interaction of the former \citep{nekkanti2023large}, which has already taken over the dominance at $x=90$, as shown in figure \ref{fig:spec_2d}(b).  These two Fourier modes at $(m,f)=(1,0.135)$ and $(2,0)$ are the most dominant ones and they will be further orthogonally expanded into spatial basis functions using SPOD in section \ref{spod}.

\begin{figure}[t!]
  \centering 
  \captionsetup[subfloat]{farskip=-2pt,captionskip=0pt}
  \subfloat{{\includegraphics[width=0.475\linewidth]{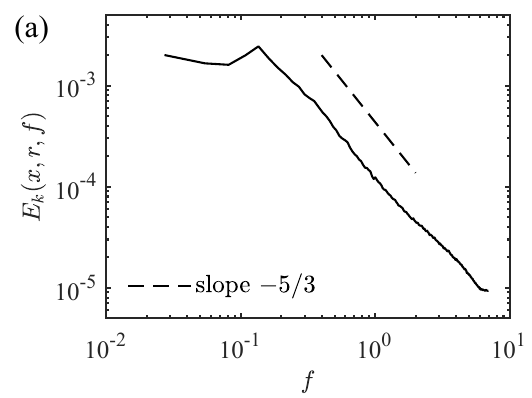}}}% 
  \subfloat{{\includegraphics[width=0.475\linewidth]{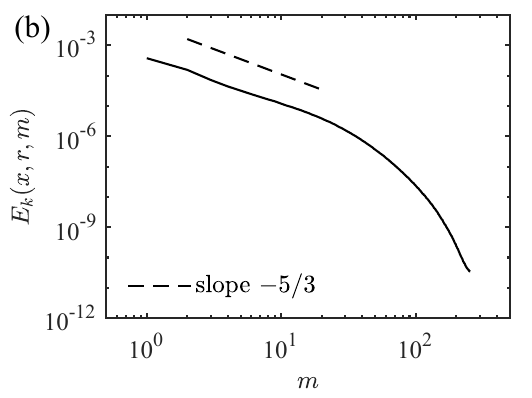}}}% 
  \caption{One-dimensional spectra of the TKE: (a) temporal spectrum at $x=30,r=1.63$, and (b) azimuthal spectrum at $x=30, r=0.60$. At $x = 30$, the outer scale is $L_k\approx 1.80\approx 630 \eta$, for reference.   } 
  \label{fig:spec_m_f} 
\end{figure}

\begin{figure}[htb]
  \centering 
  \captionsetup[subfloat]{farskip=-2pt,captionskip=0pt}
  \subfloat{{\includegraphics[width=0.475\linewidth]{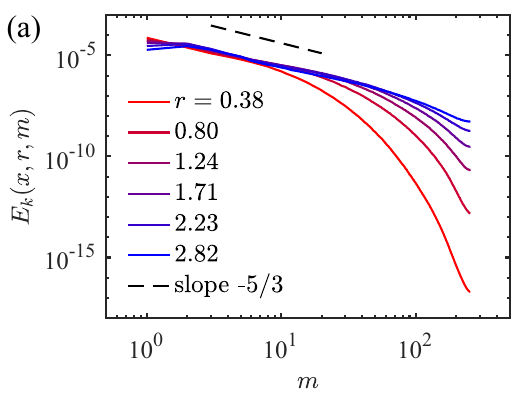}}}% 
  \subfloat{{\includegraphics[width=0.475\linewidth]{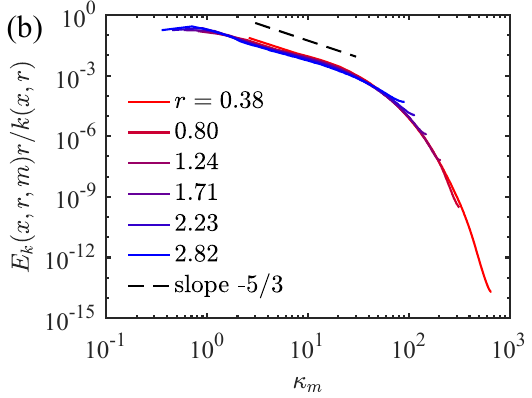}}}% 
  \\ 
  \subfloat{{\includegraphics[width=0.475\linewidth]{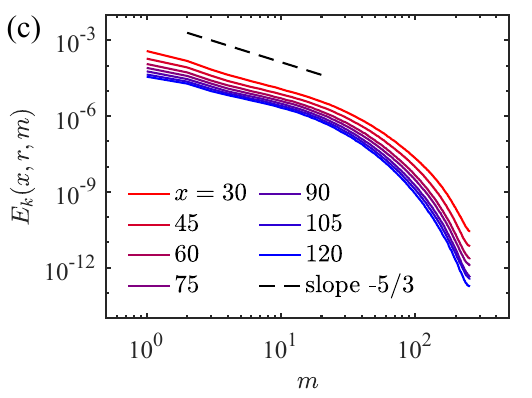}}}% 
  \subfloat{{\includegraphics[width=0.475\linewidth]{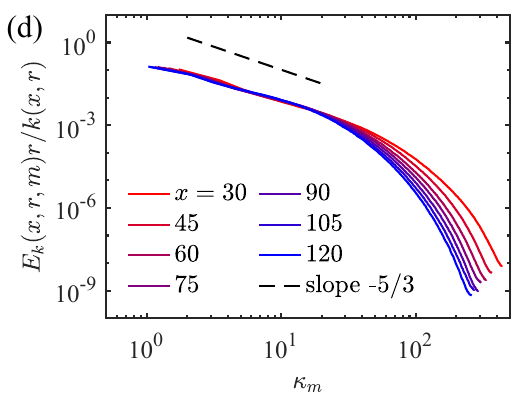}}}% 
  \caption{Azimuthal spectra: (a,b) at $x=90$ and various $r$ values below $L_k(x=90)=3.50$, and (c,d) at various $x$ and $r=L_k(x)/4$ where $L_k(x)$ is the local TKE length scale. (b,d) are the same spectra as (a,c) but as functions of the arc wavenumber $\kappa_m=m/r$ and scaled by the local TKE $k(x,r)$.} 
  \label{fig:spec_azim} 
\end{figure} 

\begin{figure}[htb]
  \centering 
  \captionsetup[subfloat]{farskip=-2pt,captionskip=0pt}
  \subfloat{{\includegraphics[width=0.475\linewidth]{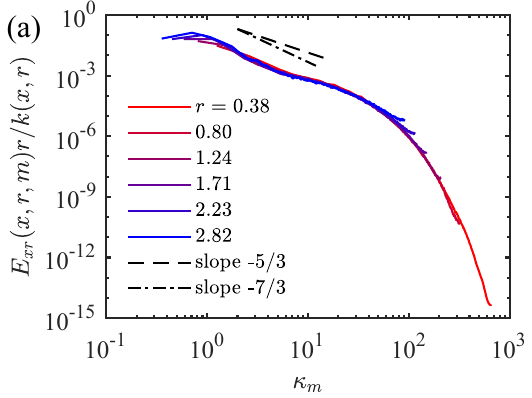}}}% 
  \subfloat{{\includegraphics[width=0.475\linewidth]{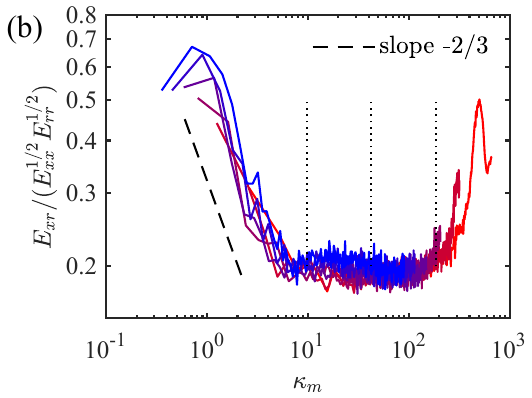}}}% 
  \caption{The Reynolds shear stress cospectra (a) and the cross-correlation coefficient spectra (b) at $x=90$.  
  %Similar results obtained at $60$ and $120$. 
  The vertical dotted lines in (b), from left to right, denote wavenumbers corresponding to $L_{12}$, $\lambda$ and $5\eta$ at $x=90$. %(c-d) The dissipation spectra and the compensated spectra at $r=L_k/4$ for various $x$. The vertical dotted line in (c) marks $40\eta$.
  \label{fig:cospec_dspec} 
  }
  % The peaks range from $m=16-22$, $k_m=35-40 \eta$ among different $x$ locations being plotted. 
\end{figure}

Figure \ref{fig:spec_m_f}(a,b) shows  1-D temporal and azimuthal spectra at a point in the core of the wake at $x=30$. Both spectra exhibit  a Kolmogorov  $-5/3$ inertial range that spans nearly a decade. Figure \ref{fig:spec_azim} further  confirms that the inertial range exists in different regions of the present wake, despite the TKE decay downstream. Notably, the spectra collapse well when the azimuthal wavenumber is scaled as $\kappa_m=m/r$, instead of $m$ alone.  The  integer wavenumber $m$ follows the symmetry of the azimuthal coordinate, comes straight from the Fourier transform, and offers the advantage of  a convenient modal label {when the large scales are concerned}. However, 
%it is dimensionless and due to the specialty of the cylindrical coordinate system,
the wavelength corresponding to a fixed $m$ is $r \Delta \theta = 2\pi r/m$, where $r$ is the radial distance so that the same value of $m$ corresponds to a different azimuthal length at different $r$. {This mismatch between the wavenumber and the physical length scale is more severe at large $r$. }
 %On the other hand, azimuthal ($\theta$) derivatives always appear as $(1/r)\partial_\theta$ in the Navier-Stokes and its Fourier counterpart is proportional to $m/r$. 
Thus,  $\kappa_m = m/r $ that  has the dimension of a wavenumber, by   removing the $r$-dependent stretching of the azimuthal arc length, is advantageous in {representing  fluctuations at a fixed length scale}.  
%In the analysis of large-scale coherent structure in Section \ref{spod}, the decomposition into $m$ is more convenient since it reflects the circular symmetry. 

The collapse shown in figure \ref{fig:spec_azim}(b) suggests that turbulence at different radial locations in the turbulent wake core follows very similar spectral density distributions but is just scaled in amplitude by  different local TKE magnitudes, $k(x,r)$. The spectra at different downstream locations are shown in figure \ref{fig:spec_azim}(c-d). When normalised with the local $k$ and expressed in terms of $\kappa_m$,   they also collapse well in the energy-containing and the inertial ranges, across almost the entire wake. 
%and only differ in the dissipation range in which the spectra differ from a $-5/3$ slope. 
The shortening of the inertial range at large $x$  reflects a decay of $Re_{\lambda}$, whose $x$-dependence is shown in figure \ref{fig:scale-sepa}. Note that the TKE scale $K_s$ decays by almost a decade from $x=30$ to 120. The radial location reported in figure \ref{fig:spec_azim}(c-d) is $L_k/4$  for each $x$ to guarantee a fully turbulent wake location. Similar results are obtained at $L_k/3$ and $L_k/2$. 

Figure \ref{fig:cospec_dspec}(a) shows the Reynolds shear stress cospectrum $E_{xr}$, which is the scale-by-scale contribution to $-\overline{u'_x u'_r}$, at various $r$ at $x = 90$. The cospectra also collapse well, when expressed as functions of $\kappa_m$ and scaled by the local $k(x,r)$. There is an inertial range in the shear spectrum as well, wherein $E_{xr}$ follows a $-7/3$ scaling on dimensional grounds, with the assumption that the shear spectrum scales linearly with the shear \citep{lumley1967similarity}. 
Such a $-7/3$ scaling of $E_{xr}$ was also found in a turbulent boundary layer \citep{saddoughi1994local}. 
According to the high-$Re$ {\em local isotropy} expectation, at sufficiently small scale or high wavenumber, the Reynolds stress tensor is close to isotropic. This expectation is evaluated here for the Reynolds shear stress by the correlation coefficient spectrum, $E_{xr}/(E_{xx}^{1/2} E_{rr}^{1/2})$, where $E_{xx}$ and $E_{rr}$ are the 1-D spectra of the streamwise and radial TKE.  Figure \ref{fig:cospec_dspec}(b) shows that the correlation coefficient is around  0.7  at low $\kappa_m$, indicating high large-scale anisotropy, and it decreases  as $\kappa_m^{-2/3}$ (consistent with theoretical power laws of $E_k$ and $E_{xr}$ in the inertial range), reaches a plateau of approximately 0.2 at $\kappa_m \approx 10$ and stays so until $\kappa_m \approx 100$ before increasing again at very high $\kappa_m$.
The plateau of  0.2 is consistently obtained, e.g.,  at  $x=30,60,90$ and 120. A  plateau of the cross-correlation spectrum  following a $-2/3$ power law decay was also found in homogeneous shear flow by \cite{shen2000anisotropy}, but at a  lower level of 0.06. In contrast, in the present wake, the cross-correlation coefficient never decays to zero --  quasi-isotropy rather than isotropy  is reached. {This is due to the effect of small-scale advection, which will be shown non-negligible in section \ref{small_advc}, that counteracts the destruction of the Reynolds shear stress by the pressure--strain correlation.}

{The scale-dependent anisotropy in terms of the normal stresses, $E_{rr}/E_{xx}$ and $E_{\theta \theta}/E_{xx}$, is also examined (not shown). Both $E_{rr}/E_{xx}$ and $E_{\theta \theta}/E_{xx}$ increase with $\kappa_m$  taking values from 0.5 to approximately 1 during the -2/3 decay of the shear stress anisotropy $E_{xr}/{E_{xx}^{1/2} E_{rr}^{1/2}}$, plateau around 1 as the cross-correlation reaches 0.2, and decrease as the small-scale anisotropy emerges. The streamwise TKE component is overall the dominant component at most wavenumbers. }

The wavenumbers corresponding to $L_{12},\lambda,$ and $5\eta$ are denoted by vertical dotted lines in figure \ref{fig:cospec_dspec}(b). It can be seen that eddies whose sizes are below $L_{12}$ contain very little Reynolds shear stress, which roughly marks the transition from anisotropic large-scale structures to quasi-isotropic inertial-range eddies. The effects of the large-scale structures do not penetrate beyond $L_{12}$.  The Taylor microscale lies in the middle of the quasi-isotropic range.  The anisotropy rises at around $5\eta$, which is approximately the smaller dimension of anisotropic dissipative structures, such as the radius of vortex tubes or the thickness of rate-of-strain sheets.
 % [+refs of small-scale anisotropy]
{The dissipation spectrum (not shown) peaks at a wavelength of approximately $40 \eta$ and has an exponential tail, such that $\kappa^{5/3}E_k \propto \exp(-\beta \kappa \eta)$ with $\beta=6$. }

\begin{figure}[thb]
  \centering 
  \captionsetup[subfloat]{farskip=-2pt,captionskip=0pt}
  \subfloat{{\includegraphics[width=0.475\linewidth]{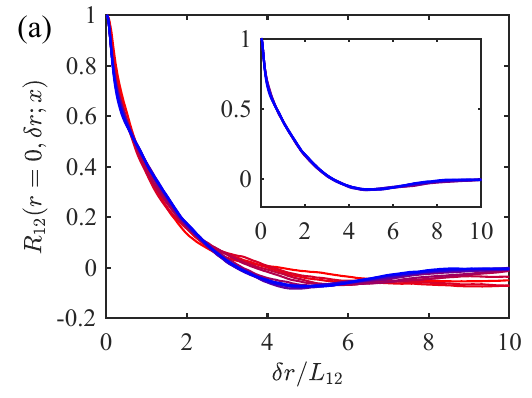}}}% 
  \subfloat{{\includegraphics[width=0.475\linewidth]{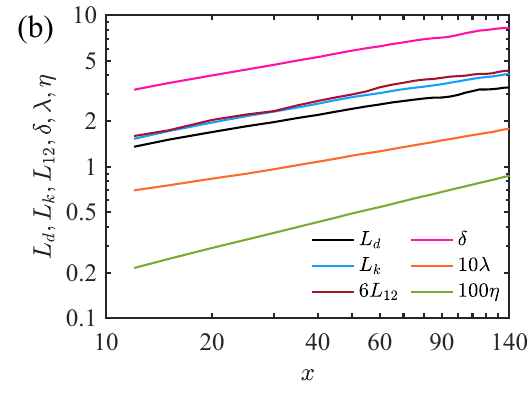}}}% 
  \caption{ (a) The transverse correlation function, $R_{12}$, at the centreline ($r=0$) for $12<x<140$. The colour code is the same as in figure \ref{fig:selfsim_profs}. The inset shows only $70<x<140$.  (b) Evolution of the length scales in the wake. The characteristic length scales $L_d,L_k,L_{12}$ and $\delta$ are as defined in the text. The Taylor and Kolmogorov length scales, $\lambda = U_k (15\nu/D_s)^{1/2}$ and $\eta =(\nu^3/D_s)^{1/4}$ are defined based on the radial maximum of dissipation, $D_s(x)$. }   
  \label{fig:lengths} 
\end{figure}

In homogeneous directions, wavenumber spectra are available to identify scales. In the inhomogeneous direction $r$, we use the transverse two-point correlation to identify the integral length scale: 
\begin{equation}
  R_{12}(r,\delta r;x) = \frac{\overline{u'_x(r;x) u'_x(r+\delta r; x)}}{\sqrt{\overline{u'^2_x(r;x) }} \sqrt{\overline{u'^2_x(r+\delta r;x) }}},  
\end{equation} where the transverse integral length scale is defined as \begin{equation}
  L_{12}(x) = \int_{0}^{r_0} R_{12}(r=0,\delta r; x)\,{\rm d}(\delta r)  \, , 
\end{equation} and $r_0$ is the first zero-crossing of $R_{12}$ similar to \cite{burattini2005normalized,kankanwadi2020turbulent}. The centreline transverse correlation functions at various $x$ that  are shown in figure \ref{fig:lengths}(a) exhibit  self-similarity at $x\ge 70$. The so-defined  $L_{12}$, shown in figure \ref{fig:lengths}(b), characterises the energy-containing eddies and grows as $L_k/6$. {This indicates that the growth of $L_{12}$ is driven by the growth of the largest eddies, for example the global SPOD modes that scale with $L_k$ (which will be shown in the next section).}
%instead of  due to the more rapid decay of the small scales as in grid turbulence.} 
%The longitudinal integral lenth will be $L_{11}=2L_{12}$ in isotropic turbulence. 
The integral wake width $\delta=[(1/U_d) \int_{0}^{\infty} (U_{\infty} - \bar{u}) \, r{\rm d} r]^{1/2}$ also increases similarly  to  $L_k$.

\begin{figure}[thb]
  \centering 
  \captionsetup[subfloat]{farskip=-2pt,captionskip=0pt}
  \subfloat{{\includegraphics[width=0.475\linewidth]{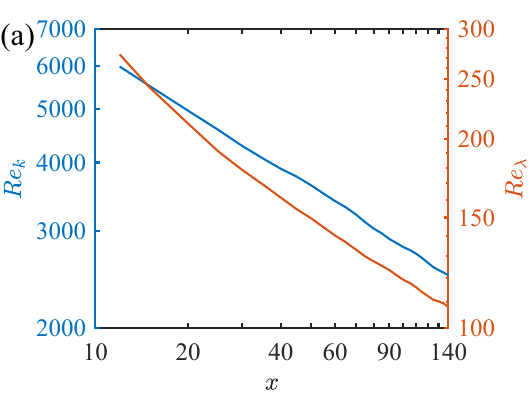}}}% 
  \subfloat{{\includegraphics[width=0.475\linewidth]{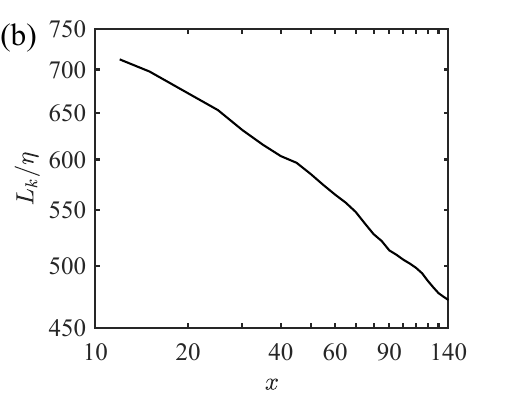}}}% 
  \\ 
  \subfloat{{\includegraphics[width=0.475\linewidth]{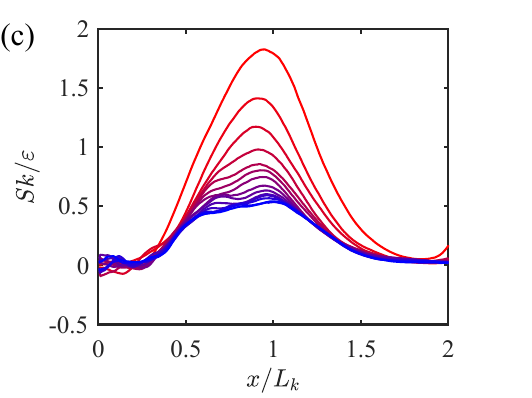}}}% 
  \subfloat{{\includegraphics[width=0.475\linewidth]{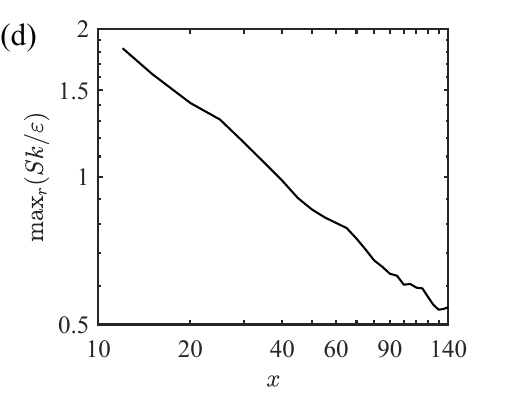}}}% 
  \caption{ (a) Streamwise evolutions of the turbulent Reynolds number $Re_k=U_k L_k/\nu$ and the Taylor Reynolds number $Re_\lambda = U_k^2 (15/\nu D_s)^{1/2}$.  (b) The separation between the largest and smallest length scales, $L_k/\eta$.  (c) The shear parameter $Sk/\varepsilon$ and its radial maximum (d)  as a function of $x$. The colour code in (c) is the same as in figure \ref{fig:selfsim_profs}.}   
  \label{fig:scale-sepa} 
\end{figure} 

Figure \ref{fig:lengths}(b) provides a comparison of different length scales as functions of $x$. All of the large-scale length scales ($L_d,L_k,L_{12},$ and $\delta$) follow similar power laws and were considered when assessing self-similarity.
% figure \ref{fig:selfsim_profs}. 
It turns out that $L_d$ yields the best self-similarity for the profiles of $U_d$, and $L_k$ for $k,-\overline{u'_x u'_r}$ and $\varepsilon$. Hence, we will use $L_k$ as the length scale for the energy-containing range in the following analyses. The Taylor microscale $\lambda$,  which reaches order 0.2$D$ at the end of the domain, is separated from the integral scale by more than a decade, while the Kolmogorov scale $\eta$ is separated from $\lambda$ by another factor greater than a decade. 
%At $x=140$, $\eta$ reaches one-hundredth of the disk diameter $D$. 
Even though  the gap between $\eta$ and $L_{12}$ decreases with streamwise distance, 
$L_k/\eta > 450$ is found until the end of the domain, as shown in figure \ref{fig:scale-sepa}(b). Such a separation of the integral and the dissipative scales  allows a decent inertial range to be established.  The turbulent Reynolds number  and the Taylor Reynolds number   are shown in figure \ref{fig:scale-sepa}(a). 
%The turbulent Reynolds number is several times higher than that in the slender-body wake of \cite{ortiz2021high} at a similar molecular Reynolds number ($Re=U_{\infty}D/\nu=$100 000 therein), due to the larger length scale $L_k$ since the value of $U_k$ is comparable in both wakes.  
The value of $Re_\lambda$ decays with $x$ but stays above 100 in the entire domain. Notably, unlike  homogeneous turbulence where $Re_{\lambda}\propto Re_k^{1/2}$ due to $\varepsilon \propto U_k^3/L_k$, here $Re_{\lambda}$ and $Re_k$ follow quite similar decay rates, with $L_k/\lambda \approx 22$--$25$ staying approximately constant throughout the wake.

Lastly,  the influence of mean shear ($S=\partial \bar{u}/\partial r$) on the fluctuations is gauged through  the shear parameter $S k/\varepsilon$ 
%\citep{corrsin1958local}
 (figure \ref{fig:scale-sepa} c,d).  The radial peak of $Sk/\varepsilon$ that  occurs near $r=L_k$ for all $x$ decays from around 2 to 0.5. At $x=90$, the radial peak of  $ Sk/\varepsilon \approx 0.63$, but there is still a half decade of $-7/3$ scaling in the cospectrum at that location in  figure \ref{fig:cospec_dspec}. The radial minimum of the Corrsin length scale \citep{corrsin1958local}, $L_c=\sqrt{\varepsilon/S^3}$, grows from approximately 0.5 to 1.0 $L_k$ in the domain (not shown in the plots).  Thus, only eddies of the size order $L_k$ or greater are able to directly feel the effect of the mean shear. The coherent structures to be discussed in detail in the next section are representative of such eddies.

% \clearpage
\subsection{Two-point, two-time correlation and SPOD global modes} \label{spod}

Despite  broadband energy spectra, 
% small-scale, sub-inertial-range turbulence, which exibits a general spectral energy concentration
 there are large-scale, anisotropic motions that display spatio-temporal coherence in the wake. They reside on the low-wavenumber end of the spectrum, contain a significant fraction of the TKE and Reynolds stress, most of the production and nearly none of the dissipation. They stay coherent for significant streamwise distances and hence play a vital role in the evolution of the mean flow. In this section, the CS will be examined based on the eigen-decomposition of the two-point, two-time correlation,
  \begin{equation}
  {R}_{ij}(\bm{x},\bm{x}',t,t') = \overline{u_i(\bm{x},t) u_j(\bm{x}',t')}. \label{2p2t}
\end{equation}  {In homogeneous turbulence, its spatial eigenfunctions are the orthogonal Fourier bases, whose energy content is described by the wavenumber-frequency spectrum. In  inhomogeneous turbulence, proper orthogonal decomposition (POD, \cite{lumley1967structure,lumley1970stochastic}) is one of the most widely used techniques to perform an empirical eigen-decomposition, leading to spatially orthogonal eigen-functions ranked by their energy contributions in the inhomogeneous directions. Spectral POD is one of the variants, which exploits the Fourier transform in the homogeneous directions and performs POD on the resulting Fourier modes in the remaining inhomogeneous directions. The detailed formulation and implementation for the present problem can be found in Appendix \ref{spod_app}.}   

\begin{figure}[t!]
  \centering 
  \captionsetup[subfloat]{farskip=-3pt,captionskip=0pt}
  \subfloat{{\includegraphics[width=0.9\linewidth]{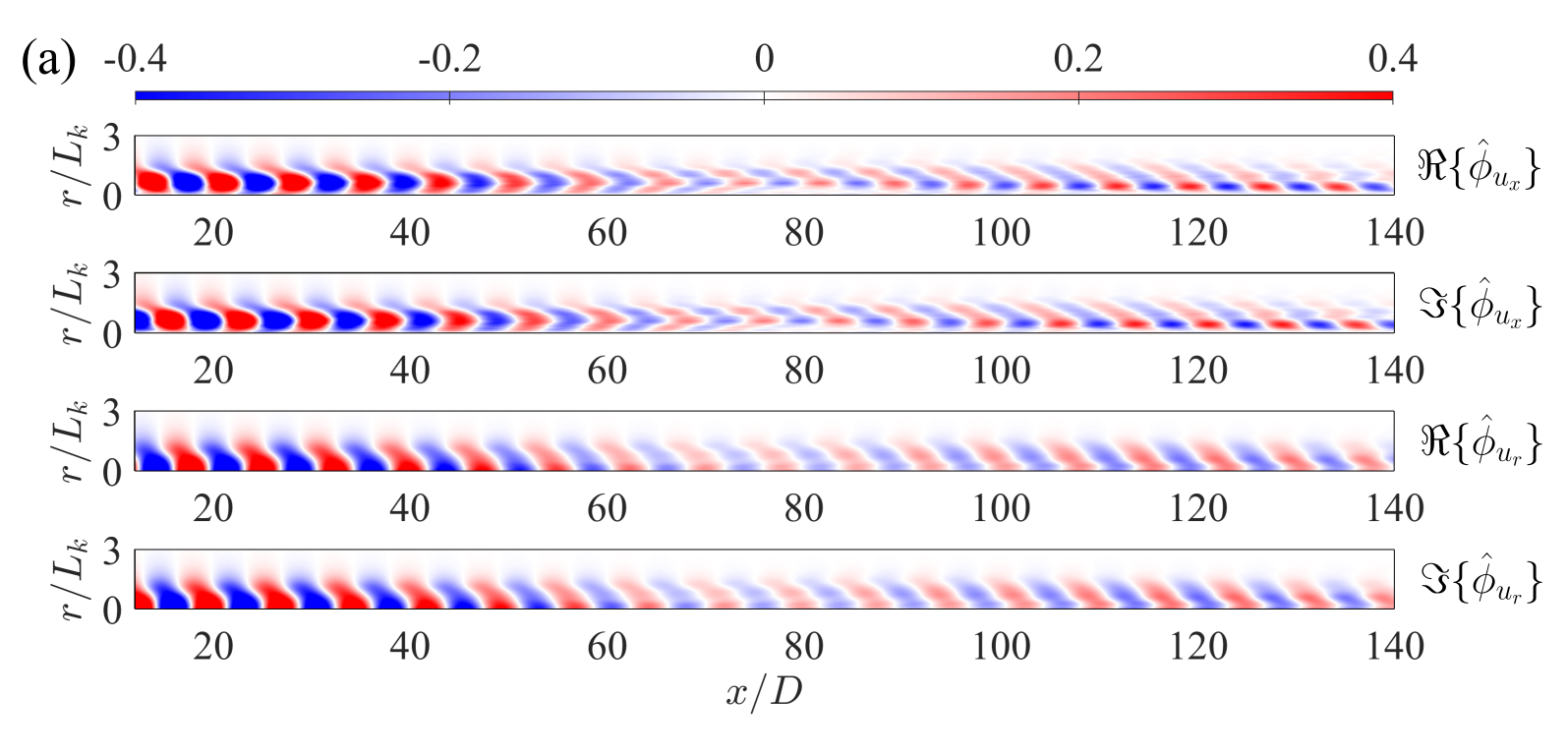}}}% 
  \\
  \subfloat{{\includegraphics[width=0.9\linewidth]{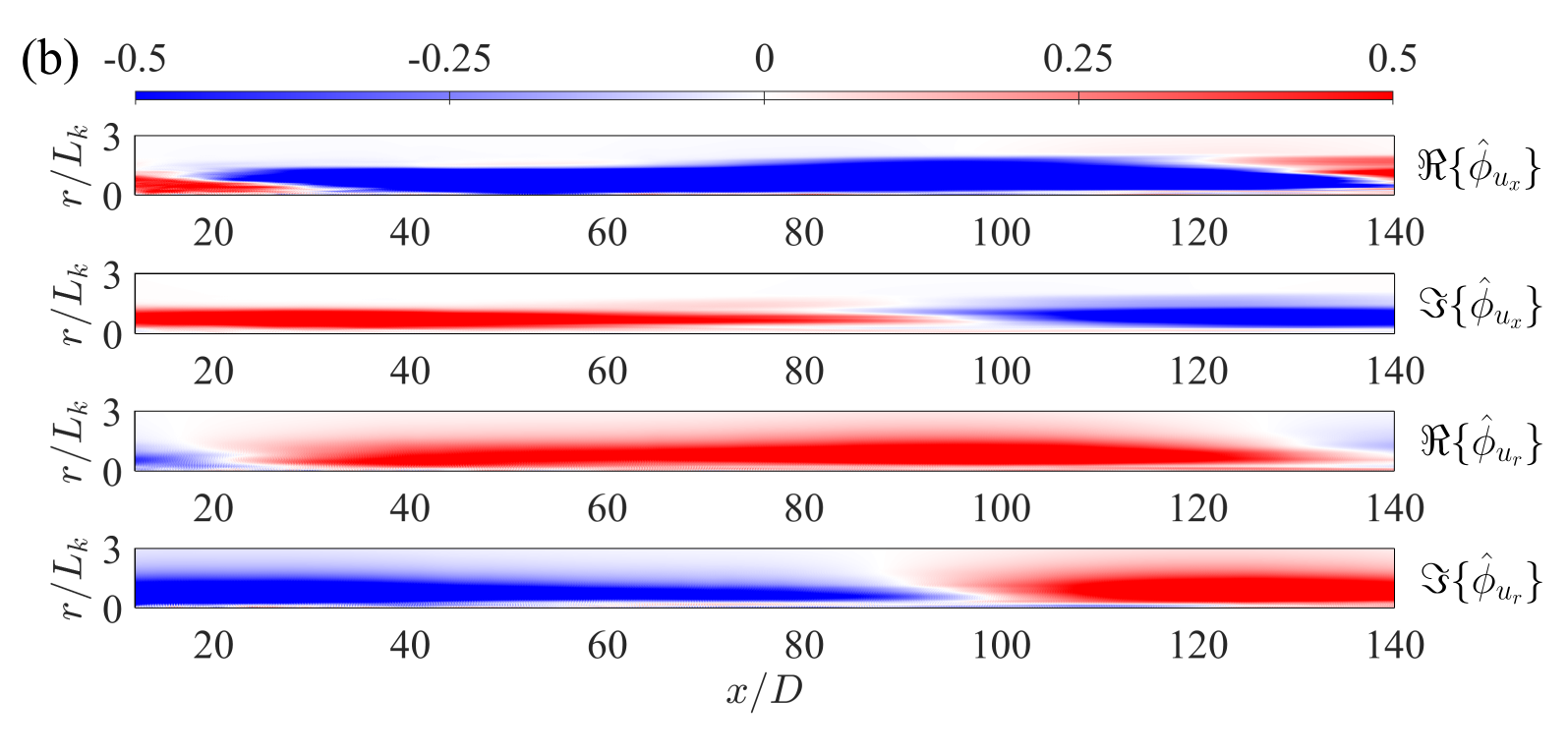}}}% 
  \caption{Global SPOD leading eigenmodes: (a) $(m,f)=(1,0.135)$ and (b) $(m,f)=(2,0)$. The real and imaginary parts of the eigenmodes of the streamwise and radial velocity components are shown.  The radial coordinate is normalised by the local TKE half-width $L_k(x)$ and the magnitudes of the SPOD eigenfunctions are normalised by the local TKE velocity scale $U_k(x)$.
  % to reflect their significance over the background turbulence. 
  The eigenfunctions are free up to a complex scalar multiple and the greatest contour magnitude in  each panel is chosen to be unity.  } 
  \label{fig:spod_xr} 
\end{figure} 

\begin{figure}[thb]
  \centering 
  \captionsetup[subfloat]{farskip=-2pt,captionskip=0pt}
  \subfloat{{\includegraphics[width=0.55\linewidth]{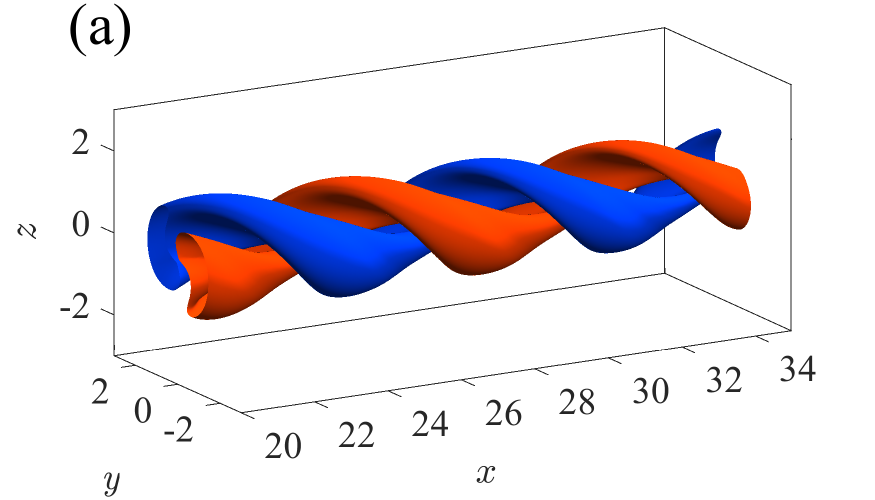}}} \\
  \subfloat{{\includegraphics[width=0.95\linewidth]{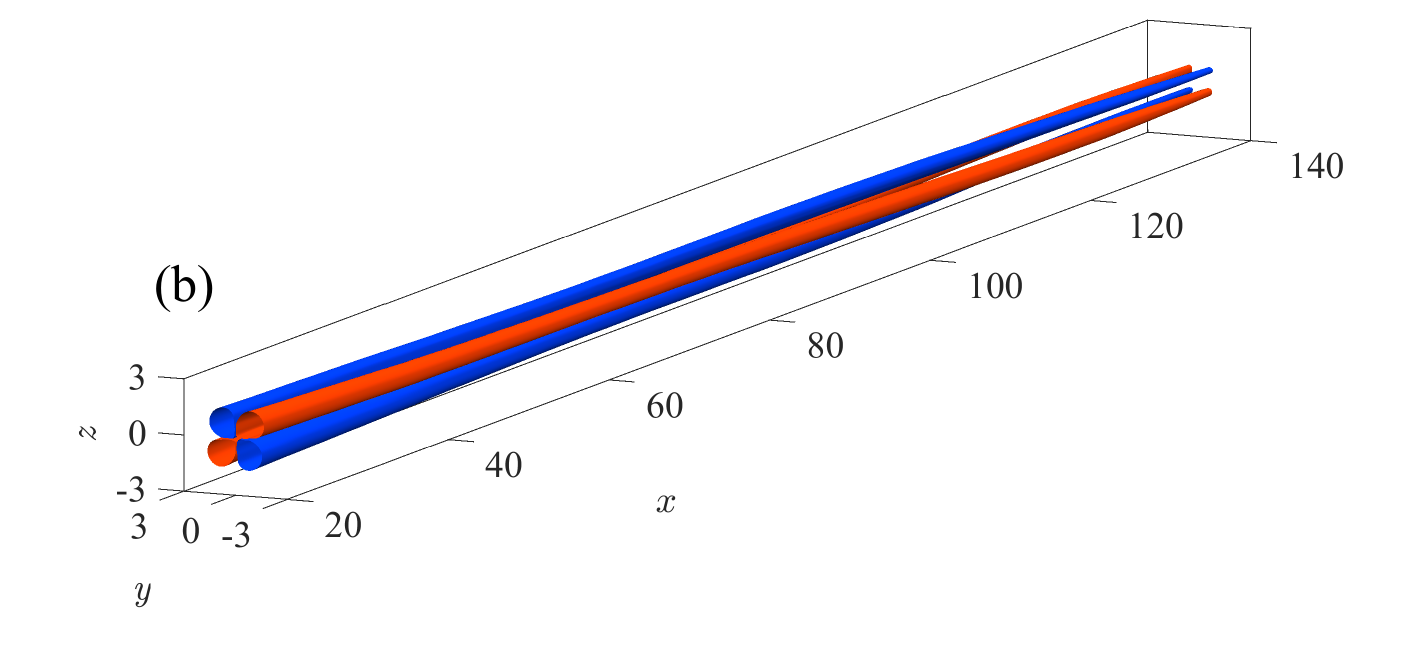}}}\\
  \caption{Three-dimensional visualisation of the SPOD modes: (a) the VS mode at $(m,f)=(1,0.135)$ and (b) the DH mode at $(m,f)=(2,0)$. Plotted quantity is the rank-1 perturbation velocity field $\Re\{\hat{\phi}_{u_r}(x,r) \exp({\rm i}m\theta) \exp({\rm i} 2\pi f t)\}$ reconstructed with only the leading SPOD mode, as a function of $(x,r,\theta)$ and at an arbitrary  time $t$. Red and blue colours represent $\pm$0.5 and $\pm$0.125 times the maximum value in the domains in (a) and (b), respectively. Streamwise coordinate $x$ is normalised by $D$ and the lateral and vertical coordinates $y,z$ are normalised by the local TKE length scale $L_k(x)$.  The domain shown in (a) only spans approximately $2U_{\infty}/f_{VS}$, where $f_{VS}=0.135$ is the VS frequency. The imaginary parts have the same shapes but are rotated in the $\theta$ direction by $\pi/2$ and are not shown. }  
  \label{fig:spod_3d} 
\end{figure} 

According to figure \ref{fig:spec_2d}, Fourier modes $(m,f)=(1,0.135)$ and $(2,0)$ are  the most dominant throughout the wake. They are referred to as the VS and DH modes, respectively. These two modes were first identified experimentally by \cite{fuchs1979large} in the near wake of a circular disk at $x/D=3$ and 9, and later by \cite{berger1990coherent}. Recent POD and SPOD analyses \citep{johansson2002proper,nidhan2020spectral} performed at $x={\rm const.}$ planes found that these two modes persist into the far wake with the dominant frequency individually identified  at each $x$ unchanged, suggesting global coherence.  Here, we rigorously extract 3-D coherent modes by performing SPOD analysis over the entire streamwise domain. We do so by utilising the inner-product of the SPOD problem \eqref{inner}, defined over $(x,r)$, since orthogonality in $\theta$ and $t$ is guaranteed by the Fourier transform. The resulting SPOD modes are hence eigenmodes of the 3-D two-point, two-time correlation tensor \eqref{2p2t} and represent spatio-temporally coherent structures.

The leading ($i=1$) SPOD modes for VS and DH are shown in figure \ref{fig:spod_xr}. It can be seen in figure \ref{fig:spod_xr}(a) that the VS mode has a streamwise wavelength $\approx 1/f$ that remains approximately the same throughout the entire domain. The imaginary part shows an almost identical structure but with a streamwise shift of $1/4$ wavelength (or $\pi/2$ in phase), which is  typical of travelling-wave modes, similar to those of the K{\'a}rm{\'a}n VS structures analysed by \cite{liu2024effect} in geophysical wakes. The modal shape of $u_r$ differs slightly but is generally similar to that of $u_x$ in terms of wavelength and direction of inclination. It is also noted that positive magnitude of $\Re\{\hat{\phi}_{u_x}\}$ is spatially correlated with negative $\Re\{\hat{\phi}_{u_r}\}$, and similarly for the corresponding imaginary parts, yielding a  positive contribution to the Reynolds stress cospectrum $\Re\{ -\hat{u}_x  \hat{u}_r^*\}$.

The 3-D SPOD modes are visualised in figure \ref{fig:spod_3d}. The VS mode at $m=1$ in figure \ref{fig:spod_3d}(a) has a clockwise spiral when viewed along positive $x$. Due to the Hermitian symmetry of velocity  spectra (see figure \ref{fig:spec_2d}) that arises since velocity signals are real-valued,  only Fourier modes in half of the $(m,f)$ plane need to be computed. We elect to keep the frequency $f$ positive and allow negative  $m$, with the sign of $m$ being interpreted as 
%parity or
handedness. The symmetric conjugate of the $(m,f)=(1,0.135)$ mode is $(-1, 0.135)$, which spirals opposite to that in figure \ref{fig:spod_3d}(a) (not plotted). The streamwise  dimming of colour in figure \ref{fig:spod_xr}(a) indicates a decay of modal energy relative to the local turbulence strength (as the eigenmodes are normalised by $U_k(x)$).
 %which will be confirmed by the streamwise evolution of relative modal energy in figure \ref{fig:k_int_r}. 

The DH mode at $(m,f)=(2,0)$, shown in figure \ref{fig:spod_xr}(b) and \ref{fig:spod_3d}(b), is obtained using $N_{FFT}=1024$, which is twice the length of the time series used in computing the temporal spectra and the SPOD mode of VS, resulting in $N_{blk}=13$. The lowest resolved frequency is $f=0.00677$, which is interpreted as $f \rightarrow 0$. It was determined by \cite{nekkanti2023large} that this mode results from the three-wave self-interaction of the VS modes and its conjugate, and has an effectively zero frequency in theory. In fact, the temporal invariance is reflected as spatial invariance of the global mode in figures \ref{fig:spod_xr}(b) and \ref{fig:spod_3d}(b), i.e., the length of the structure is almost the length of the domain. The DH structure has only progressed approximately $\pi/2$ in phase within $20<x<140$ in figure \ref{fig:spod_3d}(b). In such a case, the numerical limitation of resolving this structure becomes the length of the computational domain, which is $L_{x,o}-L_{x,i} = 130 \approx 1/0.00677$ in the present simulation.  Hence, the existence of the DH mode and  its very low frequency  require very long computational time in a very long computational domain -- a challenge to its numerical representation.

\begin{figure}[t!]
  \centering 
  \captionsetup[subfloat]{farskip=-2pt,captionskip=0pt}
  \subfloat{{\includegraphics[width=0.45\linewidth]{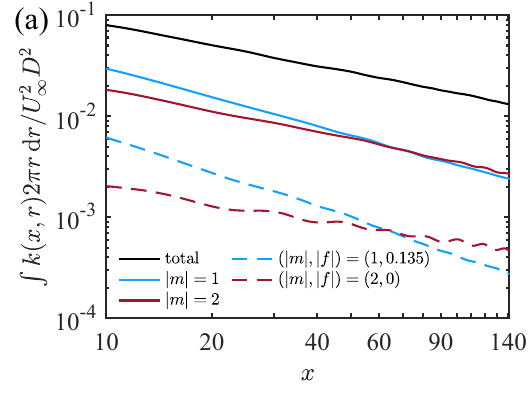}}}  
  \subfloat{{\includegraphics[width=0.45\linewidth]{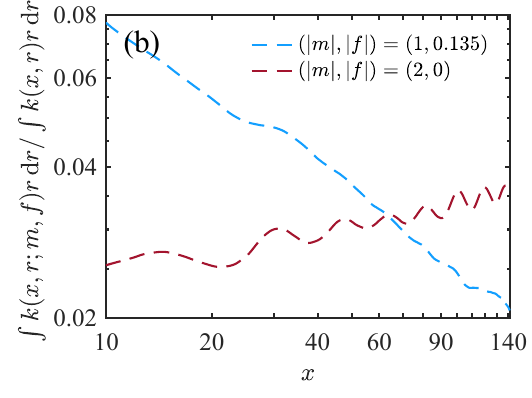}}}  
  \caption{(a,b) Evolution of the radially integrated TKE and its content in different Fourier modes as a function of $x$. In (b), the TKE contained in the VS mode $(|m|,|f|)=(1,0.135)$ and the DH mode $(2,0)$ are normalised by the total  TKE, to obtain their fractional contribution. The symmetric conjugates and the Hamming-window energy correction factor have been accounted for. }  
  \label{fig:k_int_r} 
\end{figure} 

%      x = 10   x = 20 
% m=1  37.0      18.2 
% m=2  22.9      20.3 
% VS   7.70      2.07 
% DH   2.55      3.56 

We move on to the contribution of the coherent  modes to the TKE and their relative importance. Figure \ref{fig:k_int_r} shows the streamwise evolution of the radially integrated TKE,  as well as specific Fourier modes of interest. The area-integrated TKE follows a power law and decays by a factor of six in the domain. The primary wavenumbers ($|m|=1, 2$),  each contribute approximately $O(0.2)$ to the total TKE. 
%In $10<x<140$, the energy fraction of $|m|=1$  decays from $37.0\%$ to $18.2\%$ while that of $|m|=2$ is from $22.9\%$ to $20.3\%$, together roughly half of the TKE while the $|m|=1$ mode decays much more rapidly. 
%Within $m=1$ and $m=2$, the Fourier modes of VS and DH motions, $(|m|,|f|)=(1,0.135)$ and $(2,0)$, each contributes $O(0.05)$ to the total TKE. 
It is also worth noting that the energy fraction  of the $(1,0.135)$ VS mode decays with $x$ while that of $(2,0)$ increases (figure \ref{fig:k_int_r}b), which reflects the energy transfer between the two as well as their ability to extract energy from the mean flow \citep{nekkanti2023large}. The $(2,0)$ mode eventually overtakes the $(1,0.135)$ mode at $x=65$. This exchange of relative importance between the two coherent modes coincides with the slight variation of the self-similar profiles in figure \ref{fig:selfsim_profs} as well as the slight variation of the associated power laws of the mean wake deficit, both occurring around $x=70$. 
%The influence of these two modes on the mean flow is rather small though, in accordance with their relatively weak magnitude. 
The lack of variation of relative magnitude in figure \ref{fig:spod_xr}(b) also implies that the DH mode scales weakly with the r.m.s. velocity $U_k$ while the VS mode clearly does not. {The Reynolds shear stress contributions from the VS and DH modes spatially evolve similarly to the  TKE, but with  about twice their fractional TKE contribution. The VS mode contains $17.5\%$ of the Reynolds shear stresses at $x=10$ and $4.1\%$ at $x=140$, whereas the DH mode contains $5.7\%$ and $7.1\%$ at these two locations, respectively. The Reynolds shear stress having a higher concentration in the largest scales compared to the TKE has also enabled excellent low-rank reconstruction using SPOD (see Figure 19 of \cite{nidhan2020spectral}).}

It is  remarkable how well the radial extents of the SPOD modes scale with $L_k$, which changes by a factor of three in the domain, and that these modes remain globally coherent over distances and times much larger than any local flow scale. Thus, importantly, the CS as part of two-point statistics, respect self-similarity similar to the conventional one-point statistics.  The CS  {\em do not } preclude self-similarity in the wake, as classically hypothesised, but they coexist with the self-similar wake and  are also part of the self-similarity. The variation of the strength of CS makes an impact on the details of the self-similar profiles and power laws of the mean flow. The interaction between the coherent motions and the background turbulence, especially the energy transfer that couples them, is addressed in the following section.

\section{Scale decomposition and inter-scale energy transfer} \label{scale_decomp}

Turbulence is a phenomenon of multiscale interactions mediated by a specific quadratic (in velocity) nonlinearity of the Navier--Stokes equations, which is a feature that distinguishes it from broadband random noise. Scale ranges in the wake have been individually characterised in the previous section and the focus of this section will be on quantifying the inter-scale energy transfer and its spatial and scale dependence. 

The present work takes a triple decomposition (TD) approach. A spectral sharp filter is applied in the two homogeneous directions, $\theta$ and $t$, to the fluctuations with respect to the azimuthal-temporal mean, leading to 
\begin{equation}
 u_i = \bar{u}_i + u'_i, \quad  u'_i = \tilde{u}_i + u''_i \, , \label{td_vel}
\end{equation} where $\widetilde{(\cdot)}$ denotes the filtered or large-scale portion and $(\cdot)''$ the residual or small-scale portion.  Similarly, $p' = \tilde{p} + p''$. The filtering is performed in  Fourier space $(m,f)$ by multiplying the Fourier modes with the spectral transfer function,
 \begin{equation}
  \hat{G}(m,f) = \begin{cases}
  1, & |m| \le m_{\Delta} \cap  |f| \le f_{\Delta} \\
  0, & |m| > m_{\Delta}   \cup  |f| > f_{\Delta} 
  \end{cases}\; ,  \label{transf}
\end{equation} and then transforming them back to  real space: \begin{equation}
  \tilde{u}_i(x,r,\theta,t) = \mathcal{F}^{-1}[\hat{G} \hat{u}_i(x,r,m,f)], \, 
  u''_i(x,r,\theta,t) = \mathcal{F}^{-1}[(1-\hat{G})\hat{u}_i(x,r,m,f)], 
\end{equation} where $\mathcal{F}^{-1}(\cdot)$ denotes the inverse Fourier transform.

The orthogonality of Fourier modes guarantees the idempotence and orthogonality of the projection, \begin{equation} 
 \tilde{\tilde{u}}_i  = \tilde{u}_i,\, \widetilde{u''_i} = 0, 
\end{equation} and the statistical independence between filtered and unfiltered motions, such that \begin{equation}
  \overline{\tilde{u}_i u''_j} = \overline{u''_i \tilde{u}_j} = 0, 
\end{equation} where $\overline{(\cdot)}$ is the time or ensemble average. This property and the idempotence ($\tilde{\tilde{u}}_i = \tilde{u}_i$) make the Fourier sharp filter an orthogonal projector, which the Gaussian and the box filters are not. The decomposition of the TKE evolution equation is greatly simplified with an orthogonal projector and the Fourier sharp filter \eqref{transf} will be used, unless otherwise noted.

\begin{figure}[thb]
  \centering 
  \subfloat{\includegraphics[width=\textwidth]{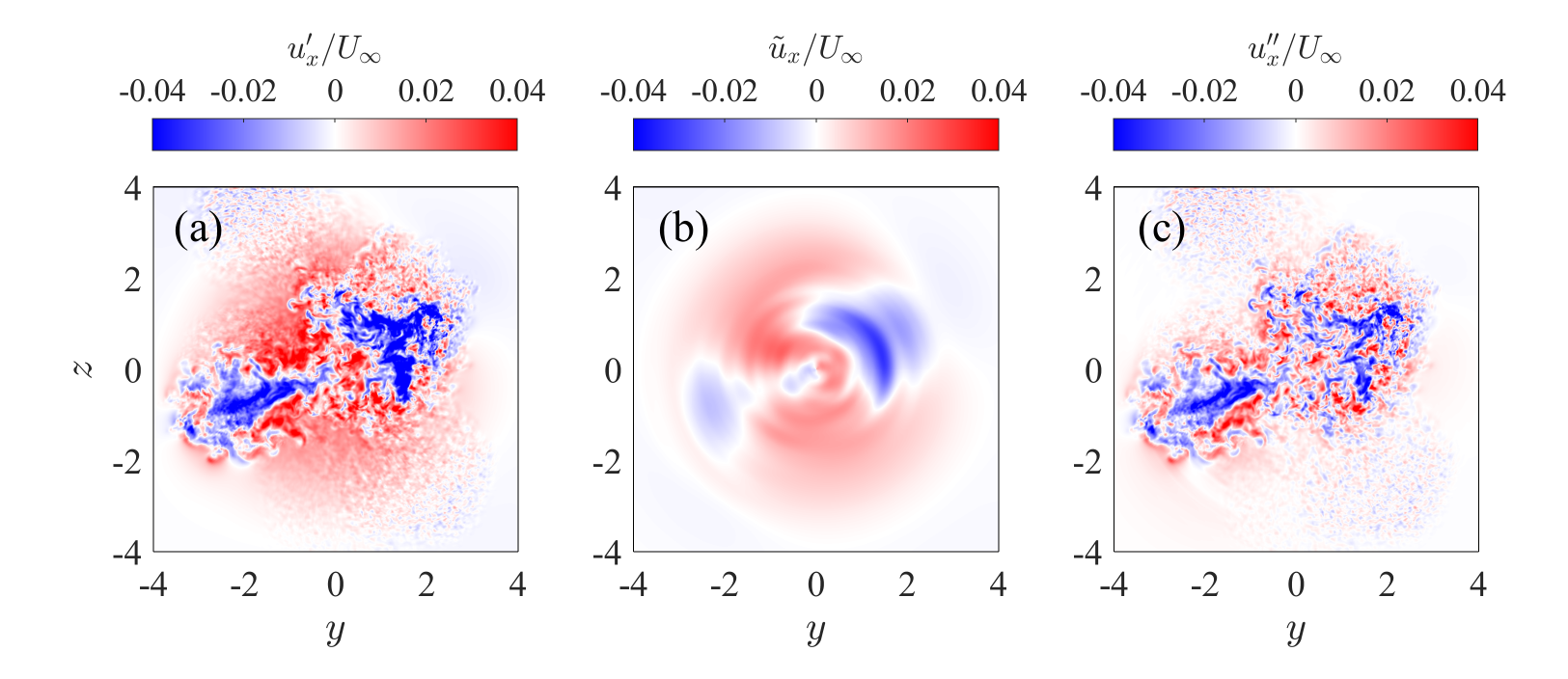}}
  \caption{Instantaneous snapshots of the full, large- and small-scale parts of $u'_x$. (a-c) $u'_x,\tilde{u}_x,u''_{x}$, respectively, at $x=60$. Cutoff wavenumbers $f_\Delta = 1, m_\Delta=6$.} % snapshot index 5
  \label{fig:eg_td} 
\end{figure}

Figure \ref{fig:eg_td} is an example of the TD. It can be seen that the low-pass fluctuation resembles the pattern of the full signal well, while exhibiting inhomogeneity and anisotropy. The high-pass fluctuations contain more fine-scale features and sharp gradients, and are less organised. 
\subsection{The triple decomposition of the TKE and its evolution equation}

Given the nature of the sharp filter, the total kinetic energy $\mathcal{K}$ is exactly 
%exclusively and exhaustively
 decomposed into three parts: \begin{align}
  \mathcal{K} = \frac{1}{2} \overline{u_i u_i} & = \frac{1}{2}  {\bar{u}_i \bar{u}_i} + \frac{1}{2}  \overline{\tilde{u}_i \tilde{u}_i} + \frac{1}{2}  \overline{u''_i u''_i} \\
  & = k_m + k_l + k_s, 
\end{align} namely the mean kinetic energy (MKE, denoted by subscript `m'), the large-scale turbulent kinetic energy (LKE, denoted by subscript `l'), and the small-scale turbulent kinetic energy (SKE, denoted by subscript `s'). The TKE \begin{equation}
  k = \frac{1}{2}  \overline{u'_i u'_i} = \frac{1}{2}  \overline{\tilde{u}_i \tilde{u}_i} + \frac{1}{2}  \overline{u''_i u''_i} = k_l + k_s, 
\end{equation} is the sum of LKE and SKE. Such a decomposition is associated with the specific  filter scale,  which will be varied systematically.
% and is therefore not unique.   

The evolution equations for the triply decomposed, turbulent kinetic energy (TD-TKE) equations are  
\begin{align}
 \underbrace{\left( \frac{\partial }{\partial t} + \bar{u}_j \frac{\partial }{\partial x_j} \right) k_l}_\text{$-A_l$}  & = \underbrace{-  \frac{\partial }{\partial x_j} 
 \left(  \overline{\tilde{p} \tilde{u}_j}   
+ \frac{1}{2} \overline{ \tilde{u}_i \tilde{u}_i \tilde{u}_j} 
 + \overline{\tilde{u}_i (\widetilde{u'_i u'_j} - \tilde{u}_i \tilde{u}_j)} 
 - \frac{1}{Re} \frac{\partial k_l}{\partial x_j} 
 \right)}_\text{$T_{p,l}+T_{t,l}+T_{q,l}+T_{v,l}$} 
\nonumber
\\
&\;\;\;\; 
\underbrace{- \overline{\tilde{u}_i \tilde{u}_j } \frac{\partial \bar{u}_i}{\partial x_j} }_\text{$P_l$} 
\underbrace{+ \overline{ (\widetilde{u'_i u'_j} -\tilde{u}_i \tilde{u}_j )\frac{\partial \tilde{u}_i}{\partial x_j} } }_\text{${\it \Pi}_l$} 
\underbrace{-\frac{1}{Re} \overline{\frac{\partial {\tilde{u}_i }}{\partial x_j} \frac{\partial {\tilde{u}_i }}{\partial x_j}}}_\text{$- \varepsilon_l$} \label{lke}  \\ 
\underbrace{\left( \frac{\partial }{\partial t} + \bar{u}_j \frac{\partial }{\partial x_j} \right) k_s}_\text{$-A_s$}  
& = \underbrace{-  \frac{\partial }{\partial x_j} 
 \left(  \overline{{p}'' {u}''_j}   
 + \frac{1}{2} \overline{ u''_i u''_i (\tilde{u}_j + u''_j)} 
 - \overline{u''_i (\widetilde{u'_i u'_j} - \tilde{u}_i \tilde{u}_j)}
 - \frac{1}{Re} \frac{\partial k_s }{\partial x_j}  
 \right)}_\text{$T_{p,s}+T_{t,s}+T_{q,s}+T_{v,s}$} 
\nonumber 
\\
& \;\;\;\; 
\underbrace{- \overline{u''_i u''_j} \frac{\partial \bar{u}_i}{\partial x_j}}_\text{$P_s$}
\underbrace{- \overline{u''_i u''_j \frac{\partial \tilde{u}_i}{\partial x_j}}
- \overline{(\widetilde{u'_i u'_j} - \tilde{u}_i \tilde{u}_j) \frac{\partial u''_i}{\partial x_j}}}_\text{${\it \Pi}_s$} 
\underbrace{ - \frac{1}{Re} \overline{\frac{\partial {u}''_i}{\partial x_j} \frac{\partial {u}''_i}{\partial x_j} }}_\text{$-\varepsilon_s$} \label{ske}
\end{align} whereas the filtered and sub-filter momentum equations from which the above equations are derived are provided in Appendix \ref{deri_td}.  

%  {The subscript of the underbraces have been changed from +epsilon to -epsilon. No need to conflict with the existing practice of positive-definite epsilon.}

The decomposition of the TKE equation into its two constituents can be viewed as  a decomposition of individual mechanisms. There are scale-local ones, which stem from the interaction of the perturbations with the mean flow and are linear in that sense, such as \begin{equation}
  \begin{aligned}
  A  = A_l + A_s , \; \;
  T_{p}  = T_{p,l} + T_{p,s} , \; \;
    % T_{t} & = T_{t,l} + T_{t,s} \\ 
  T_{v}  = T_{v,l} + T_{v,s}  , \;  \;
  P  = P_l + P_s  , \; \;
  \varepsilon & = \varepsilon_l + \varepsilon_s \, .
\end{aligned} \label{tke_decomp}
\end{equation} These terms straightforwardly split into separate contributions from the large- and the small-scale parts of the flow. The scale-nonlocal one, the turbulent transport $T_t$, represents perturbations interacting among themselves, and is nonlinear in that sense. It decomposes into  \begin{align}
  T_t &= (T_{t,l}+T_{q,l}+{\it \Pi}_l)  + (T_{t,s}+T_{q,s}+{\it \Pi}_s), \label{tt_decomp}
\end{align} in which the appearance of additional terms other than $T_{t,l}$ and $T_{t,s}$ is typical in the coarse-graining of nonlinear systems. The scale-nonlocal terms include the turbulent diffusions $T_{t,l}$ and $T_{t,s}$, the cross-fluxes $T_{q,l}$ and $T_{q,s}$, and the inter-scale fluxes ${\it \Pi}_l$ and ${\it \Pi}_s$ as viewed from the large and small scale's perspectives.  Such a scale-local/linear and scale-nonlocal/nonlinear distinction of TKE evolution mechanisms is consistent with the view from the so-called spectral TKE equations \citep{mizuno2016spectra,cho2018scale}, in which the evolution equations of Fourier modes and the associated kinetic energy were considered. The terms corresponding to $A,T_p,T_v, P, \varepsilon$ in the spectral TKE equation (at each wavenumber) only contain the contribution from the same wavenumber, whereas that of $T_t$ contains all wavenumbers through the triadic interactions. 
{For the derivation and further discussions of \eqref{tke_decomp} and \eqref{tt_decomp}, the reader is referred to Appendix \ref{deri_td}. Both decompositions \eqref{tke_decomp} and \eqref{tt_decomp} are verified numerically with the left-hand-sides obtained from the TKE equation \eqref{tke} and the right-hand-sides from the LKE and the SKE equations \eqref{lke}-\eqref{ske}.}

The decomposed TKE equations  also serve to  distinguish between physical-space transport and inter-scale transfer.  The inter-scale transfers act as a source or sink in the individual equations while moving  energy between the large and the small-scale fields. From the perspective of LKE that is  governed by  \eqref{lke}, the inter-scale transfer is  
 \begin{equation}
    {\it \Pi}_l =  \overline{ (\widetilde{u'_i u'_j} -\tilde{u}_i \tilde{u}_j )\frac{\partial \tilde{u}_i}{\partial x_j} } =  \overline{ {\tau}_{ij}^{SFS} \frac{\partial \tilde{u}_i}{\partial x_j} } , 
 \label{Pil}
\end{equation} 
where ${\tau}_{ij}^{SFS} = \widetilde{u'_i u'_j} -\tilde{u}_i \tilde{u}_j$ is the sub-filter scale (SFS) stress. It is a quantity commonly used to indicate and quantify the turbulent energy cascade.  A negative ${\it \Pi}_l$ indicates forward cascade from the LKE to the SKE  (a sink for the LKE).  With regard to physical-space transport  of LKE, it is accomplished  by the mean advection ($A_l$), by the large-scale  turbulence ($T_l$) and the cross term ($T_{q,l}$).  Unlike the inter-scale transfers, the physical-space transports, which have vanishing spatial integration over a volume, redistribute energy in space, not scale.

From the SKE's perspective, the inter-scale transfer \begin{equation}
    {\it \Pi}_s  = - \overline{u''_i u''_j \frac{\partial \tilde{u}_i}{\partial x_j}} 
 - \overline{(\widetilde{u'_i u'_j} - \tilde{u}_i \tilde{u}_j) \frac{\partial u''_i}{\partial x_j}}  
\end{equation} acts as a production mechanism that  is not pointwise identical to $-{\it \Pi}_l$, unlike the exact pointwise opposition of the mean production terms in MKE and TKE equations. The reasons will be elaborated below. It is worth noting that ${\it \Pi}_s$ has received less attention compared to ${\it \Pi}_l$ in the literature {since the latter is the quantity being modelled in LES. However, for exploiting the relation between the cascade and  dissipation in inhomogeneous turbulence, ${\it \Pi}_s$ is more suitable because both ${\it \Pi}_s$ and $\varepsilon_s \approx \varepsilon$ are dominant in  the SKE equation.}

\begin{figure}[t!]
  \centering 
  \captionsetup[subfloat]{farskip=-2pt,captionskip=0pt}
  \subfloat{{\includegraphics[width=0.45\linewidth]{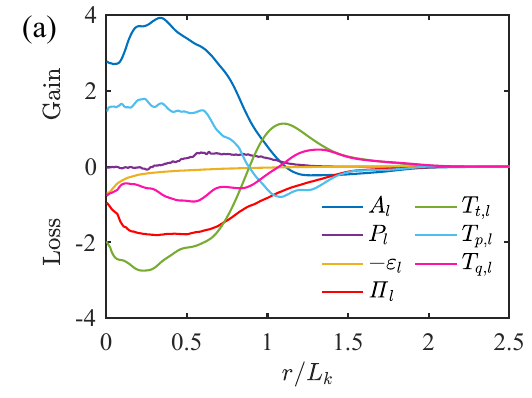}}} 
  \subfloat{{\includegraphics[width=0.45\linewidth]{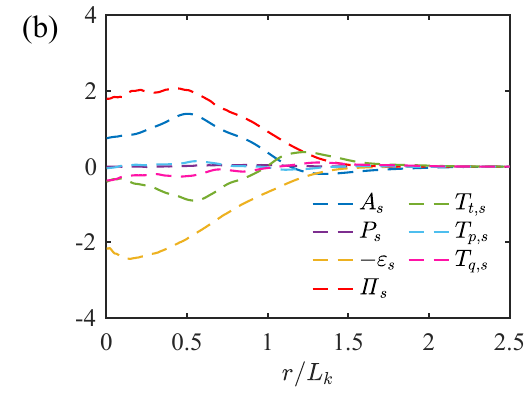}}}\\
  \subfloat{{\includegraphics[width=0.45\linewidth]{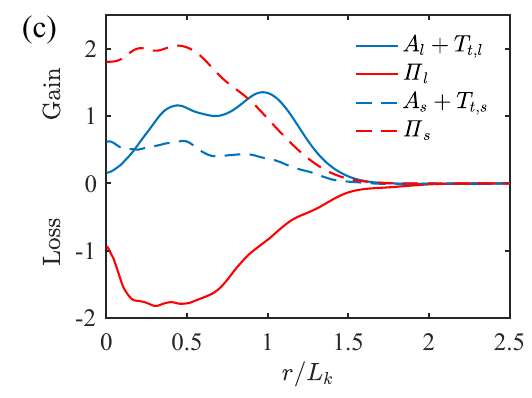}}} 
  \subfloat{{\includegraphics[width=0.45\linewidth]{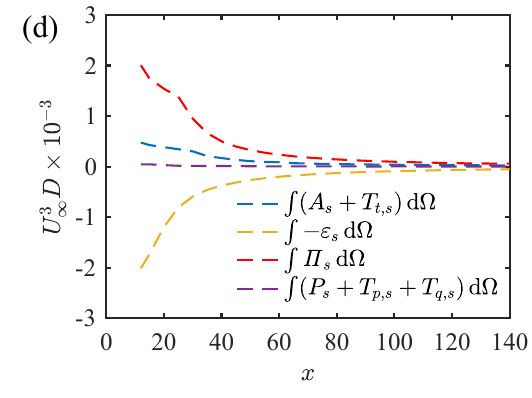}}}
  \caption{ (a-b) Balances of the LKE and the SKE equations at $x=60$. The unit of the vertical axis is $U^3_k/L_k$.  (c) Comparison between spatial transport and inter-scale transfer that were plotted in (a-b). (d) The area {(${\rm d\Omega}=2\pi r{\rm d}r$)} integrated budget terms as functions of $x$. Filter sizes $f_\Delta =1,m_\Delta =6$. }  
  \label{fig:Tt_decomp} 
\end{figure}

The decomposed LKE and SKE balances \eqref{lke}-\eqref{ske} are shown in figure \ref{fig:Tt_decomp}(a-b), using cutoff wavenumbers $f_\Delta=1,m_\Delta=6$ in \eqref{transf}. The aspect ratio $m_\Delta=6 f_\Delta$ is selected such that the frequency cutoff expressed in terms of angular frequency, $\omega_\Delta = 2 \pi f_\Delta$, gives a nearly isotropic aspect ratio $\omega_\Delta \approx m_\Delta$. Terms in \eqref{lke}-\eqref{ske} are written in tensor form for generality. In their numerical evaluation, cylindrical coordinates are used
%, in which  the above terms are expressed similarly to those in the standard TKE equation
 (see Appendix \ref{tke_full} for the one-point TKE equation in cylindrical coordinates).  It can be seen that production $P$ and pressure diffusion $T_p$ are mainly large-scale mechanisms, which are only contained in the LKE budget but not that for SKE. Dissipation $\varepsilon$, on the other hand, is a small-scale mechanism contained mainly in the SKE balance. The spatial transports, including the mean-flow advection $A$ and the turbulent diffusion $T_t$, make non-negligible contributions to both LKE and SKE.  The lower magnitudes of $A_s$ and $T_{t,s}$ relative to $A_l$ and $T_{t,l}$, are consistent with the concentration of energy at the large scales. The inter-scale transfer ${\it \Pi}$ is a sink for LKE and a source for SKE. 

The inter-space and inter-scale transfers are compared in figure \ref{fig:Tt_decomp}(c). The large scales receive energy from advection mainly, while radially transporting it outwards and locally transferring it downscale. From the LKE's viewpoint,  inter-space and inter-scale transfers have opposite signs almost everywhere, while the former generally has a lower magnitude. 
This counter-balance, obtained from a filtering approach here, agrees qualitatively and quantitatively with that found with the alternative approach of a generalised Kolmogorov equation  by  \cite{noriega2026turbulent}, who investigated the interacting wakes of two side-by-side prisms. 
The small scales also receive energy from spatial transports but the gain by spatial transports  ($A_s + T_{t,s}$) is much less than that from the inter-scale transfer ($\it \Pi_s$). Eventually, the energy dissipated at the small scales is not balanced by the inter-scale flux in the inertial range. %Such an imbalance between ${\it \Pi}$ and $\varepsilon$ was also seen in the far field of grid turbulence due to small-scale advection \citep{valente2015energy}. 

The dominant balance for the SKE is \begin{equation}
  A_s + T_{t,s} + {\it \Pi}_s - \varepsilon_s \approx 0,  \label{dom_blnc}
\end{equation} which differs from the balance in locally homogeneous turbulence: ${\it \Pi}_s = \varepsilon_s \approx \varepsilon$. This imbalance between ${\it \Pi}$ and $\varepsilon$ explains why even when ${\it \Pi}_s$ follows the inertial scaling $\mathcal{U}^3/\mathcal{L}$ (as will be shown in section \ref{pi_self-sim}), $\varepsilon$ does not. The area integrals of grouped budget terms of the SKE are shown in figure \ref{fig:Tt_decomp}(d). The same major balance in \eqref{dom_blnc} can be found at different $x$-locations. 

\subsection{Spatial dependence of the inter-scale transfer and its self-similarity} \label{pi_self-sim}

\begin{figure}[thb]
  \centering 
  \captionsetup[subfloat]{farskip=-2pt,captionskip=0pt}
  \subfloat{{\includegraphics[width=0.45\linewidth]{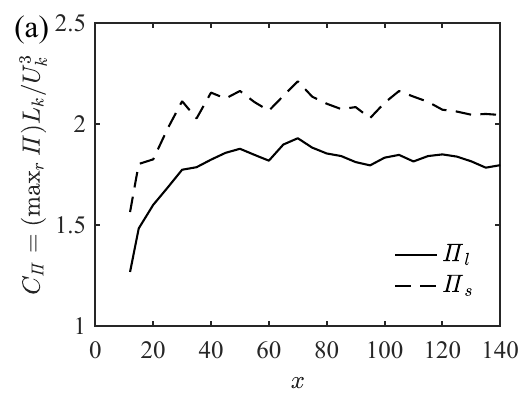}}} 
  \subfloat{{\includegraphics[width=0.45\linewidth]{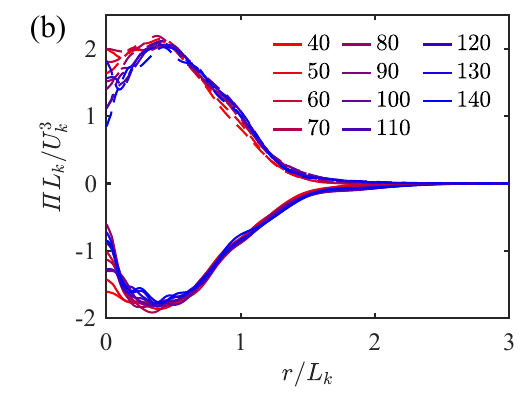}}}% 
  \caption{Inter-scale fluxes, ${\it \Pi}_l$ and ${\it \Pi}_s$, interpreted using TKE-based inertial scaling $U_k^3/L_k$: (a) the inter-scale flux coefficient $C_{\it \Pi}=(\max_r {\it \Pi})L_k/U_k^3$, and  (b) self-similar scaling of the radial profiles  where curves with negative (positive) value correspond to $\it{\Pi}_l$ ($\it{\Pi}_s$).  Filter scales $f_\Delta =1,m_\Delta =6$. }  
  \label{fig:pi_flux} 
\end{figure} 

\begin{figure}[htb]
  \centering 
  \captionsetup[subfloat]{farskip=-2pt,captionskip=0pt}
  \subfloat{{\includegraphics[width=0.45\linewidth]{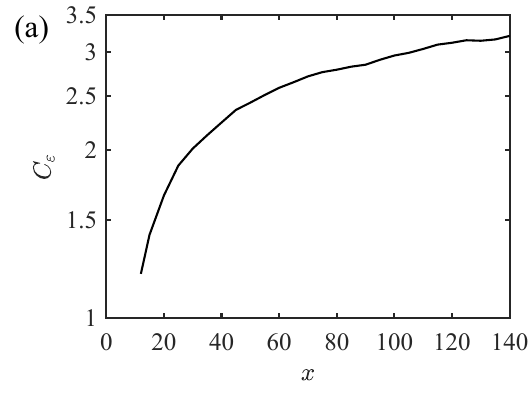}}} 
  \subfloat{{\includegraphics[width=0.45\linewidth]{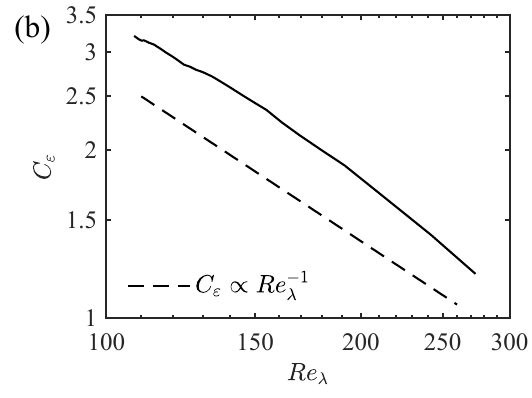}}}% 
  \caption{The dissipation coefficient $C_{\varepsilon}$, as a function of $x$ in (a) and a function of $Re_{\lambda}$ in (b). Both $C_{\varepsilon} = D_s L_k/U_k^3$ and $Re_{\lambda} = U_k^2 \sqrt{15/\nu D_s}$ are based on streamwise-local wake characteristic scales. }  
  \label{fig:ceps_relambda} 
\end{figure} 

As hypothesised in the Introduction, there could be two possible explanations of the non-equilibrium dissipation scaling in the wake, in which $\varepsilon$ does not asymptote to a constant multiple of $\mathcal{U}^3/\mathcal{L}$. The first is that  inter-scale transfer in this wake does not scale with $\mathcal{U}^3/\mathcal{L}$ unlike  homogeneous, isotropic turbulence,  hence making it less likely that $\varepsilon$ will follow such a scaling. The second, as has been demonstrated in the previous section, is that ${\it \Pi}$ is not in a  balance with $\varepsilon$ locally, due to spatial transports. That still leaves open the scaling of ${\it \Pi}$, which will be the focus of this section.

Figure \ref{fig:pi_flux} shows the evolution of the inter-scale flux coefficient $C_{\it \Pi}$, which is the radial maximum of ${\it \Pi}$ normalised by the inertial scaling $U_k^3/L_k$ based on the TKE scales. The employed filter sizes are in the inertial range. In figure \ref{fig:pi_flux}(a), $C_{\it \Pi}$ first increases with $x$, then approaches a constant value {from around $x=40$} and stays approximately  constant for a further 100$D$ downstream. Given the substantial variation of $Re_\lambda(x)$ (see figure \ref{fig:scale-sepa}) in the same range, $C_{\it \Pi}$ can be considered  to become independent of $Re_\lambda$. Figure \ref{fig:pi_flux}(b) shows the radial profiles of ${\it \Pi}$, normalised by $U_k^3/L_k$ and $L_k$. The reasonable collapse of profiles at different {$x\ge 40$} indicates self-similar evolution. The emergence of self-similarity in both ${\it \Pi}_l$ and ${\it \Pi}_s$ {is concurrent with the asymptote of $C_{\it \Pi}$ to a constant, and interestingly, occurs only slightly later than the self-similarity of the one-point statistics $k$ and $\varepsilon$ ($x\ge 30$)}. Very similar results to figure \ref{fig:pi_flux} are reproduced using neighbouring filter sizes $(f_\Delta,m_\Delta) = (0.8,  5)$,  $(f_\Delta,m_\Delta) = (1.2, 7)$, and  $(f_\Delta,m_\Delta) = (1.25, 8)$. But when the %constant-in-$x$ 
filter size is placed in the energy-containing range or the dissipation range, neither does $C_{\it \Pi}$ remain a constant independent of $x$ nor do the radial profiles become self-similar.  {Even though the radial maxima of ${\it \Pi}_l$ and ${\it \Pi}_s$ are slightly different, as seen in the constants of respective $C_{\it \Pi}$, it is mainly due to the profile shapes and the their area integrals are indeed very close (see figure \ref{fig:fsize_ftype}a). }

We can interpret the constancy of $C_{\it \Pi}$ through a timescale argument:  in the inertial range, the eddy turnover time $L_k/U_k$ sets the timescale of the energy cascade $U_k^2/{\it \Pi}$, {but not necessarily that of the dissipation $U_k^2/\varepsilon$ unless there is an equilibrium $\varepsilon \approx {\it \Pi}$ that brings about  equality of  these two  timescales. The statistical equilibrium in homogeneous, isotropic  turbulence leads to constant $C_{\varepsilon}$ \citep{sreenivasan1984scaling}. However, as shown in figure \ref{fig:ceps_relambda}, $C_{\varepsilon}$ is neither a constant in $x$ nor in $Re_{\lambda}$ in the present wake.  Since ${\it \Pi}$ follows the inertial scaling and is determined by the local large scales $U_k$ and $L_k$, the non-equilibrium dissipation  -- or the non-constancy of $C_{\varepsilon}$  --  is likely a feature of scales smaller than the inertial range.} 

% \clearpage
\subsection{Scale dependence of the inter-scale transfer and the advection-driven disequilibrium}   \label{small_advc}

\begin{figure}[htb]
  \centering 
  \captionsetup[subfloat]{farskip=-2pt,captionskip=0pt}
    \subfloat{{\includegraphics[width=0.45\linewidth]{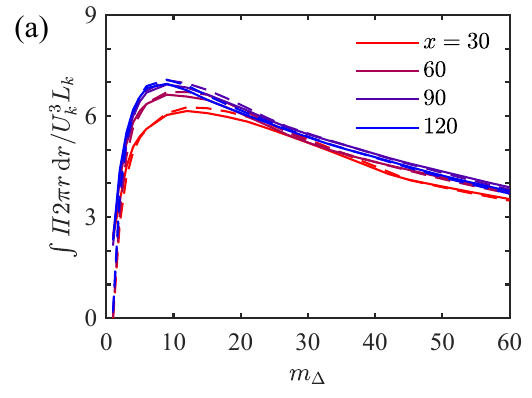}}} 
    \subfloat{{\includegraphics[width=0.45\linewidth]{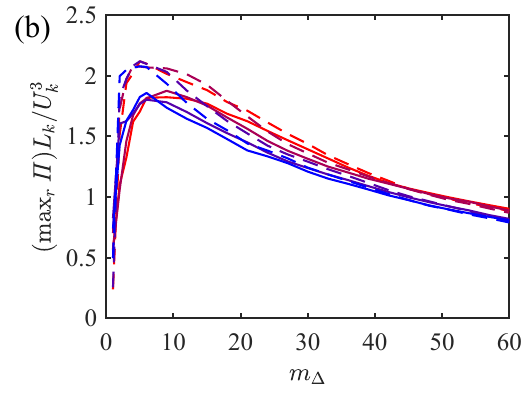}}} 
    \\ 
    \subfloat{{\includegraphics[width=0.45\linewidth]{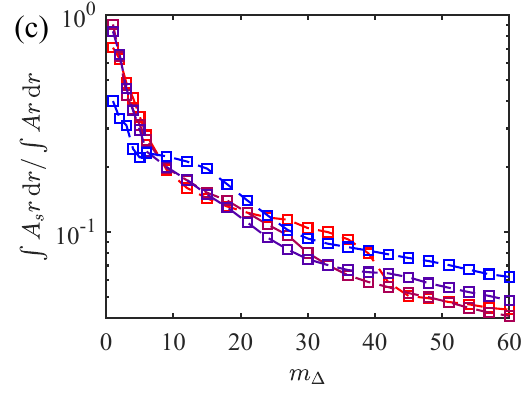}}} 
    \subfloat{{\includegraphics[width=0.45\linewidth]{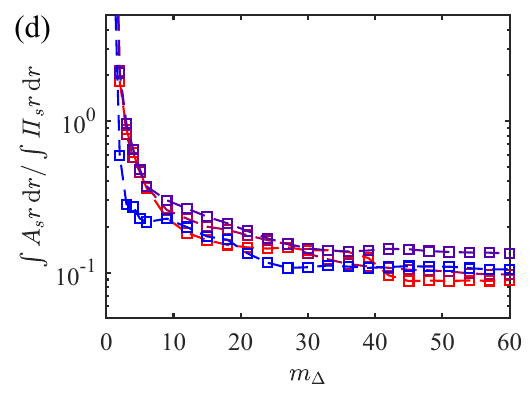}}} 
  \caption{(a,b) Area integrals and radial maxima of inter-scale fluxes, as functions of the filtering scale. (c,d) The ratios of area-integrated small-scale advection to (c) the total advection and (d) small-scale inter-scale flux. Solid lines: ${\it \Pi}_l$; dashed lines: ${\it \Pi}_s$; dashed lines with symbols: $A_s$. The same $x$-locations and colour code are used in (a-d). }  
  \label{fig:fsize_ftype} 
\end{figure} 

% --- chifan --- 

The continuous dependence of ${\it \Pi}$ on scale is a fundamental feature of turbulence. Figure \ref{fig:fsize_ftype}(a-b) shows the inter-scale fluxes at various $x$ as functions of the wavenumber $m_\Delta$, with a fixed aspect ratio $f_\Delta=m_\Delta/6$ as used before. The fluxes ${\it \Pi}_l$ and ${\it \Pi}_s$ increase from zero very rapidly as the wavenumber increases and peak in the inertial range, then start to decay slowly as the effect of viscosity becomes important. {The area integrals of ${\it \Pi}_l$ and ${\it \Pi}_s$ are very close for all filter scales, despite a small difference in  their peak values.} The filter scale corresponding to the largest radial maximum of inter-scale transfer is around $m_\Delta=6\sim 9$ and the largest fluxes (with respect to the filter scale) scale with $U_k^3/L_k$, as also shown in figure \ref{fig:pi_flux}. {The choice of $\kappa_m=m/r$ as the filtering wavenumber in place of $m$ has also been explored (see Appendix \ref{ftype} for details), which did not lead to constancy of $C_{\it \Pi}$ or self-similarity of ${\it \Pi}$. The success of filtering in $m$ suggests that the inter-scale flux is a large-scale quantity better characterised by $m$ rather than $\kappa_m$, which is more suitable for characterising the turbulence spectra in the inertial range and beyond.}

Figure \ref{fig:Tt_decomp}(b) has shown that, for the filter scale $f_\Delta=1, m_\Delta=6$, the small-scale inter-scale flux ${\it \Pi}_s$ is not balanced by the small-scale dissipation $\varepsilon_s$ due to the contribution from small-scale advection $A_s$. Provided $\int T_{t,s}\, {\rm d}\Omega \approx 0$ (found to be order $10^{-4}$ or smaller for $m_\Delta>3$ and hence not shown), the small-scale balance \eqref{dom_blnc}  is reduced to
 \begin{equation} 
  \int A_s \, {\rm d}\Omega + \int {\it \Pi}_s \,{\rm d}\Omega -  \int \varepsilon_s \,{\rm d}\Omega \approx 0 \, , \label{imbalance}
\end{equation}
in terms of spatial integrals, where ${\rm d}\Omega=2\pi r{\rm d}r$. In order to answer whether the contribution of $A_s$ is non-negligible at all scales, we plot the ratio between $\int A_s \, {\rm d}\Omega$ and $\int {\it \Pi}_s \,{\rm d}\Omega$ as a function of filter scale. First, figure \ref{fig:fsize_ftype}(c) shows that more than $10\%$ of the advection is contained in scales smaller than the wavelength corresponding to $m=20$, for all $x$ locations considered. As shown in figure \ref{fig:fsize_ftype}(d), the ratio $\int A_s \, {\rm d}\Omega / \int {\it \Pi}_s \,{\rm d}\Omega$ is large at very small $m_\Delta$ since most of the advection is still contained in $A_s$ while there is not much energy contained in the large scales to be transferred. As $m_\Delta$ increases, ${\it \Pi}_s$ increases and $A_s$ decreases more rapidly, leading to a general decrease of $\int A_s \, {\rm d}\Omega / \int {\it \Pi}_s \,{\rm d}\Omega$. However, in the scale range that the inter-scale fluxes are the largest, $m_\Delta=6$--$9$ for $\max_r {\it \Pi}$ and $m_\Delta=9$--$12$ for $\int {\it \Pi} \,{\rm d}\Omega$, the ratio $\int A_s \, {\rm d}\Omega / \int {\it \Pi}_s \,{\rm d}\Omega$ is non-negligible. It stays above $0.15\sim 0.20$ for all $x$ examined in figure \ref{fig:fsize_ftype}(d) until $m_\Delta=15$, and then decays very slowly to reach $O(0.1)$ by $m_\Delta \approx 30$, where the spectra have started to depart from  inertial-range scaling (see figure \ref{fig:spec_azim}c). Small-scale advection remains  non-negligible beyond the inertial range and is still significant at the scale of the greatest inter-scale flux, contributing to the non-equilibrium of dissipation.

{\cite{valente2015energy} employed the structure function approach to study the scale-by-scale balance among advection, inter-scale transfer, and dissipation in grid turbulence. For the near field (case RG115 therein; $1.5<x/x_{\rm peak}<3.7$, where $x_{\rm peak}$ is the streamwise location of peak TKE), the imbalance between inter-scale flux and dissipation results from small-scale advection, similar to the present observation. In the relatively far field (case RG60; $8.5<x/x_{\rm peak}<21$), however, scale-by-scale advection, inter-scale transfer, and dissipation all scale with the total dissipation and are in a three-way balance, even though small-scale advection is still non-negligible. Such an equilibrium between inter-scale transfer and dissipation has not been found in the present far wake, which marks a qualitative  difference between wake turbulence and grid turbulence. In the equilibrium far field of grid turbulence, the growth of the integral length scale, which is an average over a spectrum of scales, is mainly due to the faster decay of small scales rather than the growth of the largest eddies. In contrast, the size of the largest eddies, represented by the global SPOD modes, experiences a notable growth in $x$ direction as the wake expands (see figure \ref{fig:lengths}b) through entrainment. The growth of the external length scale is one of the factors that keep the wake from reaching an equilibrium, such that the dissipation is not solely determined by the local  scales ($U_k$ and $L_k$).}

\section{Spatially local energy transfers}  \label{cg}

The dependence of small-scale processes on the large-scale dynamics is a topic that has received  increasing attention recently, with one focus on whether the presence of CS and their energy and time scales affect the turbulent dissipation rate. The dissipation coefficient, $C_{\varepsilon} = \varepsilon \mathcal{L}/\mathcal{U}^3$, reflects the dependence of dissipation on the large-scale energy $\mathcal{U}^2$ and turnover time $\mathcal{L}/\mathcal{U}$.  
The non-constancy of either $C_{\varepsilon}(x)$ or $C_{\varepsilon}(Re_{\lambda})$ shown in figure \ref{fig:ceps_relambda} indicates a local disequilibrium in the wake that occurs  in spite of  TKE, cascade, and dissipation all obeying self-similarity,  and the flow being at sufficiently high $Re_{\lambda}$ to exclude low-Reynolds number effects. The dashed line in figure \ref{fig:ceps_relambda}(b) shows that  the  local disequilibrium has the distinguishing feature  that  the streamwise local $C_{\varepsilon}$ is anti-correlated to the local $Re_{\lambda}$, satisfying approximately $C_{\varepsilon}(x) \propto Re_{\lambda}^{-1}(x)$ instead of the equilibrium dependence $C_{\varepsilon} \approx {\rm const.}$ at high $Re_\lambda$.  As noted in the Introduction, the scaling $C_{\varepsilon} \propto Re_{\lambda}^{-1}$ has been found in the near field of grid turbulence \citep{valente2012universal}, in decaying forced isotropic turbulence \citep{goto2015energy}, and in unsteady forced isotropic turbulence \citep{valente2014origin,ghira2026non}. The existence of  the same scaling into the self-similar far wake of a bluff body up to $x/D=140$, where the effect of shear is small but advection is non-negligible beyond the inertial range, is reported here for the first time.

We note that in the power-law scaling $C_{\varepsilon} \propto Re_\lambda^{p}$, the exponent $p$ need not be $-1$, even though it is more commonly observed. Other exponents, {such as $p=-1.2$ in decaying isotropic turbulence \citep{valente2014origin} and} around $p=-1.5$ in the logarithmic layer of {turbulent channel} \citep{apostolidis2022scalings} where turbulent production is significant, were also reported. The non-equilibrium scalings of $C_{\varepsilon}$ have also been reported as power laws of the `local' Reynolds number (equivalent to $Re_k$ in this work) instead of $Re_{\lambda}$, such as \cite{dairay2015non,ortiz2021high}.  We have not explored $Re_k$-based scaling here since the homogeneous-turbulence expectation  $Re_\lambda\propto Re_k^{1/2}$ is not satisfied in the present wake (see figure \ref{fig:scale-sepa}a).

The non-equilibrium dissipation scaling $C_{\varepsilon} \propto Re_{\lambda}^{-1}$ has recently been found to hold for box-averaged local statistics in stationary forced isotropic turbulence \citep{nishimoto2026spatially}.
%suggesting that it is likely a local property general to turbulent flows. 
We are thus motivated to test in the present wake whether the non-equilibrium dissipation scaling, in particular $C_{\varepsilon} \propto Re_{\lambda}^{-1}$,  originates  from  spatially local turbulent statistics in more general turbulent flows.  Also, the evolution of local turbulence in the $C_{\varepsilon}$-$Re_{\lambda}$ phase space will be analysed to illustrate the intermittency of local non-equilibrium.

\begin{figure}[thb]
  \centering 
  \subfloat{\includegraphics[width=\textwidth]{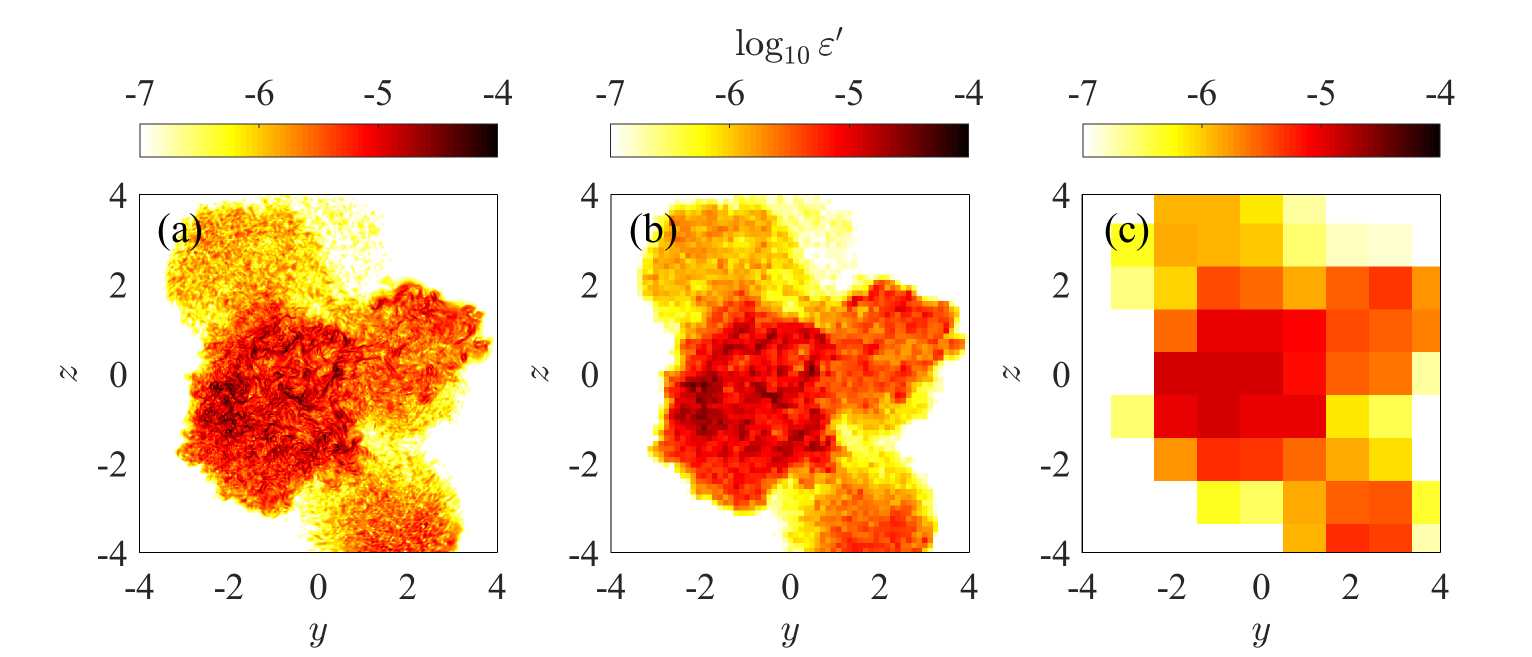}}
  \caption{(a-c) The instantaneous dissipation field at $x=60$ at different levels of CG: $l_b$, $8l_b$, and $64l_b$.} % snapshot index 5
  \label{fig:dis_cg} 
\end{figure}

A coarse-graining (CG) procedure is adopted  to divide the domain into spatially local boxes with varying scales in which local statistics are  gathered. This spatial coarse-graining complements the spectral filtering described in the previous section, serving as a method for progressive spatial averaging. The cylindrical data in the $(r,\theta)$ at constant $x$ are first interpolated using 2-D splines onto a square Cartesian grid in the $(y,z)$ domain, spanning $[-2.5L_k, 2.5 L_k] \times [-2.5L_k, 2.5 L_k]$, where $L_k$ is the local TKE length scale. Beyond $2.5L_k$, the TKE and  dissipation are negligible (figure \ref{fig:selfsim_profs}b,d). The limits of the visualisation window ($[-4, 4]\times[-4, 4]$) in figure \ref{fig:dis_cg} do not reflect the limit of the domain of investigation. The domain is then divided into $1024^2$ square boxes with the box dimension $l_b = 5L_k/1024 = O(\eta)$. The uniform grid is advantageous over the non-uniform cylindrical grid due to the equal area, making it more convenient for obtaining  CG statistics. 

The first level of CG is taken to be the box of dimension $l_b$. Coarser levels result from combining four adjacent units from the previous level in a non-overlapping manner, which raises the box scale by a factor of two and is performed recursively until $128l_b$. Three representative levels $l_b$, $8l_b$ and $64l_b$ are shown in figure \ref{fig:dis_cg}, from which it can be seen that, after each operation, small-scale details are lost while the large-scale patterns remain. Flow statistics, such as instantaneous TKE ($k'$) and dissipation ($\varepsilon'$), are only computed at level $l_b$ and then pooled and averaged during the CG, {instead of being based on  recomputed fluctuations of the coarse-grained flow.}  %coarse-grained fluctuations and velocity gradients. 
The box-local turbulent statistics, denoted by $\langle \cdot \rangle_b$, result from the algebraic average of the four neighbouring boxes being merged during CG -- an operation that conserves the area integrals of each quantity among different CG levels. 

\begin{figure}[htb]
  \centering 
  \captionsetup[subfloat]{farskip=-2pt,captionskip=0pt}
  \subfloat{{\includegraphics[width=0.45\linewidth]{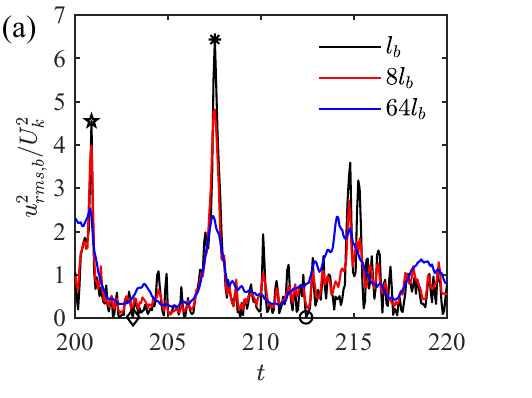}}} 
  \subfloat{{\includegraphics[width=0.45\linewidth]{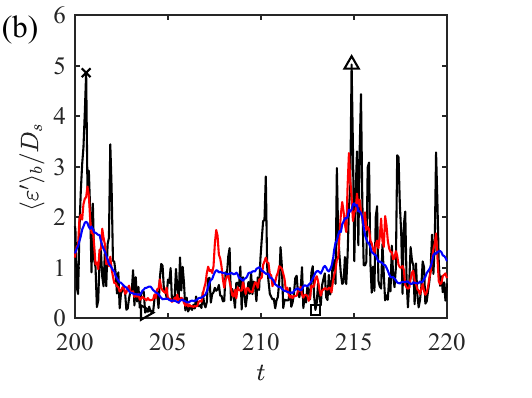}}}% 
  \\
  \subfloat{{\includegraphics[width=0.45\linewidth]{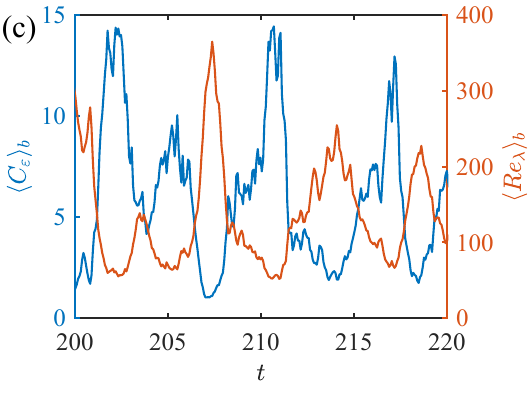}}}% 
  \subfloat{{\includegraphics[width=0.45\linewidth]{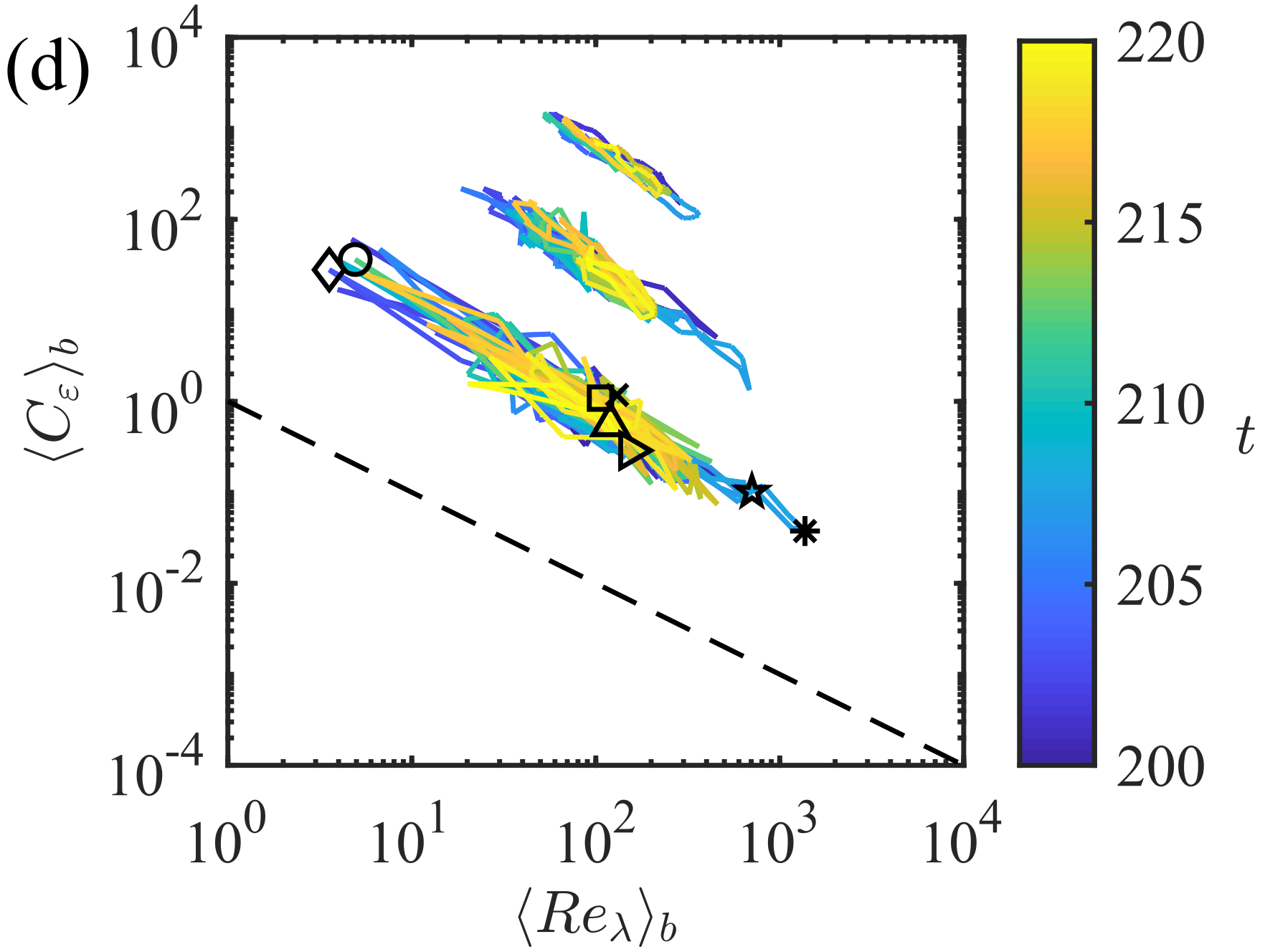}}} 
  \caption{Coarse-grained statistics, shown at  $x=60$. Time evolution: (a) centreline  $u^2_{rms,b}$, (b) centreline dissipation rate $\langle \varepsilon' \rangle_b$, and (c) dissipation coefficient and Reynolds number. (a-b) show three levels of CG ($l_b$, $8l_b$, and $64l_b$),  {whereas in (c) only level $64l_b$ is shown for clarity}. Panel (d)  shows evolution  in the phase-space of $\langle Re_\lambda \rangle_b$ and $\langle C_{\varepsilon} \rangle_b$, corresponding to the same time window as in (a-c) and the same three levels of CG. The phase portraits for levels $8l_b$ and $64l_b$ are each offset vertically by a decade for clarity. The markers in (d) indicate the same time instances as they do in (a) and (b), at which two local maxima and two local minima of $u^2_{rms,b}$  and $\langle \varepsilon' \rangle$ are captured.}  
  \label{fig:ceps_relambda_phase} 
\end{figure} 

The instantaneous box-local r.m.s. velocity $u_{rms,b}=\sqrt{2 \langle k' \rangle_b/3}$ and dissipation $\langle\varepsilon' \rangle_b$ at the centreline are shown in figure \ref{fig:ceps_relambda_phase}(a-b), at different CG levels  and over a period of $20 D/U_{\infty}$ (approximately three VS cycles). Both contain intense temporal  fluctuations that are several times greater than the mean,  which lead to large temporal variation of box-local $\langle C_{\varepsilon} \rangle_b$ and $\langle Re_{\lambda}\rangle_b$. As the CG level increases, the   trends in variability are preserved but the intensity of the fluctuations is smoothed out. 

Figure \ref{fig:ceps_relambda_phase}(d) represents the time evolution of turbulence at the centreline as trajectories in the phase space spanned by $\langle C_{\varepsilon} \rangle_b(t)$ and $\langle Re_{\lambda} \rangle_b(t)$.  The box-local dissipation coefficient is defined as $\langle C_{\varepsilon} \rangle_b = \langle \varepsilon' \rangle_b L_{k,b} / u_{rms,b}^3$ where the local TKE length scale, 
 \begin{equation}
  L_{k,b} = \sqrt{ \left( \frac{\langle k'\rangle_b}{\partial_y \langle k'\rangle_b} \right)^2 + 
  \left( \frac{\langle k'\rangle_b}{\partial_z \langle k'\rangle_b} \right)^2 } \, , 
\end{equation}
 is defined based on its gradient evaluated at its CG level. The box-local Taylor Reynolds number is $\langle Re_{\lambda} \rangle_b = u_{ rms,b}^2 \sqrt{15/\nu \langle \varepsilon' \rangle_b}$. 
% . Here, the box-local velocity and length scales are $u_{rms,b} = \sqrt{2\langle k' \rangle_b/3}$ and   with $\langle \cdot \rangle_b$ denoting the box-average and $(\cdot)'$ the instantaneous quantity.
Three levels of CG are shown. At the finest level ($l_b$), the phase trajectory exhibits the largest dynamic range -- almost three decades  --   indicating strong perturbations from equilibrium. While oscillating significantly, the values of $\langle C_{\varepsilon} \rangle_b$ and $\langle Re_\lambda \rangle_b$ still concentrate within a narrow distribution along a  line with constant slope of about  $ -1$. At levels $8l_b$ and $64l_b$, the dynamics are quite similar, albeit with increasingly shorter dynamic range of fluctuation that reflects the progressively stronger averaging. {The phase-opposed oscillatory behaviour is evident from the evolution of $ \langle C_\varepsilon \rangle_b (t)$ and $\langle Re_{\lambda}\rangle_b (t) $, shown for CG level $64l_b$  in  figure \ref{fig:ceps_relambda_phase}(c).
%after descent spatial averaging 
%corresponding to CG level $64l_b$. 
There is also good correspondence between local peaks of $u^2_{rms,b}$ in the 64$l_b$ curve of figure \ref{fig:ceps_relambda_phase}(a) and the peaks of $\langle Re_{\lambda}\rangle_b$ in figure \ref{fig:ceps_relambda_phase}(c). } 

The extremes of the phase trajectories are of particular interest, because they correspond to  greatest disequilibrium. As indicated by the markers in figure \ref{fig:ceps_relambda_phase}(a-b), which denote the local maxima and minima of $u_{rms,b}$ and $\langle \varepsilon' \rangle_b$, the extremes of $\langle C_{\varepsilon} \rangle_b(t)$ and $\langle Re_{\lambda} \rangle_b(t)$ are due to the fluctuation of the local kinetic energy $u^2_{rms,b}$ rather than the fluctuation of  the local dissipation. The time instances with extremes of $\langle \varepsilon' \rangle_b$ correspond to the centre of the phase trajectories while those with extreme $u_{rms,b}$ correspond to the corners. Therefore, it is very likely that the disequilibrium in phase space  is driven by the off-mean TKE rather than the dissipation. We draw an analogy to  the numerical experiments of \cite{ghira2026non} in  forced isotropic turbulence, in which a stationary system corresponding to one point in the $(Re_\lambda, C_{\varepsilon})$ space is perturbed with a sudden change of the power input (forcing). The system exhibits a relatively large overshoot/undershoot along $C_{\varepsilon}\propto Re_{\lambda}^{p}$ before settling back to the new stationary state $(Re_\lambda, C_{\varepsilon})$ corresponding to the new forcing (see their Figure 3). In the present example of turbulent wakes, unsteadiness or inhomogeneity, which affects the energy balance through physical-space transport, is analogous  to changing the power inputs in forced isotropic turbulence.

\begin{figure}[t!]
  \centering 
  \captionsetup[subfloat]{farskip=-2pt,captionskip=0pt}
  \subfloat{{\includegraphics[width=0.45\linewidth]{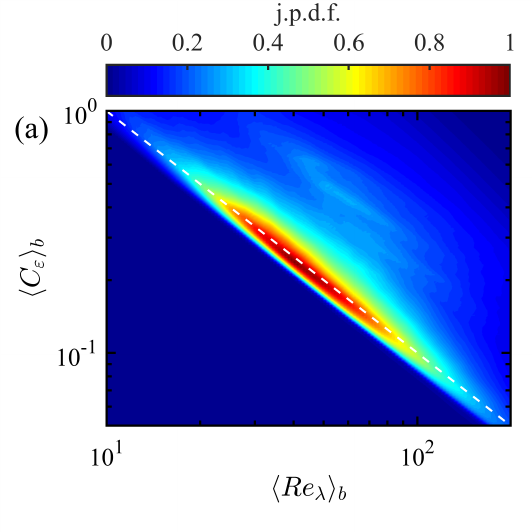}}}% 
  \subfloat{{\includegraphics[width=0.45\linewidth]{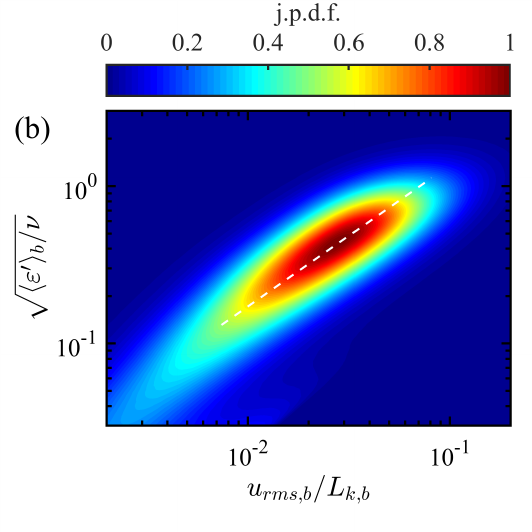}}} \\
  \subfloat{{\includegraphics[width=0.45\linewidth]{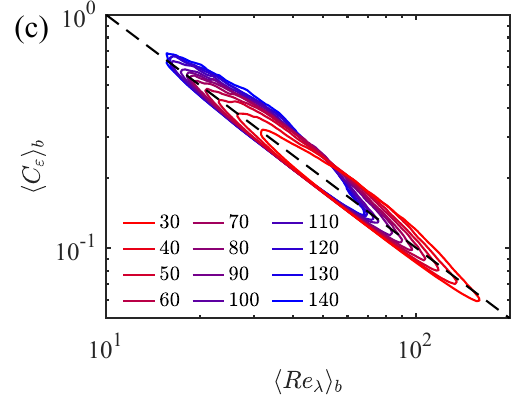}}}% 
  \subfloat{{\includegraphics[width=0.45\linewidth]{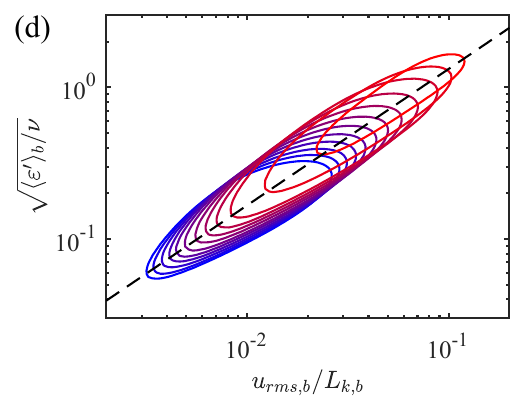}}} 
  \caption{JPDFs at the streamwise location,  $x=60$: (a) the box-local Taylor Reynolds number and dissipation coefficient; (b) the local TKE cascade frequency $u_{rms, b}/L_{k,b}$ and the local dissipation (Kolmogorov) frequency $\sqrt{\langle \varepsilon' \rangle_b /\nu }$. The JPDFs are normalised such that the maximum value is unity. Panels (c-d) show the streamwise evolution of the JPDFs, shown by the contours of $50\%$ of the maximum of each JPDF, in the region $30\le x \le140$. The diagonal dashed lines indicate a slope of $-1$ in (a,c) and 0.9 in (b,d). The CG level of  $l_b$ is employed for all plots. }  
  \label{fig:jpdfs_x_60} 
\end{figure}

The relationship between $\langle C_{\varepsilon} \rangle_b$ and $\langle Re_{\lambda} \rangle_b$  seen in the phase portrait of figure \ref{fig:ceps_relambda_phase}  motivates examination of their joint probability density function (JPDF). 
%and the clear anti-ccorelation shown by the conditional average 
%it can be  that the distribution of $\langle C_{\varepsilon} \rangle_b$ and $\langle Re_{\lambda} \rangle_b$ is narrow-banded. 
Figure \ref{fig:jpdfs_x_60}(a) shows the JPDF,  computed from the entire 2-D $y$-$z$ plane ($[-2.5L_k, 2.5L_k] \times [-2.5L_k, 2.5L_k]$) {and at the CG level $l_b$}. The $-1$ slope passes through the centre of the JPDF, which has a quite narrow distribution of $\langle C_{\varepsilon} \rangle_b$ at given $\langle Re_{\lambda} \rangle_b$. Therefore, the conditional average, $\langle C_{\varepsilon} \rangle_b |_{\langle Re_{\lambda} \rangle_b}$, when plotted against $\langle Re_{\lambda} \rangle_b$ in a log-log plot,  also shows a $-1$ slope (not shown). Indeed, this $-1$ slope of  the conditional average is {preserved as the CG level increases (not shown), and is therefore robust to spatial averaging.}   

The phase portrait in figure \ref{fig:ceps_relambda_phase}(d) suggests restoring dynamics: departures from the mean state are followed by evolution back towards it. This was described as self-regulation by \cite{yin2024dynamics,vassilicos2025scale} in which  $ C_{\varepsilon}^{-1}$ and $Re_{\lambda}$ are interpreted as the ratio of the energy loss rate of the large and the small scales, and the ratio of the energy contained in the large and small scales, respectively. They argued that these two quantities should increase and decrease together. 

However, the self-regulation hypothesis does not indicate whether this cycle is driven by intermittency in the large scale or in  the dissipative scale. The definitions of $C_{\varepsilon}$ and $Re_\lambda$ both contain $u_{rms}$ and $\varepsilon$, so they are not independent random variables. We find that this correlation between the largest and the smallest scales can be  more clearly illustrated by a timescale comparison. The cascade frequency, $u_{rms}/L_k$, and the Kolmogorov frequency, $(\varepsilon/\nu)^{1/2}$, are two (nominally) independent quantities, representing the timescales of the dynamics at the largest and the smallest scales in turbulence, respectively. The relation $C_{\varepsilon} \propto Re_\lambda^{-1}$  leads to the proportionality $(\varepsilon/\nu)^{1/2} \propto u_{rms}/L_k$. Figure \ref{fig:jpdfs_x_60}(b) shows the JPDF of the spatially local cascade frequency $u_{rms,b}/L_{k,b}$ and Kolmogorov frequency $(\langle \varepsilon' \rangle_b/\nu)^{1/2}$, at the finest CG level $l_b$. These two timescales show strong correlation with a slope of approximately 0.9. This correlation result taken together with the finding that  the driver of the phase dynamics is the kinetic energy rather than the dissipation (noted when discussing figure \ref{fig:ceps_relambda_phase}),  suggests that it is the large-scale frequency that modulates the  local Kolmogorov frequency. {This link between larger-scale structures and small-scale structures on an instantaneous basis is an appealing physical reason for the finding $C_\varepsilon \propto Re_\lambda^{-1}$ on a spatially local basis. }

The  JPDF contours at a value of  0.5 shown in figure \ref{fig:jpdfs_x_60}(c,d) show that the correlations and respective slopes in figure \ref{fig:jpdfs_x_60}(a,b) are well maintained as  $x$ increases, despite a shift of the JPDF centres toward a lower Reynolds number, consistent with the streamwise progression of the averaged state shown in figure \ref{fig:ceps_relambda}(b). This result further confirms that  the correlations $\langle C_{\varepsilon} \rangle_b \sim \langle Re_{\lambda} \rangle_b^{-1}$ and  $(\langle \varepsilon' \rangle_b/\nu)^{1/2} \sim u_{rms,b}/L_{k,b}$ are general and relatively independent of the wake Reynolds number. Although $\langle Re \rangle_b$ and both timescales decay downstream, the centres of the JPDF contours at all $x$-locations  remain approximately on the $-1$ slope line.   {Lastly, we point out that the domain considered, $[-2.5L_k, 2.5L_k]$, contains the turbulent/non-turbulent interface and the resultant intermittency. No effort has been made in this work to identify the turbulent/non-turbulent interface. Nevertheless, the scaling $\langle C_{\varepsilon} \rangle_b \sim \langle Re_{\lambda} \rangle_b^{-1}$ also appears to work at lower values of  local $\langle Re_{\lambda} \rangle_b$,  which correspond to the intermittently turbulent regions.}

\section{Conclusions} \label{conclusion} 

The high-Reynolds number wake past a circular disk has been studied numerically, up to a  downstream distance of $140$ diameters. The multiscale nature of wake turbulence is characterised and the interactions  {between large and small-scale parts}  are analysed. One-point statistics reach 
%the so-called complete 
complete self-similarity state by $x=30$, where the radial profiles of the mean velocity deficit, TKE, Reynolds shear stress, and dissipation rate all collapse in self-similar coordinates. Furthermore, two-point statistics, including the TKE and Reynolds shear stress spectra up to the inertial range, the transverse correlation function, the radial structure of the SPOD leading eigenmodes, and the inter-scale fluxes, also become self-similar.

The wake has coherent global modes but their  presence does not preclude self-similarity. In fact, the two strongest coherent modes, the VS mode and the DH mode, coexist with self-similar one-point statistics all the way to the end of the domain. 
That does not necessarily contradict with the classical suggestion that the memory of the initial and boundary conditions needs to be forgotten before self-similarity sets in \citep{townsend1976structure}. The relative intensity of the CS, particularly the VS mode at $(m,f)=(1,0.135)$ and the DH mode at $(2,0)$ with respect to the background turbulence (total TKE), has decayed to below $5\%$ by $x=40$ (figure \ref{fig:k_int_r}b), which is sufficient to satisfy,  if not the loss-of-memory assumption, a weak-influence approximation. The evolution of the CS has more influence on the mean wake than on turbulent statistics. The exchange of the relative dominance of the VS and DH modes  around $x=65$, coincides with a small variation in the self-similar profiles of the Reynolds shear stress and  the mean defect velocity, as well as a small change in the wake power-law exponents. However, the radial profiles and power laws of the TKE and dissipation do not seem to be affected. The precise effects of the CS on wake dynamics are subject to future studies that include control of the type and the strength of the CS. 

The influence of the anisotropic large scales decays rapidly with the wavenumber. The TKE spectra and Reynolds shear stress cospectra follow inertial scalings of $-5/3$ and $-7/3$, respectively, leading to a -2/3 decay of the cross-correlation coefficient. As a consequence, the inertial range of turbulence spectra at different $x$ and $r$ locations  collapse when the background turbulence intensity $k(x,r)$ is normalised out, leaving no obvious signature of the large scales. After the down-scale decay of anisotropy that accompanies the forward cascade, the cross-correlation stays at around 0.2 instead of further reaching a zero asymptote.

The evolution of the TKE is studied in terms of the standard one-point TKE balance and the filtering-based LKE and SKE balances, with a focus on the non-equilibrium scaling of dissipation ($\varepsilon \nsim U_k^3/L_k$) that is found here all the way up to $x = 140$. A triple decomposition separates the fluctuations into large- and small-scale components and their  actions on the mechanisms in the one-point TKE equation are tracked. Scale-local/linear mechanisms, such as advection, pressure diffusion, viscous diffusion, mean-flow production, and dissipation are exactly split into two parts as shown in \eqref{tke_decomp}. The scale-nonlocal/nonlinear turbulent transport, $T_t$, contains both spatial transport and inter-scale transfer, as  shown in \eqref{tt_decomp}. The inter-scale fluxes ${\it \Pi}_l$ and ${\it \Pi}_s$ viewed from the LKE and SKE's perspectives, have very similar area integrals at each $x$-location and filter size, and indicate a statistical forward cascade. When the filter scale is placed in the inertial range, the radial profiles of ${\it \Pi}_l$ and ${\it \Pi}_s$ become self-similar by $x=40$ and their radial maxima follow the classical scaling such that $C_{\it \Pi} = (\max_r {\it \Pi}) L_k / U_k^3$ is a constant. That being said, the non-constancy of $C_{\varepsilon}=\varepsilon L_k/U_k^3$ in the wake is not due to the failure of $U_k$ and $L_k$ to characterise the turbulence cascade, but a result of the disequilibrium $\varepsilon \not\approx {\it \Pi}$.

The origin of this disequilibrium between $\varepsilon$ and ${\it \Pi}$ is determined to be small-scale streamwise advection $A_s$, which remains non-negligible compared with the inter-scale transfer ${\it \Pi}_s$  {even} when the filter is in the inertial range. The statistical equilibrium between ${\it \Pi}$ and $\varepsilon$ has not been achieved by $x=140$, where the peak velocity deficit has decayed to $U_d \approx 0.01U_{\infty}$ and the peak normalised shear has decayed to $Sk/\varepsilon \approx 0.5$. It is hence unlikely that $C_{\varepsilon}$ will become a constant further downstream.  Despite $\varepsilon \not\approx {\it \Pi}$, we find   Kolmogorov $-5/3$ scaling in the inertial range of the energy spectra, whose  derivation typically invokes such an equilibrium. It is remarkable that the $-5/3$ range and the disequilibrium coexist in almost the entire wake.

The non-equilibrium dissipation coefficient is further examined and found to be anti-correlated to the Taylor Reynolds number as $C_{\varepsilon}(x) \sim Re_\lambda^{-1}(x)$, which stems from spatially localised turbulence statistics. At all $x$-locations, the coarse-grained, locally averaged statistics follow joint probability distributions centred around $\langle C_{\varepsilon} \rangle_b \sim \langle Re_\lambda \rangle_b^{-1}$, which results from a strong correlation between the local Kolmogorov frequency $\sqrt{\langle \varepsilon'\rangle_b/\nu}$ and the large-eddy frequency  $u_{rms,b}/L_{k,b}$. In this sense, the large scales are able to modulate the dissipation rate. Instantaneous non-equilibrium  oscillations are evident from the phase-space evolution of $(\langle Re_\lambda \rangle_b,\langle C_{\varepsilon} \rangle_b)$, where the perturbations are along the $\langle C_{\varepsilon} \rangle_b \sim \langle Re_\lambda \rangle_b^{-1}$ slope and driven by the off-mean local TKE. Such an instantaneous/local disequilibrium is general to turbulence and can be averaged into a statistical equilibrium in forced isotropic turbulence \citep{goto2015energy}, but remains visible in the present wake as small-scale advection prevents dissipation from settling into an equilibrium with the downscale energy cascade.

%%%%%%%%%%%%%%%%%%%%%%%%%%%%%%%%%%%%%%%%%%%%%%%%%%%%%%
\backsection[Funding]{The authors did not receive any financial support for this work. } 
\backsection[Declaration of interest]{The authors report no conflict of interest.}
\backsection[Author ORCIDs]{\\ 
\noindent J. Liu \url{https://orcid.org/0000-0003-4133-0930} \\
\noindent S. Sarkar \url{https://orcid.org/0000-0002-9006-3173}}

%%%%%%%%%%%%%%%%%%%%%%%%%%%%%%%%%%%%%%%%%%%%%%%%%%%%%%
\appendix

\section{Full expressions of the TKE equation terms} \label{tke_full}

In the present cylindrical coordinates with azimuthal homogeneity and vanishing radial mean velocity, terms in the TKE equation \eqref{tke} can be expanded as \begin{align}
  A   & = - \bar{u}_x \frac{\partial k}{\partial x} - \bar{u}_r \frac{\partial k}{\partial r}  \\
  T_p & = - \left( \frac{\partial \overline{u'_x p'}}{\partial x} + \frac{1}{r} \frac{\partial (r \overline{u'_r p'})}{\partial r} \right) \\
  T_t & = - \left( \frac{\partial \overline{k' u'_x}}{\partial x} + \frac{1}{r} \frac{\partial 
  (r \overline{k'u'_r})}{\partial r}  \right) \\
  T_v & = \frac{1}{Re} \left[ \frac{\partial^2 k}{\partial x^2} + \frac{1}{r} \frac{\partial }{\partial r} \left( r \frac{\partial k}{\partial r} \right)  \right]  \\
  P & =   - \left( \overline{u_x'u_x'} \frac{\partial \bar{u}_x}{\partial x} + \overline{u_x' u_r'} \frac{\partial \bar{u}_x}{\partial r} \right)   \\
    \varepsilon & =  \frac{1}{Re} \left[ 
  \overline{\left(\frac{\partial u'_x}{\partial x}\right)^2} +
  \overline{\left(\frac{\partial u'_x}{\partial r}\right)^2} +
  \overline{\left(\frac{1}{r} \frac{\partial u'_x}{\partial \theta}\right)^2} +
  \overline{\left(\frac{\partial u'_r}{\partial x}\right)^2} +
  \overline{\left(\frac{\partial u'_r}{\partial r}\right)^2} + \right. \nonumber \\
  & \;\;\;\; \;\;\;\; \;\;\;\;\;
  \left. 
  \overline{\left(\frac{1}{r} \frac{\partial u'_r}{\partial \theta} - \frac{u'_{\theta}}{r}\right)^2}  + 
  \overline{\left(\frac{\partial u'_{\theta}}{\partial x}\right)^2}  +
  \overline{\left(\frac{\partial u'_{\theta}}{\partial r}\right)^2}  + 
  \overline{\left(\frac{1}{r} \frac{\partial u'_{\theta}}{\partial \theta} + \frac{u'_r}{r}\right)^2} \right]
\end{align} where the components of $P$ are \begin{align}
  P_{xx} & = - \left( \overline{u_x'u_x'} \frac{\partial \bar{u}_x}{\partial x} + \overline{u_x' u_r'} \frac{\partial \bar{u}_x}{\partial r} \right)   \\
  P_{rr} & = - P_{\theta \theta} =  \frac{\overline{u_r' u_{\theta}'^2 }}{r} 
\end{align} such that $P=P_{xx}+P_{rr}+P_{\theta \theta}$. The curvature-related, inter-component transport $P_{rr}  = - P_{\theta \theta}$ re-distributes energy between $u'_r$ and $u'_{\theta}$, in addition to the pressure re-distribution, while having no net contribution to $k$. For $k$, all the production comes from $P_{xx}$.

\section{Details of the SPOD formulation} \label{spod_app}

SPOD is a modal decomposition technique that extracts coherent modes based on the two-point, two-time correlation tensor \eqref{2p2t}. In homogeneous turbulence, it reduces to $R_{ij}(\bm{x}'-\bm{x},t'-t)=R_{ij}(\bm{\xi},\tau)$, which is the Fourier counterpart of the cospectral density tensor $\hat{S}_{ij}(\bm{\kappa},\omega)$, where $\kappa_i = 2\pi/\xi_i$ is the wavenumber and $\omega = 2\pi/\tau$ is the angular frequency. In inhomogeneous turbulence, Fourier transforms are taken in the statistically homogeneous directions, e.g. in time. That leads to the cross-spectral density tensor $\hat{S}_{ij}(\bm{x},\bm{x}', f)$, where $f=\omega/2\pi$ is the dimensionless frequency. Different from the space-only inner-product in POD, the SPOD inner-product is its space-time analog, 
\begin{equation}
  (\bm{q}^{}_1, \bm{q}^{}_2)_{\bm W} = \int_{-\infty}^{\infty} \int_{\bm \Omega} \bm{q}_2^*(\bm{x},t) \bm{W}(\bm{x}) \bm{q}^{}_1(\bm{x},t) \, {\rm d}{\bm{x}} {\rm d}t  ,  \label{inner}
\end{equation} where $\bm{q}^{}_1$ and $\bm{q}^{}_2$ are two spatial functions, ${\bm{\Omega}}$ is the bounded spatial domain, and $\bm{W}(\bm{x})$ is a weight function. Based on this inner-product, a Fredholm eigenvalue problem (EVP) \begin{equation}
    \int_{-\infty}^{\infty} \int_{\bm \Omega}   R_{ij}(\bm{x},\bm{x}',t, t')  \bm{W}(\bm{x}') \bm{\psi}^{(i)}(\bm{x}',t) \, {\rm d}{\bm{x}'} {\rm d}t' = \lambda^{(i)} \bm{\psi}^{(i)}(\bm{x},t)
\end{equation} is defined, where $i$ is the eigenvalue index. With the temporal Fourier transform \begin{equation}
  \hat{S}_{ij}(\bm{x},\bm{x}', f) = \int_{-\infty}^{\infty} R_{ij}(\bm{x},\bm{x}', \tau) \exp(-{\rm i}2\pi f \tau) \, {\rm d}\tau 
\end{equation} and the eigenvector transform $\hat{\bm{\phi}}^{(i)} = \bm{\psi}^{(i)} \exp(-{\rm i}2 \pi f \tau)$, the SPOD EVP becomes \begin{equation}
  \int_{\bm \Omega}   \hat{S}_{ij}(\bm{x},\bm{x}', f)  \bm{W}(\bm{x}') \hat{\bm{\phi}}^{(i)}(\bm{x}', f) \, {\rm d}{\bm{x}'}  = \lambda^{(i)} \hat{\bm{\phi}}^{(i)}(\bm{x},f) . 
\end{equation} The solution of the EVP leads to eigenmodes $\{\hat{\bm{\phi}}^{(i)}(\bm{x},f)\}_{i=1}^{\infty}$ that are spatially orthogonal at each frequency.  

In the present wake, axisymmetry is exploited via an azimuthal Fourier decomposition since POD modes degenerate to Fourier modes in periodic directions. The extraction of 3-D global modes is enabled by an on-the-fly azimuthal Fourier decomposition, followed by storage of the leading $|m| \le 3$ modes at the same sampling frequency as other stored data planes. The spatial part of the inner-product \eqref{inner} is defined in the $(x,r)$ domain such that the SPOD modes are 3-D global modes, which is an advance on  the 2-D modal analysis by \cite{nidhan2020spectral}. Each SPOD eigenmode is indexed by the wavenumbers $(m,f)$ and the modal index $i$, and is a function of $(x,r)$.

The numerical implementation of SPOD follows \cite{towne2018spectral,schmidt2020guide,nidhan2020spectral}. The azimuthal-temporal Fourier modes $\{\hat{\bm{u}}_{mf}\}=[\hat{u}_{x,mf}^{T} \; \hat{u}_{r,mf}^{T} \; \hat{u}_{\theta,mf}^{T}]^{T}$  are arranged into the snapshot matrix \begin{equation}
  \hat{\bm{U}}_{mf} = [\hat{\bm{u}}_{mf}^{(1)} \,  \hat{\bm{u}}_{mf}^{(2)}\,  \, ...\,   \hat{\bm{u}}_{mf}^{(N_{blk})}], 
\end{equation} where $N_{blk}=27$ is the number of blocks in the temporal direction and is equal to the number of eigenmodes. The cross-spectral density matrix is constructed as \begin{equation}
  \hat{\bm{S}}_{mf} = \hat{\bm{U}}_{mf}^{H} {\bm{W}} \hat{\bm{U}}^{}_{mf}, 
\end{equation} where $(\cdot)^{H}$ denotes the Hermitian transpose and the weight matrix ${\bm{W}}$ contains the numerical quadrature coefficients of the non-uniform grid. By solving the method-of-snapshot eigenvalue problem \citep{sirovich1987turbulence}  \begin{equation}
  \hat{\bm{S}}_{mf} \hat{\bm{\Psi}}_{mf} = \hat{\bm{\Psi}}_{mf} \hat{\bm{\Lambda}}_{mf}, 
\end{equation} the eigenvalues $\hat{\bm{\Lambda}}_{mf}={\rm diag}(\lambda^{(1)}_{mf}, \lambda^{(2)}_{mf}, ...,\lambda^{(N_{blk})}_{mf})$ and the method-of-snapshot eigenvectors $\hat{\bm{\Psi}}_{mf}$ are obtained. The eigenvectors of the cross-spectral density tensor ($\hat{S}_{ij}$) are then recovered according to $\hat{\bm{\Phi}}_{mf} = \hat{\bm{U}}_{mf} \hat{\bm{\Psi}}_{mf} \hat{\bm{\Lambda}}_{mf}^{-1/2} = [\hat{\bm{\phi}}_{mf}^{(1)}\,\hat{\bm{\phi}}_{mf}^{(2)}\, ... \, \hat{\bm{\phi}}_{mf}^{(N_{blk})}]$, where the mutually orthonormal SPOD modes $\{ \hat{\bm{\phi}}^{(i)}_{mf}\}$ are ranked according to energy optimality.

\section{The triply decomposed momentum and kinetic energy equations} \label{deri_td}

In this Appendix, we derive the momentum and the kinetic energy equations \eqref{tke}, \eqref{lke}, \eqref{ske} under the TD ($u_i = \bar{u}_i + u'_i, \,  u'_i = \tilde{u}_i + u''_i $), and discuss the energy transfer mechanisms revealed by such a decomposition. Tensor notations are presented for generality, whereas the detailed expansions of individual terms in the cylindrical coordinates are similar to those in Appendix \ref{tke_full}.

The mean continuity and momentum equations are  \begin{align}
  \frac{\partial \bar{u}_i}{\partial x_i} & = 0 \\ 
  \frac{\partial \bar{u}_i}{\partial t} + \bar{u}_j \frac{\partial \bar{u}_i}{\partial x_j} 
  & = -  \frac{\partial \bar{p}}{\partial x_i} - \frac{\partial }{\partial x_j}(    \overline{u_i u_j} -\bar{u}_i \bar{u}_j ) + \frac{1}{Re} \frac{\partial^2 \bar{u}_i}{\partial x_j^2} ,
\end{align} where the Reynolds stress $\tau^{Rey}_{ij}  = \overline{u'_i u'_j} = \overline{u_i u_j} -\bar{u}_i \bar{u}_j$ is written as generalised residual stress \citep{germano1992turbulence}.   

The fluctuation equations are obtained by subtracting the mean equations from the full equations: \begin{align}
   \frac{\partial {u}'_i}{\partial x_i} & = 0 \label{wino00} \\ 
  \frac{\partial {u}'_i}{\partial t} + \bar{u}_j \frac{\partial {u}'_i}{\partial x_j} + u'_j\frac{\partial \bar{u}_i}{\partial x_j} 
  & = - \frac{\partial p'}{\partial x_i}
  - \frac{\partial }{\partial x_j} \left( u'_i u'_j - \overline{u'_iu'_j} \right)  
  + \frac{1}{Re} \frac{\partial^2 u'_i}{\partial x_j^2}. \label{wino0}
\end{align} When filtered, they lead to the large-scale equations: \begin{align}
  \frac{\partial \tilde{u}_i}{\partial x_i} & = 0  \\ 
  \frac{\partial \tilde{u}_i}{\partial t} + \bar{u}_j \frac{\partial \tilde{u}_i}{\partial x_j}
  + \tilde{u}_j \frac{\partial \bar{u}_i}{\partial x_j}
  & = -  \frac{\partial \tilde{p}}{\partial x_i} 
  - \frac{\partial }{\partial x_j} \left(   \widetilde{u'_i u'_j} - {\overline{u'_iu'_j}} \right)   
  +  \frac{1}{Re} \frac{\partial^2 \tilde{u}_i}{\partial x_j^2} , 
  \label{wino}
\end{align} with the latter also written as \begin{equation}
  \begin{aligned}
  \frac{\partial \tilde{u}_i}{\partial t} + \bar{u}_j \frac{\partial \tilde{u}_i}{\partial x_j}
  + \tilde{u}_j \frac{\partial \bar{u}_i}{\partial x_j}
  + \tilde{u}_j \frac{\partial \tilde{u}_i}{\partial x_j}
  & = - \frac{\partial \tilde{p}}{\partial x_i} 
  - \frac{\partial }{\partial x_j} \left(   (\widetilde{u'_i u'_j} - \tilde{u}_i \tilde{u}_j) - {\overline{u'_iu'_j}} \right)   
  + \frac{1}{Re} \frac{\partial^2 \tilde{u}_i}{\partial x_j^2}  . 
  \label{wino1}
\end{aligned}
\end{equation}
Here $\tilde{u}_i=\widetilde{(u'_i)}$ denotes the filtered velocity and $u''_i=u'_i - \tilde{u}_i$ is the sub-filter residual. Same notations are used for the pressure. In addition to the Reynolds stress, there is an additional sub-filter-scale (SFS) stress $\tau^{SFS}_{ij}  = \widetilde{u'_i u'_j} - \tilde{u}_i \tilde{u}_j  
$ resulting from coarse-graining (filtering) the nonlinear term.  

The small-scale equations are obtained by subtracting the large-scale equations from the fluctuation equations: 
\begin{align}
  \frac{\partial {u}''_i}{\partial x_i} & = 0 \\ 
    \frac{\partial {u}''_i}{\partial t} 
  + \bar{u}_j \frac{\partial {u}''_i}{\partial x_j} 
  + u''_j\frac{\partial \bar{u}_i}{\partial x_j} 
   & = - \frac{\partial p''}{\partial x_i} - \frac{\partial }{\partial x_j} ( u'_i u'_j - \widetilde{u'_i u'_j}) 
   + \frac{1}{Re} \frac{\partial^2 u''_i }{\partial x_j^2} , 
    \label{wino2}
\end{align} with the latter also written as 
\begin{equation}
    \frac{\partial {u}''_i}{\partial t} 
  + \bar{u}_j \frac{\partial {u}''_i}{\partial x_j} 
  + u''_j\frac{\partial \bar{u}_i}{\partial x_j} 
  + \tilde{u}_j \frac{\partial {u}''_i}{\partial x_j} 
  + u''_j\frac{\partial \tilde{u}_i}{\partial x_j} 
   = - \frac{\partial p''}{\partial x_i} 
  % \\   & \;\;\;\;
   - \frac{\partial }{\partial x_j} 
  ( u''_i u''_j + \tilde{u}_i \tilde{u}_j - \widetilde{u'_i u'_j}  ) 
  + \frac{1}{Re} \frac{\partial^2 u''_i}{\partial x_j^2 } \label{sform}
\end{equation}

The nonlinear stress-divergence term in \eqref{sform} takes the same form as that in the $u'_i$ momentum equation \eqref{wino0}, where both contain the self-interaction stress, $u'_i u'_j$ or $u''_i u''_j$,  and the (negative-signed) cross-scale interaction stress, $-\tau_{ij}^{Rey}$ or $-\tau_{ij}^{SFS}$. Multiplying each equation with respective velocity components and taking the ensemble average, the triply decomposed kinetic energy equations \eqref{tke}, \eqref{lke}, and \eqref{ske} in the main text are derived. In the two limiting cases, $u'_i=\tilde{u}_i$ or $u'_i = u''_i$, the two-point LKE equation \eqref{lke} or the SKE equation \eqref{ske} reduces to the one-point TKE equation \eqref{tke}.  

It'is straightforward to individually verify that corresponding terms in \eqref{lke} and \eqref{ske} add into those in \eqref{tke}. But a more constructive approach will be keeping track of the actions of the coarse-graining operation on each term of \eqref{tke}. Consider the advection term $\bar{u}_j \partial_j u'_i$ in the perturbation equation, which is coarse-grained into $\bar{u}_j \partial_j \tilde{u}_i$ and the residual $\bar{u}_j \partial_j u''_i$ (filtering then subtracting). The equations of ${u}'_i$, $\tilde{u}_i$, and $u''_i$ are then multiplied by ${u}'_i$, $\tilde{u}_i$, and $u''_i$ respectively, then ensemble averaged to the TKE, LKE, and SKE equations, during which the advection terms become $\bar{u}_j \partial_j (\overline{u'_i u'_i }/2)$, $\bar{u}_j \partial_j (\overline{\tilde{u}_i \tilde{u}_i }/2)$, and $\bar{u}_j \partial_j (\overline{u''_i u''_i }/2)$, respectively. That leads to $A=A_l+A_s$ and similar in the rest of \eqref{tke_decomp}.

The quadratic nonlinear operator $\mathcal{N}(u'_i) \triangleq -\partial_j(u'_i u'_j)$ in the fluctuation momentum equation is what leads to the extra dynamics reflected in \eqref{tt_decomp}.  
% which can be shown by considering the
% which is not a stand-alone tensor equality but a result of the fluctuating, large-, and small-scale momentum constraints involving 
It is coarse-grained into $\mathcal{N}_l = \tilde{\mathcal{N}} = - \partial_j(\widetilde{u'_i u'_j})$ and $\mathcal{N}_s = \mathcal{N} - \mathcal{N}_l = - \partial_j (u'_i u'_j - \widetilde{u'_i u'_j})$ in the right-hand-side of \eqref{wino} and \eqref{wino2}. Since $ \overline{u''_i \mathcal{N}_l} = \overline{\tilde{u}_i \mathcal{N}_s} = 0$  due to the spectral sharp filter ($u''_i \mathcal{N}_l$ and $\tilde{u}_i \mathcal{N}_s$ cannot contribute to the zero wavenumber), the statistical condition \begin{equation}
  \overline{u'_i \mathcal{N}} = \overline{\tilde{u}_i \mathcal{N}_l} + \overline{u''_i \mathcal{N}_s}, 
\end{equation} leads to \eqref{tt_decomp}, where \begin{align}
  \overline{u'_i \mathcal{N}}  & = - \overline{u'_i \frac{\partial u'_i u'_j}{\partial x_j}} = - \frac{\partial}{\partial x_j} \left(\frac{1}{2}\overline{ u'_i u'_i u'_j } \right) = T_t \\ 
  \overline{\tilde{u}_i \mathcal{N}_l} & = - \overline{\tilde{u}_i \frac{\partial \widetilde{u'_i u'_j}}{\partial x_j}}  =  - \overline{\tilde{u}_i \frac{\partial }{\partial x_j}(\widetilde{u'_i u'_j} + \tilde{u}_i \tilde{u}_j 
  - \tilde{u}_i \tilde{u}_j) } \nonumber \\
  & = - \frac{\partial}{\partial x_j} \left( \frac{1}{2} \overline{\tilde{u}_i \tilde{u}_i \tilde{u}_j} \right) - \overline{\tilde{u}_i \frac{\partial }{\partial x_j} \left(\widetilde{u'_i u'_j} - \tilde{u}_i \tilde{u}_j \right)}  \nonumber \\   & = T_{t,l} + T_{q,l} + {\it \Pi}_l \\ 
  \overline{u''_i \mathcal{N}_s} & = - \overline{u''_i \frac{\partial }{\partial x_j} \left( u'_i u'_j - \widetilde{u'_i u'_j} \right)}  \nonumber \\ 
  & = - \frac{\partial }{\partial x_j} \frac{1}{2} \overline{u''_i u''_i (\tilde{u}_j + u''_j)} - \overline{u''_i u''_j \frac{\partial \tilde{u}_i}{\partial x_j}} -  \overline{u''_i \frac{\partial }{\partial x_j} \left( \tilde{u}_i \tilde{u}_j - \widetilde{u'_i u'_j} \right)} \nonumber \\
  & = T_{t,s} + {\it \Pi}_s + T_{q,s} 
\end{align} The implication of \eqref{tt_decomp} is that the turbulent transport $T_t$ involves spatial transport and inter-scale transfer, and both are nonlinear. 

Lastly, we point out the difference between the present filtering-based TD and the conventional phase-average-based TD  \citep{hussain1970mechanics,reynolds1972mechanics}. In the  filtering approach, the filtered nonlinear term is
 \begin{equation}
  \widetilde{u'_i u'_j} = \widetilde{\tilde{u}_i \tilde{u}_j} + \widetilde{u''_i u''_j} + \widetilde{u''_i \tilde{u}_j} + \widetilde{\tilde{u}_i u''_j} \ne \tilde{u}_i \tilde{u}_j + \widetilde{u''_i u''_j}  \label{llhh}
\end{equation} because low-low wavenumber interactions can alias to wavenumbers higher than the cutoff and low-high wavenumber interactions can also alias back to wavenumbers below the cutoff. The last inequality in \eqref{llhh} is an equality in \cite{hussain1970mechanics,reynolds1972mechanics}, with $\widetilde{\tilde{u}_i \tilde{u}_j} =  \tilde{u}_i \tilde{u}_j$ and $ \widetilde{u''_i \tilde{u}_j} = \widetilde{\tilde{u}_i u''_j}=0$.  This is because when $\widetilde{(\cdot)}$ is taken to be the phase average, the TD is a deterministic-stochastic decomposition of the signal similar to the Reynolds (ensemble) average, and  the phase average of the correlation of the deterministic and the stochastic components is zero.  
% In a spectral view, signals oscillating at the frequency of the reference signal and its higher harmonics will be phase-averaged into the deterministic, or the `wave' part, while the rest are termed `stochastic'. The present TD is based on filtering, which is more a projection in the spectral space, especially when a Fourier sharp filter is used.  
When $\widetilde{u'_i u'_j} = \tilde{u}_i \tilde{u}_j + \widetilde{u''_i u''_j}$ is satisfied, \eqref{wino1} recovers the large-scale equation in \cite{reynolds1972mechanics} and the nonlinear terms in \eqref{lke}-\eqref{ske} also recover corresponding terms in the kinetic energy equations therein. 

\section{Effect of filter selection} \label{ftype}

\begin{figure}[thb]
  \centering 
  \captionsetup[subfloat]{farskip=-2pt,captionskip=0pt}
    \subfloat{{\includegraphics[width=0.45\linewidth]{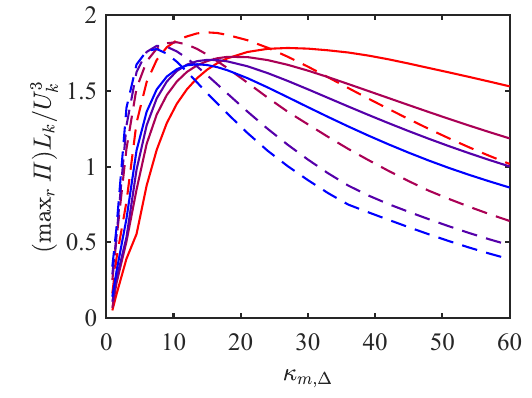}}} 
  \caption{Similar to figure \ref{fig:fsize_ftype}(b), but filtering on $\kappa_m$ instead of $m$ and using a Gaussian filter. At $x=30,60,90,120$, with the same colour code as figure \ref{fig:fsize_ftype}. }  
  \label{fig:fsize_ftype_vfr_gau} 
\end{figure} 

As suggested by the scaling of the spectra and cospectra shown in  figures \ref{fig:spec_azim}-\ref{fig:cospec_dspec}, {turbulence beyond the inertial range is better characterised in the spectral space by $\kappa_m=m/r$}, as well as the anisotropy. It becomes necessary to examine the inter-scale transfer when filtering is performed in the $(\kappa_m,f)$ space. Since the filtering arc wavenumber needs to be converted into the standard azimuthal wavenumber as $m_\Delta = \kappa_{m,\Delta} r$, which is no longer a guaranteed integer, a Gaussian filter is used instead of a spectral sharp one, to avoid sudden jumps in $m_\Delta(r)$. Moreover, a Gaussian filter has the advantage of non-oscillating, localised supports in both spatial and spectral domains. The 2-D Gaussian filter used in this work has the spectral transfer function of \begin{equation}
  \hat{G}(m,f) = \exp \left[ -\frac{(\Delta_m m)^2}{24} - \frac{(\Delta_f f)^2}{24} \right], 
\end{equation} where $\Delta_m=2\pi/m_{\Delta}$ and $\Delta_f = 2\pi/f_\Delta$. The same filter aspect ratio $f_{\Delta}=\kappa_{m,\Delta}/6$ is kept. 

Figure \ref{fig:fsize_ftype_vfr_gau} shows the filter-size dependence of the inter-scale fluxes when filtering the arc wavenumber $\kappa_{m}=m/r$. In this case, the filter scales corresponding to the largest inter-scale fluxes are different for ${\it \Pi}_l$ and ${\it \Pi}_s$, and have a clear $x$-dependence. The wavenumbers $\kappa_m$ at which the greatest inter-scale fluxes are achieved, depending on whether it is ${\it \Pi}_l$ or ${\it \Pi}_s$ and the streamwise location, roughly fall in the range $\kappa_m \approx 7 \sim 25$, which are still sufficiently into the inertial range and past the end of the $-2/3$ decay of anisotropy (see figure \ref{fig:cospec_dspec}). However, unlike the spectral sharp filter in figure \ref{fig:fsize_ftype}(a-b), the magnitudes of ${\it \Pi}_l$ and ${\it \Pi}_s$ continue to disagree even at high wavenumbers. This is possibly due to greater spatial transport as the result of the Gaussian filter and that the turbulent transport decomposition \eqref{tt_decomp} is no longer satisfied precisely because of the spectral overlap between $\hat{G}$ and $1-\hat{G}$. 

Moreover, since a Gaussian filter is not idempotent and not a projector, the TKE cannot be exactly decomposed into LKE and SKE, and the split budget analysis performed above does not strictly apply. {In all, we find that filtering on the azimuthal wavenumber $m$ instead of $\kappa_m=m/r$ provides the most consistent results.} 

%%%%%%%%%%%%%%%%%%%%%%%%%%%%%%%%%%%%%%%%%%%%%%%%%%%%%%
\bibliographystyle{jfm}
\bibliography{jjl} 

@article{ortiz2021high,
  title={High-{Reynolds}-number wake of a slender body},
  author={Ortiz-Tarin, JL and Nidhan, S and Sarkar, S},
  journal={Journal of Fluid Mechanics},
  volume={918},
  pages={A30},
  year={2021},
  publisher={Cambridge University Press}
}

@article{towne2018spectral,
  title={Spectral proper orthogonal decomposition and its relationship to dynamic mode decomposition and resolvent analysis},
  author={Towne, Aaron and Schmidt, Oliver T and Colonius, Tim},
  journal={Journal of Fluid Mechanics},
  volume={847},
  pages={821--867},
  year={2018},
  publisher={Cambridge University Press}
}

@article{schmidt2020guide,
  title={Guide to spectral proper orthogonal decomposition},
  author={Schmidt, Oliver T and Colonius, Tim},
  journal={AIAA Journal},
  volume={58},
  number={3},
  pages={1023--1033},
  year={2020},
  publisher={American Institute of Aeronautics and Astronautics}
}

@article{yang2006embedded,
  title={An embedded-boundary formulation for large-eddy simulation of turbulent flows interacting with moving boundaries},
  author={Yang, Jianming and Balaras, Elias},
  journal={Journal of Computational Physics},
  volume={215},
  number={1},
  pages={12--40},
  year={2006},
  publisher={Elsevier}
}

@article{balaras2004modeling,
  title={Modeling complex boundaries using an external force field on fixed {Cartesian} grids in large-eddy simulations},
  author={Balaras, Elias},
  journal={Computers \& Fluids},
  volume={33},
  number={3},
  pages={375--404},
  year={2004},
  publisher={Elsevier}
}

@book{lumley1970stochastic,
  title={Stochastic Tools in Turbulence},
  author={Lumley, John L},
  year={1970},
  publisher={Academic Press},
  location={New York}
}

@article{germano1992turbulence,
  title={Turbulence: the filtering approach},
  author={Germano, Massimo},
  journal={Journal of Fluid Mechanics},
  volume={238},
  pages={325--336},
  year={1992},
  publisher={Cambridge University Press}
}

@article{cimarelli2021spatially,
  title={Spatially evolving cascades in temporal planar jets},
  author={Cimarelli, A and Mollicone, J-P and Van Reeuwijk, M and De Angelis, E},
  journal={Journal of Fluid Mechanics},
  volume={910},
  pages={A19},
  year={2021},
  publisher={Cambridge University Press}
}

@article{de2014large,
  title={Large eddy simulation of the near to intermediate wake of a heated sphere at {$Re$} = 10,000},
  author={de Stadler, Matthew B and Rapaka, Narsimha R and Sarkar, Sutanu},
  journal={International Journal of Heat and Fluid Flow},
  volume={49},
  pages={2--10},
  year={2014},
  publisher={Elsevier}
}

@article{saunders2022decay,
  title={Decay of the drag wake of a sphere at {Reynolds} number $10^5$},
  author={Saunders, D Curtis and Britt, Justen A and Wunsch, Scott},
  journal={Experiments in Fluids},
  volume={63},
  number={4},
  pages={71},
  year={2022},
  publisher={Springer}
}

@article{cho2018scale,
  title={Scale interactions and spectral energy transfer in turbulent channel flow},
  author={Cho, Minjeong and Hwang, Yongyun and Choi, Haecheon},
  journal={Journal of Fluid Mechanics},
  volume={854},
  pages={474--504},
  year={2018},
  publisher={Cambridge University Press}
}

@article{mizuno2016spectra,
  title={Spectra of energy transport in turbulent channel flows for moderate {Reynolds} numbers},
  author={Mizuno, Yoshinori},
  journal={Journal of Fluid Mechanics},
  volume={805},
  pages={171--187},
  year={2016},
  publisher={Cambridge University Press}
}

@article{hussain1970mechanics,
  title={The mechanics of an organized wave in turbulent shear flow},
  author={Hussain, Abul Khair Muhammad Fazle and Reynolds, William C},
  journal={Journal of Fluid Mechanics},
  volume={41},
  number={2},
  pages={241--258},
  year={1970},
  publisher={Cambridge University Press}
}

@article{reynolds1972mechanics,
  title={The mechanics of an organized wave in turbulent shear flow. Part 3. Theoretical models and comparisons with experiments},
  author={Reynolds, William C and Hussain, A K M Fazle},
  journal={Journal of Fluid Mechanics},
  volume={54},
  number={2},
  pages={263--288},
  year={1972},
  publisher={Cambridge University Press}
}

@article{yin2024dynamics,
  title={Dynamics of turbulent energy and dissipation in channel flow},
  author={Yin, Le and Hwang, Yongyun and Vassilicos, John Christos},
  journal={Journal of Fluid Mechanics},
  volume={996},
  pages={A12},
  year={2024},
  publisher={Cambridge University Press}
}

@incollection{vassilicos2025scale,
  author    = {Vassilicos, John Christos and Laval, Jean-Philippe},
  title     = {Scale-by-Scale Nonequilibrium in Turbulent Flows},
  booktitle = {Coarse Graining Turbulence: Modeling and Data-Driven Approaches and their Applications},
  editor    = {Grinstein, Fernando F. and Pereira, Filipe S. and Germano, Massimo},
  publisher = {Cambridge University Press},
  year      = {2025},
  chapter   = {11},
  pages     = {333--354},
  doi       = {10.1017/9781009377379.014}
}

@article{seoud2007dissipation,
  title={Dissipation and decay of fractal-generated turbulence},
  author={Seoud, RE and Vassilicos, JC},
  journal={Physics of Fluids},
  volume={19},
  number={10},
  year={2007},
  publisher={AIP Publishing}
}

@article{valente2014origin,
  title={Origin of the imbalance between energy cascade and dissipation in turbulence},
  author={Valente, P C and Onishi, R and Da Silva, CB},
  journal={Physical Review E},
  volume={90},
  number={2},
  pages={023003},
  year={2014},
  publisher={APS}
}

@article{taylor1935statistical,
  title={Statistical theory of turbulence},
  author={Taylor, Geoffrey Ingram},
  journal={Proceedings of the Royal Society of London. A},
  volume={151},
  pages={421--444},
  year={1935},
  publisher={The Royal Society London}
}

@article{valente2015energy,
  title={The energy cascade in grid-generated non-equilibrium decaying turbulence},
  author={Valente, P C and Vassilicos, J C},
  journal={Physics of Fluids},
  volume={27},
  number={4},
  year={2015},
  publisher={AIP Publishing}
}

@article{hearst2014scale,
  title={Scale-by-scale energy budget in fractal element grid-generated turbulence},
  author={Hearst, R Jason and Lavoie, Philippe},
  journal={Journal of Turbulence},
  volume={15},
  number={8},
  pages={540--554},
  year={2014},
  publisher={Taylor \& Francis}
}

@article{hearst2014decay,
  title={Decay of turbulence generated by a square-fractal-element grid},
  author={Hearst, R Jason and Lavoie, Philippe},
  journal={Journal of Fluid Mechanics},
  volume={741},
  pages={567--584},
  year={2014},
  publisher={Cambridge University Press}
}

@article{valente2012universal,
  title={Universal dissipation scaling for nonequilibrium turbulence},
  author={Valente, Pedro Cardoso and Vassilicos, John Christos},
  journal={Physical Review Letters},
  volume={108},
  number={21},
  pages={214503},
  year={2012},
  publisher={APS}
}

@article{noriega2026turbulent,
  title={Turbulent diffusion--cascade interaction},
  author={Noriega, Ernesto Fuentes and Vassilicos, John Christos},
  journal={Journal of Fluid Mechanics},
  volume={1028},
  pages={A3},
  year={2026},
  publisher={Cambridge University Press}
}

@article{nishimoto2026spatially,
  title={Spatially local dissipation scaling in grid turbulence from direct numerical simulations},
  author={Nishimoto, Yohei and Nagata, Koji and Watanabe, Tomoaki and Zhou, Yi},
  journal={Physics of Fluids},
  volume={38},
  number={4},
  year={2026},
  publisher={AIP Publishing}
}

@article{sreenivasan1984scaling,
  title={On the scaling of the turbulence energy dissipation rate},
  author={Sreenivasan, Katepalli R},
  journal={The Physics of Fluids},
  volume={27},
  number={5},
  pages={1048--1051},
  year={1984},
  publisher={AIP Publishing}
}

@article{sirovich1987turbulence,
  title={Turbulence and the dynamics of coherent structures. {I}. {Coherent} structures},
  author={Sirovich, Lawrence},
  journal={Quarterly of Applied Mathematics},
  volume={45},
  number={3},
  pages={561--571},
  year={1987}
}

@article{lumley1967similarity,
  title={Similarity and the turbulent energy spectrum},
  author={Lumley, J L},
  journal={The physics of fluids},
  volume={10},
  number={4},
  pages={855--858},
  year={1967},
  publisher={AIP Publishing}
}

@article{corrsin1958local,
  title={Local isotropy in turbulent shear flow},
  author={Corrsin, Stanley},
  journal={Res. Memo. NACA},
  volume={58B11},
  year={1958}
}

@article{apostolidis2022scalings,
  title={Scalings of turbulence dissipation in space and time for turbulent channel flow},
  author={Apostolidis, Argyrios and Laval, Jean-Philippe and Vassilicos, J C},
  journal={Journal of Fluid Mechanics},
  volume={946},
  pages={A41},
  year={2022},
  publisher={Cambridge University Press}
}

@article{saddoughi1994local,
  title={Local isotropy in turbulent boundary layers at high {Reynolds} number},
  author={Saddoughi, Seyed G and Veeravalli, Srinivas V},
  journal={Journal of Fluid Mechanics},
  volume={268},
  pages={333--372},
  year={1994},
  publisher={Cambridge University Press}
}

@article{nekkanti2023large,
  title={Large-scale streaks in a turbulent bluff body wake},
  author={Nekkanti, Akhil and Nidhan, Sheel and Schmidt, Oliver T and Sarkar, Sutanu},
  journal={Journal of Fluid Mechanics},
  volume={974},
  pages={A47},
  year={2023},
  publisher={Cambridge University Press}
}

@article{welch1967use,
  title={The use of fast {Fourier} transform for the estimation of power spectra: a method based on time averaging over short, modified periodograms},
  author={Welch, Peter},
  journal={IEEE Transactions on Audio and Electroacoustics},
  volume={15},
  number={2},
  pages={70--73},
  year={1967},
  publisher={IEEE}
}

@article{ghira2026non,
  title={Non-dimensional dissipation at strong unsteady transitions in isotropic turbulence},
  author={Ghira, Afonso Avelar and Elsinga, Gerrit E and Da Silva, Carlos Bettencourt},
  journal={Journal of Fluid Mechanics},
  volume={1028},
  pages={A46},
  year={2026},
  publisher={Cambridge University Press}
}

@article{goto2015energy,
  title={Energy dissipation and flux laws for unsteady turbulence},
  author={Goto, Susumu and Vassilicos, J Christos},
  journal={Physics Letters A},
  volume={379},
  number={16-17},
  pages={1144--1148},
  year={2015},
  publisher={Elsevier}
}

@article{nidhan2020spectral,
  title={Spectral proper orthogonal decomposition analysis of the turbulent wake of a disk at {$Re= 50 000$}},
  author={Nidhan, Sheel and Chongsiripinyo, Karu and Schmidt, Oliver T and Sarkar, Sutanu},
  journal={Physical Review Fluids},
  volume={5},
  number={12},
  pages={124606},
  year={2020},
  publisher={APS}
}

@article{lumley1967structure,
  title={The structure of inhomogeneous turbulent flows},
  author={Lumley, John Leask},
  journal={Atmospheric Turbulence and Radio Wave Propagation},
  pages={166--178},
  year={1967},
  publisher={Nauka}
}

@article{chongsiripinyo2020decay,
  title={Decay of turbulent wakes behind a disk in homogeneous and stratified fluids},
  author={Chongsiripinyo, Karu and Sarkar, Sutanu},
  journal={Journal of Fluid Mechanics},
  volume={885},
  pages={A31},
  year={2020},
  publisher={Cambridge University Press}
}

@article{vandine2018hybrid,
  title={Hybrid spatially-evolving {DNS} model of flow past a sphere},
  author={VanDine, Alexandra and Chongsiripinyo, Karu and Sarkar, Sutanu},
  journal={Computers \& Fluids},
  volume={171},
  pages={41--52},
  year={2018},
  publisher={Elsevier}
}

@article{liu2024effect,
  title={Effect of rotation on wake vortices in stratified flow},
  author={Liu, Jinyuan and Puthan, Pranav and Sarkar, Sutanu},
  journal={Journal of Fluid Mechanics},
  volume={999},
  pages={A44},
  year={2024},
  publisher={Cambridge University Press}
}

@book{tennekes1972first,
  title={A First Course in Turbulence},
  author={Tennekes, Hendrik and Lumley, John Leask},
  year={1972},
  publisher={MIT Press}
}

@article{uberoi1970turbulent,
  title={Turbulent energy balance and spectra of the axisymmetric wake},
  author={Uberoi, Mahinder S and Freymuth, Peter},
  journal={The Physics of Fluids},
  volume={13},
  number={9},
  pages={2205--2210},
  year={1970},
  publisher={American Institute of Physics}
}

@article{obligado2016nonequilibrium,
  title={Nonequilibrium scalings of turbulent wakes},
  author={Obligado, Martin and Dairay, T and Vassilicos, John Christos},
  journal={Physical Review Fluids},
  volume={1},
  number={4},
  pages={044409},
  year={2016},
  publisher={APS}
}

@article{shen2000anisotropy,
  title={The anisotropy of the small scale structure in high {Reynolds} number {($R_\lambda \sim$1000)} turbulent shear flow},
  author={Shen, X and Warhaft, Z},
  journal={Physics of Fluids},
  volume={12},
  number={11},
  pages={2976--2989},
  year={2000},
  publisher={American Institute of Physics}
}

@book{frisch1996turbulence,
  title={Turbulence: The Legacy of A.N. Kolmogorov},
  author={Frisch, Uriel},
  year={1995},
  publisher={Cambridge University Press}
}

@article{johansson2002proper,
  title={Proper orthogonal decomposition of an axisymmetric turbulent wake behind a disk},
  author={Johansson, Peter BV and George, William K and Woodward, Scott H},
  journal={Physics of Fluids},
  volume={14},
  number={7},
  pages={2508--2514},
  year={2002},
  publisher={American Institute of Physics}
}

@article{berger1990coherent,
  title={Coherent vortex structures in the wake of a sphere and a circular disk at rest and under forced vibrations},
  author={Berger, Eberhard and Scholz, D and Schumm, Michael},
  journal={Journal of Fluids and Structures},
  volume={4},
  number={3},
  pages={231--257},
  year={1990},
  publisher={Elsevier}
}

@article{fuchs1979large,
  title={Large-scale coherent structures in the wake of axisymmetric bodies},
  author={Fuchs, Helmut V and Mercker, Edzard and Michel, Ulf},
  journal={Journal of Fluid Mechanics},
  volume={93},
  number={1},
  pages={185--207},
  year={1979},
  publisher={Cambridge University Press}
}

@article{portela2017turbulence,
  title={The turbulence cascade in the near wake of a square prism},
  author={Alves Portela, F and Papadakis, G and Vassilicos, J C},
  journal={Journal of Fluid Mechanics},
  volume={825},
  pages={315--352},
  year={2017},
  publisher={Cambridge University Press}
}

@article{portela2018turbulence,
  title={Turbulence dissipation and the role of coherent structures in the near wake of a square prism},
  author={Alves Portela, F and Papadakis, G and Vassilicos, J C},
  journal={Physical Review Fluids},
  volume={3},
  number={12},
  pages={124609},
  year={2018},
  publisher={APS}
}

@article{burattini2005normalized,
  title={On the normalized turbulent energy dissipation rate},
  author={Burattini, Paolo and Lavoie, Phillipe and Antonia, Robert Anthony},
  journal={Physics of Fluids},
  volume={17},
  number={9},
  year={2005},
  publisher={AIP Publishing}
}

@article{kinjangi2023characterization,
  title={Characterization of energy transfer and triadic interactions of coherent structures in turbulent wakes},
  author={Kinjangi, Dinesh Kumar and Foti, Daniel},
  journal={Journal of Fluid Mechanics},
  volume={971},
  pages={A7},
  year={2023},
  publisher={Cambridge University Press}
}

@article{thiesset2013scale,
  title={Scale-by-scale turbulent energy budget in the intermediate wake of two-dimensional generators},
  author={Thiesset, F and Antonia, RA and Danaila, L},
  journal={Physics of Fluids},
  volume={25},
  number={11},
  year={2013},
  publisher={AIP Publishing}
}

@article{thiesset2014dynamical,
  title={Dynamical interactions between the coherent motion and small scales in a cylinder wake},
  author={Thiesset, F and Danaila, L and Antonia, RA},
  journal={Journal of Fluid Mechanics},
  volume={749},
  pages={201--226},
  year={2014},
  publisher={Cambridge University Press}
}

@article{baj2017interscale,
  title={Interscale energy transfer in the merger of wakes of a multiscale array of rectangular cylinders},
  author={Baj, Pawel and Buxton, Oliver RH},
  journal={Physical Review Fluids},
  volume={2},
  number={11},
  pages={114607},
  year={2017},
  publisher={APS}
}

@article{kankanwadi2020turbulent,
  title={Turbulent entrainment into a cylinder wake from a turbulent background},
  author={Kankanwadi, Krishna S and Buxton, Oliver RH},
  journal={Journal of Fluid Mechanics},
  volume={905},
  pages={A35},
  year={2020},
  publisher={Cambridge University Press}
}

@article{sreenivasan1998update,
  title={An update on the energy dissipation rate in isotropic turbulence},
  author={Sreenivasan, Katepalli R},
  journal={Physics of Fluids},
  volume={10},
  number={2},
  pages={528--529},
  year={1998},
  publisher={American Institute of Physics}
}

@book{townsend1976structure,
  title={The Structure of Turbulent Shear Flow},
  author={Townsend, A A},
  year={1976},
  edition = "Second", 
  publisher={Cambridge University Press}
}

@article{nedic2013axisymmetric,
  title={Axisymmetric turbulent wakes with new nonequilibrium similarity scalings},
  author={Nedi{\'c}, J and Vassilicos, John Christos and Ganapathisubramani, Bharathram},
  journal={Physical Review Letters},
  volume={111},
  number={14},
  pages={144503},
  year={2013},
  publisher={APS}
}

@article{dairay2015non,
  title={Non-equilibrium scaling laws in axisymmetric turbulent wakes},
  author={Dairay, Thibault and Obligado, Martin and Vassilicos, John Christos},
  journal={Journal of Fluid Mechanics},
  volume={781},
  pages={166--195},
  year={2015},
  publisher={Cambridge University Press}
}

@incollection{george1989self,
  title={The self-preservation of turbulent flows and its relation to initial conditions and coherent structures},
  author={George, William K.},
  booktitle={Advances in Turbulence},
  editor={George, W. K. and Arndt, R.},
  pages={39--73},
  year={1989},
  publisher={Hemisphere New York}
}

@article{vassilicos2015dissipation,
  title={Dissipation in turbulent flows},
  author={Vassilicos, J Christos},
  journal={Annual Review of Fluid Mechanics},
  volume={47},
  number={1},
  pages={95--114},
  year={2015},
  publisher={Annual Reviews}
}

@article{burattini2005scale,
  title={Scale-by-scale energy budget on the axis of a turbulent round jet},
  author={Burattini, Paolo and Antonia, RA and Danaila, L},
  journal={Journal of Turbulence},
  volume = {6},
  number = {},
  pages={N19},
  year={2005},
  publisher={Taylor \& Francis}
}

@article{hill2002exact,
  title={Exact second-order structure-function relationships},
  author={Hill, Reginald J},
  journal={Journal of Fluid Mechanics},
  volume={468},
  pages={317--326},
  year={2002},
  publisher={Cambridge University Press}
}

\end{document}